\documentclass[a4paper,fleqn]{cas-sc}

\usepackage{lscape}
\usepackage{float}
\usepackage{lineno}
\usepackage[authoryear,longnamesfirst]{natbib}

\def\tsc#1{\csdef{#1}{\textsc{\lowercase{#1}}\xspace}}
\tsc{WGM}
\tsc{QE}
\tsc{EP}
\tsc{PMS}
\tsc{BEC}
\tsc{DE}

\begin{document}
\let\WriteBookmarks\relax
\def\floatpagepagefraction{1}
\def\textpagefraction{.001}
\shorttitle{Orbit of the comet in AD 760 just from historical sources}
\shortauthors{D.L. Neuh\"auser et~al.}

\title [mode = title]
        {Orbit determination just from historical observations\,? \\
        Test case: the comet of AD 760 is identified as 1P/Halley}
%
%

\author[1]{D.L. Neuh\"auser}


\address[1]{Independent scholar, 39012 Meran/o, Autonome Provinz Bozen, S\"udtirol/Provincia Autonoma di Bolzano, Alto Adige, Italy}

\author[2]{R. Neuh\"auser}
\ead{ralph.neuhaeuser@uni-jena.de}
\cormark[1]

\author[2]{M. Mugrauer}


\address[2]{Astrophysikalisches Institut und Universit\"ats-Sternwarte Jena, Universit\"at Jena,
        Schillerg\"a\ss{}chen 2-3, 07745 Jena, Germany}

\author[3]{A. Harrak}

\address[3]{Department of Near and Middle Eastern Civilizations,
        University of Toronto, 4 Bancroft Avenue, Toronto, Ontario, Canada M5S 1C1}

\author[4]{J. Chapman}

\address[4]{Department of East Asian Studies, New York University, New York NY, 10003, United States}

\cortext[cor1]{Corresponding author}
%

\begin{abstract}
Recent advances in techniques of critical close reading of historical texts
can now be applied to records 
of pre-telescopic celestial observations -- 
allowing significant progress for analyzing and solving orbits of past comets:
here, as a blueprint test case, we exemplify our method by solving the orbit of the comet 
in AD 760 {\it only} with historical observations and {\it then} identify it with 1P/Halley.
A detailed eyewitness record with drawing of a comet in AD 760 in the Syriac 
{\it Chronicle of Zuqn}\textit{\={\i}}{\it n} (finished AD 775/6) was not yet included in the study 
of its orbit -- the Chinese reports alone do not yield a sufficient number of dated positions. 
We analyze the Syriac and Chinese sources with critical methods for quantitative astronomical usage,
we also consider a few further records from the Mediterranean and West Asian area. 
With our conservatively derived dated positions we can determine 
the best fitting Keplerian orbital solution 
by least squares fitting 
yielding the orbital elements 
($\chi^{2}_{\rm red}$ {\textless} 2 based on 1 million runs);
the parameter ranges for non-periodic solutions and highly eccentric periodic solutions are consistent with each other.
The allowed parameter ranges for perihelion distance and inclination
are sufficiently small to identify the comet with 1P/Halley.  
Although 1P/Halley is the only comet, where the telescopic orbit is credibly linked to pre-telescopic returns,
e.g. to AD 760, our identification confirms
claims from extrapolating telescopic observations backward in time --
here independently based on historical data.
In particular, we obtained a precise perihelion time (AD 760 May $19.1 \pm 1.7$).
The inferior conjunction between comet and Sun as on the previously published orbit (AD 760 May 31.9, Yeomans \& Kiang 1981) 
is shifted by about one day compared to 
our new orbit (June 1.8), only the new one is consistent with the last observation (June 1.0) 
before conjunction as reported in the {\it Chronicle of Zuqn}\textit{\={\i}}{\it n}.
Such a precision would be most critical for studying non-gravitational forces on comets.
By studying the comet's brightness evolution, we also compute its absolute brightness and activity parameter for AD 760
and found indications that the comet was quite dusty that year.
As the last return before a close encounter with Earth in AD 837,
the AD 760 perihelion is particularly important for extrapolation further back in time
(at AD 837 and 800, Yeomans \& Kiang 1981 had to introduce
corrections in their standard orbit).
Our improved methods developed in a multidisciplinary cooperation offer possibilities also to solve more 
orbits and to identify more comets from the rich and widespread pre-telescopic transmissions.
\end{abstract}




\maketitle

\section{Introduction}


Historical (pre-telescopic) observations were utilized for studying cometary orbits
until a few decades ago (e.g. Kiang 1972, Yeomans \& Kiang 1981, and Stephenson \& Yau 1985 for 1P/Halley, 
also Marsden et al. 1993 for
109P/Swift-Tuttle).
The last years and decades, however, have seen strong advances in both editions
of historical sources from various civilizations,
and in methods of historical-critical close reading;
this enables very literal translation by taking into account that
reports, in particular those written by professional court astronomers, 
are composed by special technical terms.
In-depth source- and text-critique
led to a much improved understanding of historical observations, which we can
use for modern astrophysical applications (see the proceedings of Focus Meeting 5
on this topic during the International Astronomical Union General Assembly 2018,
e.g. R. Neuh\"auser et al. 2020).
Therefore, it is now timely to revisit not only previously studied texts,
but also those which became newly or better available, in order to solve comet
orbits from historical observations alone.
Here, we apply our methods to the comet of AD 760,
to determine its orbit with a sufficient number of dated positions -- 
without assuming which comet it was and without linking it to other perihelia by a period.
Previously, only by extrapolating telescopic data backward, it was found  
that the comet of AD 760 should be 1P/Halley (see Sect. 1.1 for details).
Halley's comet is particularly useful to demonstrate our technique --
as independent evaluation of the backward extrapolated orbit.
Our methods offer possibilities also to identify other historical comets not yet
linked (or with as yet uncertain links) with telescopically observed returns,
and also, if not enough historical positions are available, to identify them
by linking to backward extrapolated well-known comets.

First, let us briefly exemplify some problems that were encountered in previous attempts, e.g.,
109P/Swift-Tuttle: based on the 1862 observations alone, the comet was expected to return around 1981.  
The possibility of identity with the comet of 1737 was briefly mentioned by Lynn (1902) 
and fully investigated by Marsden (1973), who wrote that if Swift-Tuttle was not 
recovered by late 1983, searches should resume for it in 1992 on the assumption
that the 1737 comet was Swift-Tuttle (predicted T = 1992 Nov $25.3 \pm 2$ months).  
The comet was rediscovered at a $\Delta$T of +17 days.
Marsden et al. (1993) considered further historical returns (see also Yau et al. 1994).
With the best fit period of $\sim$125 yr (Marsden et al. 1993), 
the identification with reported sightings in AD 188 and BC 69 remains uncertain for a number of reasons: \\
(i) for AD 188 July 28, the Chinese source reports a ``guest star as large
as a vessel with capacity of 3 pints'' (Ho 1962), the whole report is not typical for comets,
no duration is given, it could just be
a fireball (Kronk 1999, p. 44, quoting R. Stothers); \\
(ii) for BC 69 Aug 20 and 27, the Chinese source gives only two very rough positions
for a ``guest star'',
not far from each other, within few days, not sufficient for an
independent orbital solution with a period of some 125 yr (Marsden et al. 1993); \\
(iii) numerically integrating the 1992 orbit back in time shows that 
the comet had a closer approach to Earth in AD 698,
at least for certain values of the non-gravitational parameter A$_{2}$,
so that orbital parameters may have changed significantly,
e.g. by AD 442 about $\pm 1$ yr perihelion time --
in AD 188 the approach is much closer, ``but the -68 (BC 69 return) can be made to fit'',
(Marsden et al. 1993); \\
(iv) only the ``acceptance of an observed perihelion time during July-Sep 188
essentially restricts'' the effect of non-gravitational forces (Marsden et al. 1993); \\
(v) for BC 69, the overall orbit yields a minimum separation of 0.62 au,
which may be too large for naked-eye detection (Marsden et al. 1993). \\
Hence, only by {\it assuming} that the objects of BC 69 and AD 188 were the
same comet as the one seen in 1737, 1862, and 1992, one can arrive at an overall orbital 
solution; this solution surprisingly would not need any non-gravitational forces
(Marsden et al. 1993), maybe in contradiction to the strong activity of Swift-Tuttle.

In Hasegawa (2002) orbits for ten comets are determined --
except AD 1554, none were derived only from the historical positions,
but by other assumptions from the roughly reported comet path.
Of those ten, there are three from the first millennium, e.g. AD 839:
Hasegawa (2002) does not reflect on the dating problems in the Chinese transmission 
(see Ho 1962 and Pankenier et al. 2008).
While the presumable comet path is shown in his figure 2, 
he does not distinguish between the various textual specifications of the two comet locations given --
and he also does not determine or consider the measurement uncertanties (point coordinates without error bars).
He also does not consult other, e.g. European, observations existing for this comet. 

When Hasegawa \& Nakano (2003) consider to identify comet 153P/Ikeya-Zhang
with historical comets, they assume the orbital elements
from linking the 2002 and 1661 perihelia and then extrapolate backward to AD 877.
The sources from Japan and Europe are brief and contradict each other;
the late European source (no comet in contemporaneous texts) is used to fix the duration, 
but they assume that the ``name of the constellation [Libra] would be mistaken''. 
The position of the ``guest star'' derived from the Japanese source is questionable:
it was not considered whether the given ``(Dong-)Bi'' means the asterism or the lunar mansion.
The Japanese text is not included any more in the new compilations of guest stars
in Xu et al. (2000) or of comets in Pankenier et al. (2008); 
e.g., Schove (1984) considered this object as potential nova. 

Let us now introduce our approach:
a strong radiocarbon variation around AD 775 motivated us to survey celestial observations for
some hundred years around this time, in order to find possible sources of its origin. To distinguish between various
transient phenomena and to recognize and avoid misidentifications, we developed methods and clear criteria (see, e.g.,
R. Neuh\"auser \& D.L. Neuh\"auser 2015a and D.L. Neuh\"auser et al. 2018a, 2018b on sunspots and aurorae borealis;
more general in R. Neuh\"auser et al. 2020,
see also www.astro.uni-jena.de/index.php/terra-astronomy.html). 
We studied in detail not only solar activity proxies
(R. Neuh\"auser \& Hambaryan 2014, D.L. Neuh\"auser \& R. Neuh\"auser 2015, R. Neuh\"auser \& D.L. Neuh\"auser 2015b),
but also comets in the AD 760s and 770s,
e.g. it was suggested that the $^{14}$C and $^{10}$Be enhancement observed for around AD 775 could have
been delivered by a comet impact on Earth in early AD 773, but we have shown that this claim was based
on an incorrect translation of the historical Chinese text (see Chapman et al. 2014, 2015).

As important
eyewitness source, we noticed the Syriac {\it Chronicle of Zuqn}\textit{\={\i}}{\it n} (already mentioned by
Dall'Olmo 1978 and Schove 1984), which features northern lights, comets, and other celestial phenomena in the decades
until AD 776 (Harrak 1999, D.L. Neuh\"auser et al. 2018b, 2018c); we discussed its aurorae (text and drawings)
and fixed their dating to AD 772 and 773 in R. Neuh\"auser \& D.L. Neuh\"auser (2015a); Hayakawa et al. (2017) considered
only observations with drawings, however with questionable identifications and interpretations 
(see footnotes 10 and 17).

The West Asian {\it Chronicle of Zuqn}\textit{\={\i}}{\it n} (finished AD 775/6) includes an eyewitness report with drawing of a
comet for AD 760, now known to be the perihelion passage of comet 1P/Halley; this is complementary
to and at least as valuable as the transmitted records from China. By literal technical rendering (new translation) of the
historical transmissions from Syriac (West-Aramaic) and Classical Chinese as well as text-critical analysis and
close reading, we might obtain more precise and additional observing dates and positions. Such a new set of dated
positions with conservative error bars may enable for the first time to solve all orbital elements for AD 760 just from
the historical observations. Since the East-Asian observations alone were not sufficient to solve for all elements,
previous works had to assume or extrapolate some orbital elements from other perihelia
probably belonging to the same comet (e.g. Laugier 1846); more recent work extrapolated the 
telescopically observed orbits to the past and then tried to fix the perihelion time with
historical observations (e.g. Kiang 1972, Yeomans \& Kiang 1981, henceforth YK81), Sect. 1.1. 
All previous work stressed the importance of historical 
observations, in particular perihelion times,
to determine orbital elements 
and to study non-gravitational forces (e.g. Sitarski 1988).

Historical observations of comets are useful to reconstruct their orbital
elements, which in turn are needed to study
their secular variations, connections between comets and meteor showers, to identify additional sightings in the
past, and to predict future appearances. According to the previous orbital
reconstruction, the comet had its closest
approach to Earth (0.0334 au) on AD 837 Apr 11, the closest for the last few millennia (YK81).
The motion of 1P/Halley is highly sensitive to the circumstances of that encounter.
Hence, the study of the orbit of AD 760, the next earlier perihelion
before AD 837, is also important in this regard. 

A comparison of the Syriac with the Chinese observations can also help
improving the understanding of the transmitted records, and it is also relevant for various aspects of the history of
astronomy, e.g. how the observations were performed, which precision was achieved, whether and which precession
constant was applied. Here, we re-determine the AD 760 perihelion orbit only from historical observations. 
In a future publication, we will then also
re-consider the observations in AD 837 (in prep.). With very precise reconstructions of several perihelia, 
one can study the influence of non-gravitational effects and cometary activity on the orbit
just with historical observations.

\smallskip

First, we introduce comet 1P/Halley and its previous orbital reconstructions (Sect. 1.1) as
well as the historical sources used, mainly the Syriac {\it Chronicle of Zuqn}\textit{\={\i}}{\it n} (1.2) and
Chinese compilations about celestial phenomena (1.3). 
Then, in the context of our improved methods, we present our new technical translations from {\it Zuqn}\textit{\={\i}}{\it n}
and China, discuss the sources and the text in detail (Sects. 2.1-2.5 and 3.1), and obtain dated positions
from the critically evaluated material (Sects. 2.6 and 3.2). 
A few other observations from the East Mediterranean and West Asia are examined more briefly (Sect. 4). We solve
for the orbit just with historical observations of AD 760 with least squares fitting, 
present the intermediate results -- we can confirm that it is comet 1P/Halley (Sect. 5);
we also discuss our results in comparison with previous work.
We finish with a consideration about the precession constant as was used implicitly in AD 760 
in the {\it Chronicle of Zuqn}\textit{\={\i}}{\it n} (Sect. 6) as well as a summary and future perspective (Sect. 7).
Some of the results of this work were first presented at the IAU General Assembly 2018
(Focus Meeting 5) by D.L. Neuh\"auser et al. (2018c) and Mugrauer et al. (2018).

\subsection{Previous orbit reconstructions of comet 1P/Halley}

The first periodic (1P) comet, for which an orbital solution was found, is 1P/Halley.
Edmund Halley (1705) noticed that three of the comets for which he had computed orbits -- namely those in
AD 1531, 1607, and 1682 -- had similar orbits and had appeared at roughly 75 years.  He assumed
them to be the same object, made a rough calculation of the planetary perturbations from 1682 onwards 
and predicted the comet's return in 1758 (Halley 1749).
Mostly by extrapolating the orbital period (and/or elements) from
telescopic observations backward, it was noticed that all perihelia of the last two millennia were recorded in East
Asia, some only marginally (e.g. Kiang 1972, YK81, Stephenson \& Yau 1985, see also Kronk 1999). Observations of perihelion
passages also include Babylonian observations in the last two centuries BC (Stephenson et al. 1985, see also Landgraf 1986).

The orbital elements change from perihelion to perihelion due to perturbations by solar system
bodies (e.g. Cowell \& Crommelin 1908 for AD 837 and 760) and non-gravitational acceleration (cometary
outgasing), e.g. the period changes between 74 and 79 yr for the last 29 returns (YK81). Reconstructions of those
osculating orbital elements were done by studying telescopic observations and extrapolating backward -- fixed by the
perihelion time from historical observations. 

For the comet of AD 760, Pingr\'{e} (1783) 
lists just the Chinese observations and also summarizes
the short report by Theophanes (see our Sect. 4a). With the orbital period of 1P/Halley as 75.3 yr, Laugier (1846)
calculated backward from AD 1152, expected its return around AD 775, and identified the comet of AD 760 with
1P/Halley, because some orbital elements were consistent with the presumable observational dates and positions recorded
in AD 760 -- he did not have sufficient historical data from AD 760 to solve for the orbit. (When Cowell \& Crommelin
1908 wrote that ``Laugier {\dots} identified the apparitions of (AD) 451 and 760 from the observations alone'', one could
misunderstand that Laugier would have used only historical observations from AD 760 to identify the comet of 760 with
1P/Halley.) Then, Hind (1850), also by assuming a certain orbital period, studied several more perihelia in detail and
solved for orbital elements; regarding AD 760, he fully agreed with Laugier (1846) including the perihelion date on AD
760 June 11. This date was also accepted by Cowell \& Crommelin (1908), who linked observations from 2-3 perihelia into
one orbital solution; planetary perturbations from AD 837 Feb 25 calculated backward with the variation-of-elements
technique resulted in a perihelion of AD 760 June 15 (period 28013 days), so that the significant changes in orbital
elements over centuries are mainly due to perturbations by planets, 
while the {\it ``combined effect [of non-gravitational causes] does not amount to more than a week per revolution''}
(Cowell \& Crommelin 1908, p. 514).

Kiang (1972) extrapolated backward the orbit again with the variation-of-elements method,
rectified his calculated perihelion passage times with revisited Chinese observations, and obtained perihelion passages
on AD 837 Feb $28.27 \pm 0.05$ (very small error bar due to large motion on sky at close approach to Earth) and 760
May 22.5 (and many others back to BC 240). Hasegawa (1979) corrected some observing dates in Kiang (1972) by one
calendar day, see below; he obtained perihelion times of AD 760 June $5 \pm 4$ and 837 Feb $28.15 \pm 0.1$.

Yeomans (1977) extrapolated backward the orbit by numerical integration, but only back to AD
837, because of a very close approach with a planet, namely Earth (0.04 au) in AD 837. His work was revised and
extended back to BC 1404 by YK81, limited again by a close approach to Earth in BC 1404 by
0.03 au; the perihelion times of the two recent millennia were fixed by historical observations with partly revised
data from Kiang (1972); for AD 760, they obtained, e.g., a period of 77.00 yr and a perihelion on May 20.67, which is
1.83 days earlier than the May 22.5 date obtained by Kiang (1972); note that YK81
did not revise the Kiang (1972) data on the AD 760 perihelion. 

YK81 noticed that the best-fitting orbit found in Yeomans (1977), namely orbit no. 2 in table 3 in YK81,
did not work sufficiently well for AD 837, so that they use a different one, orbit no. 3 (table 3 in YK81).
This means that one of the non-gravitational parameters (considered to be constant in time)
were forced to change instantaneously by YK81 at AD 837, see table 3 in YK81.
Furthermore: {\it ``Before the integration of orbit 3 was continued backward, the osculating perihelion passage
time was given an empirical correction of $-0.88$ day at epoch 2026840.5 (JD)''}, i.e. AD 837 Mar 14 (YK81, p. 641).
YK81 also had to do {\it ``an empirical adjustment of the osculating eccentricity''}
at epoch AD 800 (used for technical reasons instead of AD 837), so that orbit no. 3 works 
for the perihelia before AD 800 (YK81, pp. 641-642). 

Sitarski (1988) wrote that YK81 ``had to make some subjective changes in 
orbital elements for AD 837 when the comet closely approached Earth'';
he parameterized the non-gravitational forces as a secular change in the semi-major axis ($da/dt$) and obtained
$da/dt$ for 24 orbits from AD 1986 to BC 87. Twenty five historically observed perihelion times (from YK81) were
used to fit a parabolic function $da/dt(t)$; however, without the two data points in the first two centuries AD,
the fit would probably be very different (constant or sinusoidal), see his figures 1 \& 2;
Sitarski (1988) also presents a solution with $da/dt(t)$ being constant (e.g., his table 5). 
The acceleration parameter $da/dt$ has small error bars back to the 10th century 
(smaller than $0.52 \cdot 10^{-8}$), 
but is always larger than $(1.3-3.1) \cdot 10^{-8}$ 
for the 3rd to 9th century
AD (Sitarski 1988). With the parabolic fit to $da/dt$, Sitarski (1988) obtained as perihelion times, e.g., AD 837 Feb
28.31 and 760 May 20.53 (our perihelion date is AD 760 May $19.1 \pm 1.7$).

In table 5 in Sitarski (1988), one can see that both the non-gravitational parameters as well as the
difference between computed and (presumably) observed perihelion passage time show particular large changes at AD 912,
the largest from AD 1986 to BC 86. 
For the return in AD 912, it is not even certain whether and which of two short transmissions on a comet 
from China and Japan pertains to 1P/Halley (Xu et al. 2000), so that extrapolations back in time are also more
uncertain due to the problem of identifying the correct report.

YK81 stress that perihelion passage times are best to fix orbital elements in backward extrapolations;
for doing so, they use historical observations of one or two more or less accurate dated positions
in AD 141, 374, and 837, when the comet was closest to Earth.
In our paper, we can present for the first time a very well
observed perihelion time (to within $\pm 1.7$ days) for AD 760 observed by the author of the 
Syriac {\it Chronicle of Zuqn}\textit{\={\i}}{\it n} recorded in his eyewitness report
(while the Chinese records give a precise dated position only for their last observation of the comet
several weeks after perihelion passage).
The passage in AD 760 is of particular importance for further backward extrapolations because
of the uncertainties in AD 912, the close approach in AD 837, and also the orbit changes 
enforced by YK81 for around AD 800 and 837.

For more details on the history of orbital solutions of 1P/Halley, see, e.g., YK81 and Yeomans et al. (1986). 
The most recent orbital elements for many perihelia are published
in Marsden \& Williams (2008), based on YK81 (and then precessed to J2000.0), which we use below for comparison.

For the next sections, in particular the determination of dated positions and a new orbit, we
do not need to assume that the comet of AD 760 is 1P/Halley.

\subsection{The 8th century Syriac Chronicle of Zuqnin as source for a comet in AD 760}

For an 8th century manuscript, the Syriac {\it Chronicle of Zuqn}\textit{\={\i}}{\it n} is exceptional:
the single manuscript that exists is very likely the autograph,
i.e. the original manuscript hand-written by the author (Harrak 1999); 
it includes a detailed report with drawing of a comet in AD 760 with stars and planets nearby (Fig. 1).

The {\it Chronicle of Zuqn}\textit{\={\i}}{\it n} offers a world history starting with `creation' as in the Bible and
ending at around the time of writing, AD 775/776. It survived in one manuscript of 173 folios located as Codex
Zuqninensis at the Vatican Library (Vat. Syr. 162), and the remaining six folios are found in the British Library (Add.
14.665 folio 2-7); in Codex Zuqninensis, 129 folios are palimpsest, one even a double-palimpsest (Harrak 1999). Some of
the folios in the British Library which cover the last years are partly worm-eaten and very fragmentary. Its first and
last folios are lost together with the name of the author (Harrak 1999). The Chronicle is divided into four parts, all
translated to English (Harrak 1999, 2017) and French (Chabot 1895). 

The author of the chronicle was
probably the stylite monk Joshua (Harrak 1999); a stylite is an early Byzantine or Syrian Christian ascetic living and
preaching on a pillar in the open air, so that many celestial observations can be expected in his work. The author of
the {\it Chronicle of Zuqn}\textit{\={\i}}{\it n} may have lived on a pillar for some time (Harrak 1999).
During the time of writing of the
{\it Chronicle of Zuqn}\textit{\={\i}}{\it n},
the area was outside the border of the Byzantine empire and already under $^{c}$Abbasid rule.

The {\it Chronicle of Zuqn}\textit{\={\i}}{\it n} is not known to be copied and disseminated; 
sometime during the 9th century it was transferred to the Monastery of the Syrians in the Egyptian desert;
see Sect.\,4b for a possible use by Nu$^{c}$aym ibn \d Hamm\={a}d. Shortly after the manuscript was found and bought for
the Vatican, it was considered to be written by the West Syrian patriarch Dionysius I of Tell-Ma\d hr\=e, so that
this chronicle was long known as Chronicle of Dionysius of Tell-Ma\d hr\=e (Assemani 1719-1728). Dionysius did write
an otherwise lost world chronicle, but lived later (died AD 845). Since this mistake was noticed, the chronicle has
been called the Chronicle of Pseudo-Dionysius of Tell-Ma\d hr\=e (Chabot 1895, Abramowski 1940) or, better, the
{\it Chronicle of Zuqn}\textit{\={\i}}{\it n} (Harrak 1999),
because the text mentions the monastery of {\it Zuqn}\textit{\={\i}}{\it n} as the living 
place of the author; {\it Zuqn}\textit{\={\i}}{\it n} was located near
Amida, now Diyarbak{\i}r in Turkey near the border to Syria.

The {\it Chronicle of Zuqn}\textit{\={\i}}{\it n} is made of four parts: Part I runs from the creation to Emperor
Constantine (AD 272-337), Part II from Constantine to Emperor Theodosius II (AD 401-450) plus a copy of the so-called
{\it Chronicle of Pseudo-Joshua the Stylite} (AD 497 to 506/7), Part III from Theodosius to Emperor Justinian (AD 481-565),
and Part IV to the time of writing, AD 775/776. The Chronicler used a variety of sources, some of them otherwise lost
(Harrak 1999, 2017). The author knew that some of his sources did not provide a perfect chronology;
for him, it is more important to convey his message (to learn from history) than to give perfect datings.

The events reported in the text are dated using the Seleucid calendar; the Seleucid Era (SE)
started on 
October 7, BC 312 (= Dios 1). 
There are several versions of the Seleucid calendar, including the Babylonian
(Jewish), Macedonian, and West Syrian (Christian) ones. The author of our chronicle systematically used the latter
version for reports during his lifetime -- a solar calendar, in which the year ran from
Tishri/October 1 to Elul/September 30, applied since at least the fifth century AD (Hatch 1946).

The {\it Chronicle of Zuqn}\textit{\={\i}}{\it n} reports about a variety of celestial phenomena, which can be
classified as northern lights, meteor showers, meteorites, a bolide, comets, halo displays, a solar eclipse,
and other (atmospheric) darkenings. Observations of auroral (with drawings for AD 502, 772, and 773) as well as
meteoric phenomena (showers, meteorites, bolide) were published before by R. Neuh\"auser \& D.L. Neuh\"auser (2015a) and
D.L. Neuh\"auser et al. (2018b).\footnote{Hayakawa et al. (2017) cited some of the celestial
observations in this chronicle, including the comet observation in AD 760, but in an unsatisfactory manner: they
misinterpreted the Chronicle's text and drawing, they left out the Chinese observations, and they did not attempt to
fit a new orbit (for details, see footnotes 10 and 17).} Eyewitness reports from the
author of the Chronicle or sources close to him start around folio 128 in AD 743 and end in AD 775/6, probably shortly
before the death of the author (Harrak 1999). The reports are more extended for the last years of the Chronicle.
For his lifetime, he has reported from source material (e.g. letters,
Easter tables, etc.) and, even for the very detailed comet report discussed here, probably not only from his own memory
as eyewitness (years later), but also from written notes. 

\subsection{Understanding historical observations from China}

In imperial China, court astronomers observed the sky all day and night
in order to notice changes;\footnote{Such observations were performed, because it was thought that
they identify dangerous political trajectories (astrology, but also weather rules etc., e.g. from the
{\it Han} dynasty: ``320 stars can be named.
There are in all 2500 ... All have their influence on fate'', Needham \& Wang 1959, p. 265), or can indicate misgovernment (``any
anomalous happenings in nature ... were construed as signs of warnings by heaven toward the misbehaviour or
misgovernment of the ruler of man'', also from the {\it Han}, Wang Y\"u-chuan 1949, Bielenstein
1984). The dramatic appearance of comet Halley in BC 12, for
example, was interpreted by both Gu Yong and Liu Xiang as a sign that the Western {\it Han}
dynasty was in danger of collapse; the two writers each identified different court factions as responsible for the
peril the dynasty faced, and both held that if the right actions were undertaken the sign would vanish and the dynasty
would likely survive; neither writer saw the future as fixed or determined, though both associated it with an elevated
likelihood of disastrous political events (Chapman 2015).} owing to this
practise -- among other transients -- comets were recorded in observing
logs. While the original night reports for the 8th century are not extant, later compilations or copies thereof
are available, which are shortened and may suffer from scribal errors.
These include: Jiu Tang shu (JTS) by Liu Xu et al. from AD 945, Tang hui yao (THY) by Wang Pu et al. from AD 961, 
and Xin Tang shu (XTS) by Ouyang Xiu et al. from AD 1061, i.e. the astronomical chapters of the {\it History of the Tang} dynasty
({\it Tang shu}), as well as the collection Wenxian tongkao (WHTK) by Ma Duanlin from AD 1317. 
Extracts for comets were published
by Pingr\'{e} (1783) in French as well as by Hsi (1957, only for 1P/Halley AD 837), 
Ho (1962, see also Hasegawa 1980 for comments and additions), Kiang (1972), Xu et al. (2000),
and Pankenier et al. (2008), all in English.

For general information about astronomy in imperial China,
please refer to the detailed monographs by Needham \& Wang (1959) and
Sun \& Kistemaker (1997, henceforth SK97), and short summaries also in Kiang
(1972), Clark \& Stephenson (1977), Stephenson (1994),
Xu et al. (2000), Stephenson \& Green (2002), 
and Pankenier et al. (2008).

Since the {\it Han} dynasty (BC 206 to AD 220), 
the sky was structured into about 283 asterisms of various sizes with almost 1500 stars in total (down to
6th mag and a few fainter ones, these are of course incomplete); 
a Chinese asterism\footnote{Groups of stars ({\it xing cang}) were given certain
names, which do not normally reflect their appearance on sky, even if connected with skeleton lines; this is similar
for Babylonian, Western, and Chinese constellations. To discriminate from Western constellations, Chinese star groups are
often called {\it asterism}. However, this term derives from the Greek {\it asterismos} as was used by Ptolemy in his
Almagest for what we now call {\it constellations} (now defined as fields on sky by IAU mostly based on Ptolemy's Almagest). 
{\it Xing qun} is the modern Chinese term for constellation; 
literally, it means {\it group of stars}.} can contain one, few or many stars; the stars of an asterism were
combined by lines (skeletons). While this system had a strong continuity since the
{\it Han}, some details
changed later (not only in Korea and Japan, also in China). 

The term {\it xing}, often
rendered as {\it star(s)}, can be combined to, e.g. {\it ke xing}
as {\it guest star(s)} or {\it hui xing} as {\it broom star(s)}. 
Classical Chinese word morphology does not distinguish between singular and plural.

The names of 28 asterisms are also used for the 28 lunar mansions (LM), which are right ascension ranges
from the determinative (or leading) star of one LM to the next,
omitting the south polar region which was not visible from the Chinese mainland,
while the north circumpolar region was of special importance
known as the enclosure ({\it yuan}) named {\it `Ziwei'} or {\it `Zigong'},
see Stephenson (1994), SK97, and Ho (2003, p. 144).
For a list of the 28 LMs and their determinative stars, see, e.g., SK97, Xu et al. (2000), 
Stephenson \& Green (2002), or Pankenier et al. (2008). 
Given this equatorial system, hour angles of objects can be given as a certain number 
of {\it du} ($0.9856^{\circ}$) East of the respective determinative star.

There also exist Chinese star charts from the time of the {\it Tang} dynasty, 
namely the Dunhuang maps (manuscript Stein 3326 dated AD 649-684 by style of characters, mentioning of an
astronomer of that time, style of clothing shown in a figure, and usage of two taboo characters), where more than 1300
stars in 257 asterisms are drawn with skeleton lines, apparently in azimuthal projection (Bonnet-Bidaud et al.
2009).\footnote{Stars and asterisms on the 13 charts are drawn only in a crude
way with rough positions and several mistakes, e.g. the asterism name
{\it Lou} in Aries is missing (but the three stars apparently are drawn), 
the colour-convention for stars is not followed strictly (Chinese charts show the stars and
asterisms from three {\it Han} dynasty
schools in different colour: red for those from Shi Shen, black from Gan De, and white/yellow from Wu Xian), twice the
Chinese characters for ``right'' {\it zuo} and ``left'' {\it you}, which are very similar, are
mixed up, the asterism {\it Sangong} near
the pole is shown twice (Bonnet-Bidaud et al. 2009). Given that the maps are drawn on expensive pure mulberry fibres
(3940 mm by 244 mm scroll), this atlas may be a copy produced by a wealthy but not well-talented student of Li
Chunfeng, one of the main astronomers of the 7th century, 
who is mentioned in the accompanying text and could have done the (now lost) original map based on observations
and/or the astronomical chapters of the {\it Jin shu}, which he had written.}

Separations on sky including comet tail lengths are given in
certain old Chinese linear measures, which can
be converted to angles such as 1 {\it chi} being about $1^{\circ}$
(Stephenson \& Green 2002; Kiang 1972 gave 1 chi = $1.50 \pm 0.24^{\circ}$),
1 {\it cun} being 0.1 {\it chi}, and 1 {\it zhang} $=$ 10 {\it chi} 
(see Ho 1966, Kiang 1972, Wilkinson 2000, Stephenson \& Green 2002).

Sometimes, in addition to or instead of a celestial position given as one coordinate, angle, or separation, 
the compilations of observing records list the general direction as azimuth, which can be specified in terms 
of several different compasses; the precision of the compass used (e.g. 4- or 24-point) then defines the 
uncertainty or azimuth range of such a position.  

The observing dates are specified by name of the emperor, year with a
multi-year reign period, lunar month, and then
usually the day count in a 60-day-cycle ({\it ganzhi}) -- a continuous counting
was achieved prior to the advent of the imperial period in 221 BC; sometimes, instead of or in addition to the day
count (1-60), the age of the Moon is given; the luni-solar calendar had 12 lunar months starting on the second new-moon 
after winter solstice (i.e. in January or February), plus seven intercalary months in 19 years (called just
``x-th intercalary month'' located after the ``x-th'' month), like the Meton cycle; 
these rules were in use since a calendar reform during the {\it Han}.

The normal Chinese 24h-day ran from midnight to midnight, but in astronomical records, for observations after midnight, the former date is
given (some late sources may have modified the date
to the new civil date). The night was separated into five {\it watches} of equal lengths per night,
which changed during the year.

\section{The comet of AD 760 as reported in the Syriac Chronicle of Zuqnin}

\subsection{Syriac text and text-critique}

In the {\it Chronicle of Zuqn}\textit{\={\i}}{\it n} for SE 1071, i.e. AD 759 Oct~1 to AD 760 Sep 30 (part IV,
folio 136v with drawing and caption, see Fig. 1 with the Syriac hand-writing), 
we find the following report (brackets from us), here our own literal translation
(line breaks by us); 
the author of the Chronicle used different kinds of punctuation,
all marked in the appendix with the transliterated Syriac text --
below, his rhombs and colons are given as in the Syriac text,
while for his bold points we mostly give full stops and for his weak points we mostly give comma,
following English language rules (in the Syriac autograph, it is sometimes difficult to
distinguish between a bold and a weak point):
\begin{quotation}
``The year [SE] one thousand seventy one (AD 759/760). \\
In the month of {\it iyy\=or}
(May)\footnote{Chabot (1895), Harrak (1999), and Hayakawa et al. (2017) read
``\=ad\=ar/\=od\=or'' and gave ``March'' here (``\=od\=or'' is the correct West-Syrian transliteration here, while
``\=ad\=ar'' is East-Syrian); in Syriac, the words for March ({\it \=od\=or}) and May
({\it iyy\=or}) are written very similar:
{\it `DR} and {\it `YR}, respectively. We came to the conclusion that
{\it iyy\=or} is given here in the MS: (a)
epigraphically, the Syriac letter /d/ (as in
{\it \=od\=or}) should have a tail, which is
not found in the MS, (b) there is no space between /y/ and the following /r/, the two letters are ligatured, but if it
were /d/ (as in {\it \=od\=or}) there should
be a space (as seen in all occurrences of this letter in the month name 
{\it `DR} = {\it \=od\=or}), and (c) because of a dot
underneath the /y/, the letter was thought to be /d/, i.e. reading {\it `DR} = {\it \=od\=or}, 
however, in five occurrences of the month name {\it `YR} in the
MS, four do not have this diacritical dot, one (folio 150v) has it as a thick one, which should be thin -- the
chronicler was by no means consistent in using diacritics and symbols. Michael the Syrian also gives
{\it iyy\=or} as month of the first sighting (Sect. 4d).}
a white sign was seen in the sky, \\
before early twilight (Syriac: {\it \v{s}afr\=o}), in the north-east [quarter], \\
in the Zodiac [sign] which is called Aries ({\it emr\=o}), 
to the north from these three stars ({\it kawkb\=e}) in it, which are very shining. \\
And it resembled in its shape a broom, 
while it was still in the same Aries
({\it emr\=o}) at its edge/end/furthest part ({\it r\=\i\v{s}eh})\footnote{The
Syriac word {\it r\=\i\v{s}eh} mainly means ``its head'', but ``its tip, its edge, its end, its furthest part'' 
etc. and such meanings are also attested in
dictionaries (e.g. Sokoloff 2002). See below for a discussion of position 2.}: \\
in/at the initial degree ({\it m\=ur\=o})\footnote{The Syriac {\it m\=ur\=o} from
Greek {\it moira} for {\it degree} is also attested in Ptolemy's Almagest for
{\it degree}.}
[of] the second\footnote{Harrak (1999) gave ``in
the first degree (of the Zodiacal circle), the second''; 
Hayakawa et al.: ``in the first
degree (of the sign), two (degrees)''; see below for a discussion of position 2.}
[sign] (i.e. Taurus) 
from these wandering stars ({\it kawkb\=e}), Kronos (Saturn) and Ares (Mars), like somehow 
a bit to the south, 
on [day] 22 in the same month. \\
And the sign itself remained for fifteen nights, until dawn
({\it n\=ogah})\footnote{Chabot: ``la veille''; respectively ``eve'' in Harrak (1999); 
the comet was seen in the morning, as mentioned before; for {\it n\=ogah}, see footnote 11.}
of the feast of Pentecost. \\
And [at] its one end/tip
({\it r\=\i\v{s}\=oh}), the narrow one, a very bright star
({\it kawkb\=o}) was seen at its head/end/tip
({\it r\=\i\v{s}eh})\footnote{An
alternative translation could be ``and its one end/tip, the narrow one, was very bright; a
star was seen at its head/end/tip'', but it does not work because in the MS there
is a punctuation between {\it qa\d t\textit{\={\i}}n\=o} (``narrow'')
and {\it yat\textit{\={\i}}r bah\=ur\=o} (``very bright''). Hayakawa et al. (2017) brings a
punctuation in their transliteration that is in many places inconsistent with the autograph, 
in particular they overlooked the punctuation by translating 
``And one end of it was narrow and duskier, one star was seen in its tip'', 
and they confused the meaning by rendering ``duskier'' instead of ``very bright'': 
the original word {\it bah\=ur\=o}
means ``dim'' in old Syriac, but later also ``bright'' after Arabic influence;
Chabot (1933) emended {\it bah\=ur\=o} into {\it n\=oh\=ur\=o},
which just means ``bright'', but this emendation is not
necessary; the first letters (/b/ and /n/) are also quite different in Syriac. The translation by Hayakawa et al.
(2017) is not satisfactory: ``duskier'' would be in contrast to the ``star'' at this end
(comet head), and it would not be in contrast to what is later given as ``wide and very dark'' (the other end); the
drawing also clearly shows a ``very bright star'', the comet head; see below for our discussion of the drawing.}. 
And it was tilting to the north side, 
but the other wide and very dark one was tilting to the south side, \\
and it was going bit by bit to the North-East [direction]. \\
Its shape is as follows [now 4
points forming a rhomb meant as pointing to the drawing,
which is embedded in the next lines,~Fig.~1]. \\
However, at
the beginning ({\it n\=ogah})\footnote{For the Syriac {\it n\=ogah}, instead
of ``beginning'', Hayakawa et al. (2017) gave ``dusk'', which is
not attested in Syriac dictionaries; the word {\it n\=ogah}
does mainly mean ``dawn'' (see above), but this is not possible here, because the observation was in the
``evening time''. Harrak (1999) gave ``eve''. Our translation ``at the beginning'' follows
oriental calendars, where the 24h-day begins with sunset, e.g.
{\it n\=ogah d-shapt\=o} meaning ``Sabbath vespers'',
which happen in the evening after sunset. 
In the report on a bolide in AD 754, the {\it Chronicle of Zuqn}\textit{\={\i}}{\it n} gave 
the timing as ``on Tuesday, when Wednesday was dawning ({\it n\=ogah}) ... In the same evening ...'', 
i.e. it uses {\it n\=ogah} here for the beginning of the oriental 24h-day 
(D.L. Neuh\"auser et al. 2018b, event 5, p. 77, Harrak 1999, p. 196).}
of [the] third [day] after Pentecost, it was
seen again at evening time, from the north-west [quarter] \\
and it remained for twenty-five evenings. \\
And it was going bit by bit to the south:: [actually 4
points forming a rhomb meant here as a break] \\
And it again disappeared. \\
And then it returned [and] was seen in the
south-west\footnote{Lit. west southern} [quarter], \\
and thus there it remained for many days.''
\end{quotation}
Here, our own very literal
technical translation ends, we continue with the translation from Harrak (1999, p. 198):
\begin{quotation}
``During this time, many schisms took
place in the church because of leadership. The eastern monasteries made John Patriarch, while neither the cities of the
{\it Jaz\textit{\={\i}}ra} nor all the monasteries approved him. The people of the West and Mosul approved George. Because of this the
entire Church became troubled.''
\end{quotation}

Several paragraphs later, in year SE 1075, AD 763/4, there is more text on this comet (Harrak 1999, pp. 200-1):
\begin{quotation}
``A severe plague among horses took place in the whole land. ... This disease spread throughout all the
nations and kingdoms of the earth, to the point that people were left without horses. The effect of 'the broom' seen a
short while before, was clearly seen in reality, as it swept the world like a broom that cleans the house.''
\end{quotation}

\subsection{Drawing}

A drawing (folio 136v, our Fig. 1) embedded in the text shows the ``broom'',
which was first called ``white sign'': to the left, we can see it with a small
circle (``very bright star'') and cone-like lines directed away from it to the upper right, more
``narrow'' closest to the star-circle; neither the space between these lines nor the
star-circle are filled: the impression is that the sign is here brighter than at its wider end 
(``wide''), where the space between the lines are mostly filled with ink (``very
dark''). The objects drawn are the observed comet with tail (left part), the three brightest stars of
Aries ($\alpha$, $\beta$, $\gamma$ Aries, shown in position relative to each other as on sky) in the centre,
shown as three empty circles (labelled {\it emr\=o} for
Aries), and the planets Mars and Saturn to the right (labelled {\it Ares} and {\it Kronos}, respectively), also drawn as empty
circles (apparently to indicate them as sources of light) -- 
all roughly aligned, probably meant to be at about the same altitude above horizon. In the text, the Syriac
word {\it kawkb\=o} is used for each of the six empty circles,
i.e. for the three different kinds of objects (comet head, three fixed stars, and two planets), so that
{\it kawkb\=o} stands here in general for ``celestial
object'' appearing round (similar to the Arabic {\it kawkab}), all these different objects were considered as
some kind of {\it star}. This drawing is also shown on the front cover of the English translation (Harrak 1999), and it
is the only figure that was also mentioned and redrawn in the French translation (Chabot 1895),
see also footnote 17.

The record in our Chronicle is probably the author's eyewitness report, because the
drawing is embedded in the text, labelled by the same hand, philological arguments (same terms), 
and because other sources are not mentioned (otherwise, it often gives the source).
Since the drawing compares well with the real situation on sky in particular regarding Saturn and Mars
relative to each other and
relative to the stars of Aries (see Fig. 2), we can date the drawing to about AD 760 May 25 (early morning).
Saturn and Mars had a close conjunction in the night of AD 760 May 22/23 with a separation of only $\sim 40^{\prime}$.
Our new orbit (Fig. 2) shows the position of the comet for May 25, where the angular separation between the comet and
$\alpha,\beta,\gamma$ Ari is quite similar to the angular separation between $\alpha,\beta,\gamma$ Ari and the planets Mars and Saturn --
similar as drawn in the Chronicle. While the text has for the first detection ``a white sign'' (May 18),
it is stated for May 22 ``it resembled in its shape a broom''. Regarding the depiction it is said: ``Its shape is as follows''.
Indeed, we clearly see ``a broom'' ``tilting to the south side'' (while ``going ... to the north-east'').
Thus, the scenario is fitting well, but we did not use this drawing for the orbit reconstruction.
The text information is more precise.

\begin{figure}[ht]
\begin{center}
\includegraphics[angle=0,width=13cm]{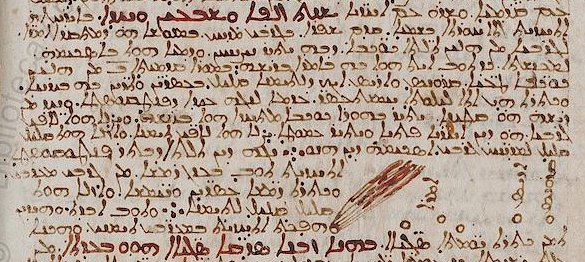}
\end{center}
\caption{
{\bf Syriac text and drawing:}
{\rm The relevant Syriac text from the {\it Chronicle of Zuqn}\textit{\={\i}}{\it n}
(finished AD 775/6) on the comet AD 760 (1P/Halley) --
from the middle of the first line shown to the middle of the last line --
with a drawing embedded in the text (Vatican Library, Vat. Syr. 162, folio 136v):
the comet to the left,
the three brightest stars of Aries ($\alpha$, $\beta$, and $\gamma$ Aries) in the center,
and the planets Mars and Saturn as {\it Ares} and {\it Kronos} to the right,
as identified in the Syriac caption.
The drawing fits best for around May 25 given the relative position of Ares/Mars east (left) of Kronos/Saturn,
both west of Aries.
See Fig. 2 for a comparison with a computed position of 1P/Halley for May 25 at 0h UT.}
}
\end{figure}

\begin{figure}[h]
\begin{center}
\includegraphics[angle=0,width=13cm]{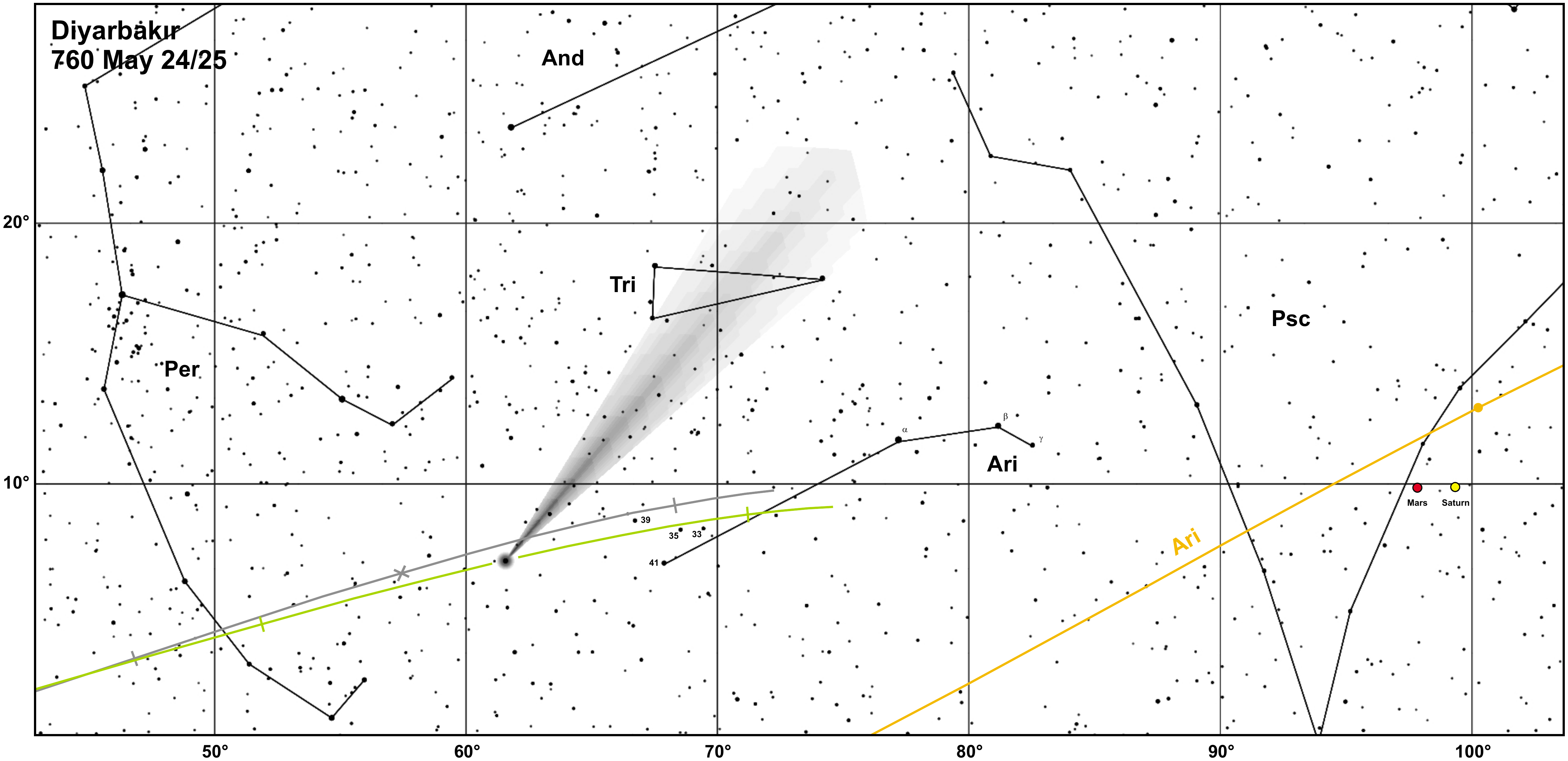}
\end{center}
\caption{
{\bf A comparison with Fig. 1 shows that the original drawing is for a date
around AD 760 May 25 in the early morning} {\rm (2:40h local time, 0h UT):
both Sun (NE) and Moon were under the horizon at that time, 
1P/Halley was $\sim 7^{\circ}$ above the NE horizon.
IAU constellations are indicated in black, the ecliptic in orange with dots at
the borders of zodiacal signs.
The planets Mars (0.7 mag, red dot) and Saturn (0.5 mag, yellow dot) are still close to each other. 
The position of the comet is indicated
on our own new (green) orbit (and as grey cross on the old orbit, JPL, YK81); both orbits start on May 17.
The drawing (Fig. 1) was not used for orbital reconstruction.
Here and in all other figures, the comet plasma tail pointing away from the Sun is displayed,
while observation and drawing regard the dust tail.
This and all other such figures are drawn with Cartes du Ciel (v3.10).}
}
\end{figure}

\subsection{Dating}

The text reports the ``white sign'' to be seen ``for
fifteen nights, until dawn of the feast of Pentecost'', i.e. on Pentecost Sunday. For the year given (SE
1071, i.e. AD 759/760), most Christian churches (including churches under the Byzantine Patriarchate) celebrated Pentecost on AD 760 May
25, but some eastern churches celebrated Pentecost one week later on June 1. Hence, 15 nights earlier is definitely
May, so that in any case, it is obvious that the {\it Chronicle of Zuqn}\textit{\={\i}}{\it n} meant May here (Syriac
{\it iyy\=or}), see footnote 5.

The reason for the two different Pentecost (and Easter) dates in AD 760 is the difference between two
ecclesiastical Easter calendars: in AD 760, the first computed (cyclic) full moon after the start of spring (defined
for March 21 at the AD 325 Council of Nicaea) was on Saturday Apr 5 according to the 532-year 
cycle\footnote{19 years Meton cycle $\times$ 4 years leap year cycle $\times$ 7 days per week are 532 years,
with the number of days in 532 Julian solar years being identical to 235 average synodic months
within less than one day, so that Easter falls onto the same date and weekday after 532 years.}
constructed by Irion in AD 562 for the Byzantine Patriarchate
(based on a previous 200-year cycle by Andreas of Byzantium for AD 353-552), 
so that the Byzantines celebrated Easter on Apr 6 
(like also the Roman church following the 532-year Easter calendar by Dionysius Exiguus starting in AD 532), 
while the Armenian, Jacobite, and Nestorian churches followed a different
532-year Easter table, namely the Armenian scholar Anania Sirakaci's (AD 610-685) 
reform (early AD 660ies) of Andreas' Easter table, 
according to which the paschal full moon in AD 760 would be on Sunday Apr 6, 
so that Easter has to be dated Apr 13 (see Sanjian 1966, Mosshammer 2008, pp. 257-277).
This dispute is also
reflected in the {\it Chronicle of Zuqn}\textit{\={\i}}{\it n} (Harrak 1999): 
\begin{quotation}
``The year (SE) 1070: Lent was confused. Some of the Easterners
introduced Lent on the 18th of {\it \v{S}eb\=a\d t} (Feb)
and ended it on the 6th of {\it N\textit{\={\i}}s\=an} (Apr). Others
introduced Lent on the 25th of {\it \v{S}eb\=a\d t} (Feb)
and ended it on the 13th of {\it N\textit{\={\i}}s\=an} (Apr). All of the
Christians were confused, when in one place they celebrated Easter, in another place Palm Sunday; 
in one place it was Passion week, in another place Easter.''
\end{quotation}
(With the above expression ``some of the Easterners'' for the other churches,
our author probably refereed to the Byzantine Patriarchate or other churches west of the Euphrates.)
Our Chronicle reported the Easter dating problem for SE 1070, i.e. AD 758/759; 
in AD 759, Easter Sunday was on April 22, in AD 760 on April 6 or 13 (see above);
hence, the above given end date of lent (Apr 6 or 13) points to AD 760; 
the given introduction of lent on Feb 18 or 25 would be a Monday in 760, i.e. the correct weekday for
the start of lent in the Syriac churches (where there is no Ash Wednesday). 
There is also a brief mention of this problem by Theophanes, who dates it to AD 760.
Hence, all the evidence points to AD 760 for the report on the Easter dating problem
misdated to AD 759 in the Chronicle of {\it Zuqn}\textit{\={\i}}{\it n}.
The same problem also happened in AD 570 and 665 (Mosshammer 2008, pp. 276-277).\footnote{The {\it Chronicle of
Zuqn}\textit{\={\i}}{\it n} does not report any Easter dating problems for 
AD 570 nor 665;
this problem, called ``crazatik'' or ``Erroneous Easter'', was resolved only in AD 1824 (Mosshammer 2008, p. 277).
Our Chronicle narrates one other Easter confusion for SE 857 (i.e. Easter AD 546, 
but correct year is AD 547, see Mosshammer 2008, p. 256), 
when three different dates for lent and Easter are mentioned to have been followed by different parts of the
population. 
For a discussion of the Easter problem and Easter tables, see McCluskey (1998, pp. 84-87) 
and for the Eastern churches also Sanjian (1966) and Mosshammer (2008).}

The monastery of {\it Zuqn}\textit{\={\i}}{\it n} belonged to the Syriac Orthodox church, informally known as the Jacobite Church;
this is known, because our chronicler listed bishops and patriarchs, which were also listed by the 12th century 
Michael the Syrian (e.g. Chabot 1899-1910), who clearly identified them as to belong to the Syriac Orthodox patriarchate (Jacobite).
Hence, it is clear that Easter was on Apr 13 and Pentecost on June 1 
at the monastery of {\it Zuqn}\textit{\={\i}}{\it n}: since it is reported that the comet was seen ``for fifteen nights,
until dawn of the feast of Pentecost'', it was first detected on May 18 ``before early twilight''.
This is well consistent with the fact that the Chinese sources give May 17 for the first detection (Sect. 3.1).  

\subsection{The ``white sign'' as comet: criteria}

In transmitted texts on celestial transients, using a pheno-typical description, 
it is often uncertain which kind of celestial phenomenon is meant:
in our text the phenomenon is not called ``comet'', and even if it would be called that way,
it may still be uncertain whether a comet in today's sense is meant.
Five criteria are developed (timing, position/direction, color/form, motion/dynamics,
and duration/repetition) for various kinds of celestial phenomena,
see, e.g., R. Neuh\"auser \& D.L. Neuh\"auser (2015a) and D.L. Neuh\"auser et al. (2018a) for criteria for
aurora borealis and D.L. Neuh\"auser et al. (2018b) for meteor showers (and aurorae).

The ``white sign'' or ``broom'' reported in the Chronicle of {\it Zuqn}\textit{\={\i}}{\it n}
fulfils all five criteria for comets: \\
(i) timing, observed at night-time or twilight: ``before early twilight'', ``fifteen
nights'', ``at evening time'', ``twenty-five evenings'', and stars and planets are mentioned (and shown in the drawing); \\
(ii) Position of first and/or last sighting: often close to Sun, in or near the ecliptic:
``before early twilight, in the north-east'' 
``seen again at evening time, from the north-west'', 
and ``in the Zodiac [sign] which is called Aries ({\it emr\=o})'';
also tail direction away from the Sun: 
``[at] its one end/tip,
the narrow one, a very bright star ({\it kawkb\=o}) 
was seen at its head/end/tip. 
And it was tilting to the north side, but the other wide
and very dark one was tilting to the south side''; \\
(iii) colour and form (extension): ``white sign'', ``resembled in its shape a broom'', the
{\it white broom} points to the comet dust tail appearing white
due to reflection of sunlight (while the plasma tail would appear bluish and much fainter); \\
(iv) dynamics, i.e. moving on
sky relative to the stars: first ``north from these three stars'', ``it was
going bit by bit to the North-East'', seen until Pentecost (June 1 morning), then again soon later after
conjunction with the Sun, ``it was seen again {\dots} from the north-west'',
``it was going bit by bit to the south'', etc.; and \\
(v) duration: ``remained for fifteen nights'', ``remained for twenty-five evenings'', etc. \\
Furthermore, our Chronicler connects the sighting of this transient object as negative portent
with unfortunate events (e.g. ``many schisms''), as was not unusual at this time.

\subsection{Nomenclature of transient celestial objects}

The {\it Chronicle of Zuqn}\textit{\={\i}}{\it n} describes the object of AD 760 as ``white sign'' 
and as {\it kawkb\=o} (``star'') with or in the shape of a ``broom'' (for {\it kawkb\=o}, see Sect. 2.2), 
but it did not use the Syriac term {\it nayzk\=o} -- usually translated with ``comet'', 
literally meaning ``short spear'' or ``lance'';
maybe, the term used here by our Chronicler is motivated by the real form of the phenomenon 
on sky resembling more a {\it broom} than a {\it lance}.  

Our Chronicler also called an object reported for AD 768/9 (probably AD 770 May) 
``sign in the likeness of a broom'', also a comet in today's sense (Harrak 1999, pp. 226/7). 
For the 6th century, the {\it Chronicle of Zuqn}\textit{\={\i}}{\it n} describes three objects as both
{\it kawkb\=o} and {\it nayzk\=o} (Harrak 1999, p. 136, n. 5), for the first two it is explicitly mentioned that 
they are called ``kometes'' by the Greek (e.g. Harrak 1999, p. 93) -- 
the term ``kometes'' is taken from its source, the otherwise mostly lost Chronicle of John of Ephesus 
(based on John Malalas). 

The author of the {\it Chronicle of Zuqn}\textit{\={\i}}{\it n} 
should have noticed that all these objects are of the same class (comet in our sense) given similar drawings.
Greek terms like ``kometes'' may have been outmoded in the $^{c}$Abbasid caliphate, 
but acceptable when used in quotation. The terms {\it nayzk\=o} in Syriac, {\it nayzak} in Arabic, 
and ``kometes'' in Greek formerly all meant the same -- 
not only a comet in today's sense, but more generally a transient, extended celestial object;
bright supernovae were sometimes also called ``kometes'' or {\it nayzak}, as they appeared to be
extended due to strong scintilation, see R. Neuh\"auser et al. (2016). 

\subsection{Dated positions from the Chronicle of Zuqnin}

The observations of the {\it Chronicle of Zuqn}\textit{\={\i}}{\it n} were made near Amida, 
now Diyarbak{\i}r (Turkey) at longitude $40^{\circ} 13^{\prime}$ East and latitude $37^{\circ} 55^{\prime}$ North.
The Syriac record gives details regarding the comet path on sky (dated positions compiled in Table 1).

(Z1) On AD 760 May 18 in the morning, ``a white sign was seen in the sky, before
early twilight, in the north-east [quarter], in the Zodiac [sign] which is called Aries, to the north from these three
stars in it, which are very shining''. The Chronicler specifies that the ``white sign'' is seen ``in
the Zodiac [sign] {\dots} called Aries'', 
which is either the constellation figure of Aries or the ecliptic longitude range;
this does not need to be decided here,
because there is a more stringent constraint next: the ``three'' ``very shining'' ``stars'' in ``Aries''
are obviously ${\alpha}$, ${\beta}$, and ${\gamma}$ Ari (2.0 to 4.5 mag), which are also depicted in the drawing (Fig. 1). 
Regarding the drawing, our Chronicler's description is here clearly in the horizontal system,
``to the north'' means from ${\alpha}$,
${\beta}$, and ${\gamma}$ Ari toward the azimuth north at about the same altitude above horizon, 
being ``in the Zodiac [sign] ... called Aries''.
To specify that the comet appeared ``in the north-east'' is correct, 
since the Chronicler here means the whole NE quadrant ($0-90^{\circ}$) 
-- later he specified that the comet ``was going bit by bit to the North-East'',
which is more strictly the direction toward azimuth NE, $\sim 45^{\circ}$ (see Z3). 

\begin{figure}
\includegraphics[angle=0,width=16cm]{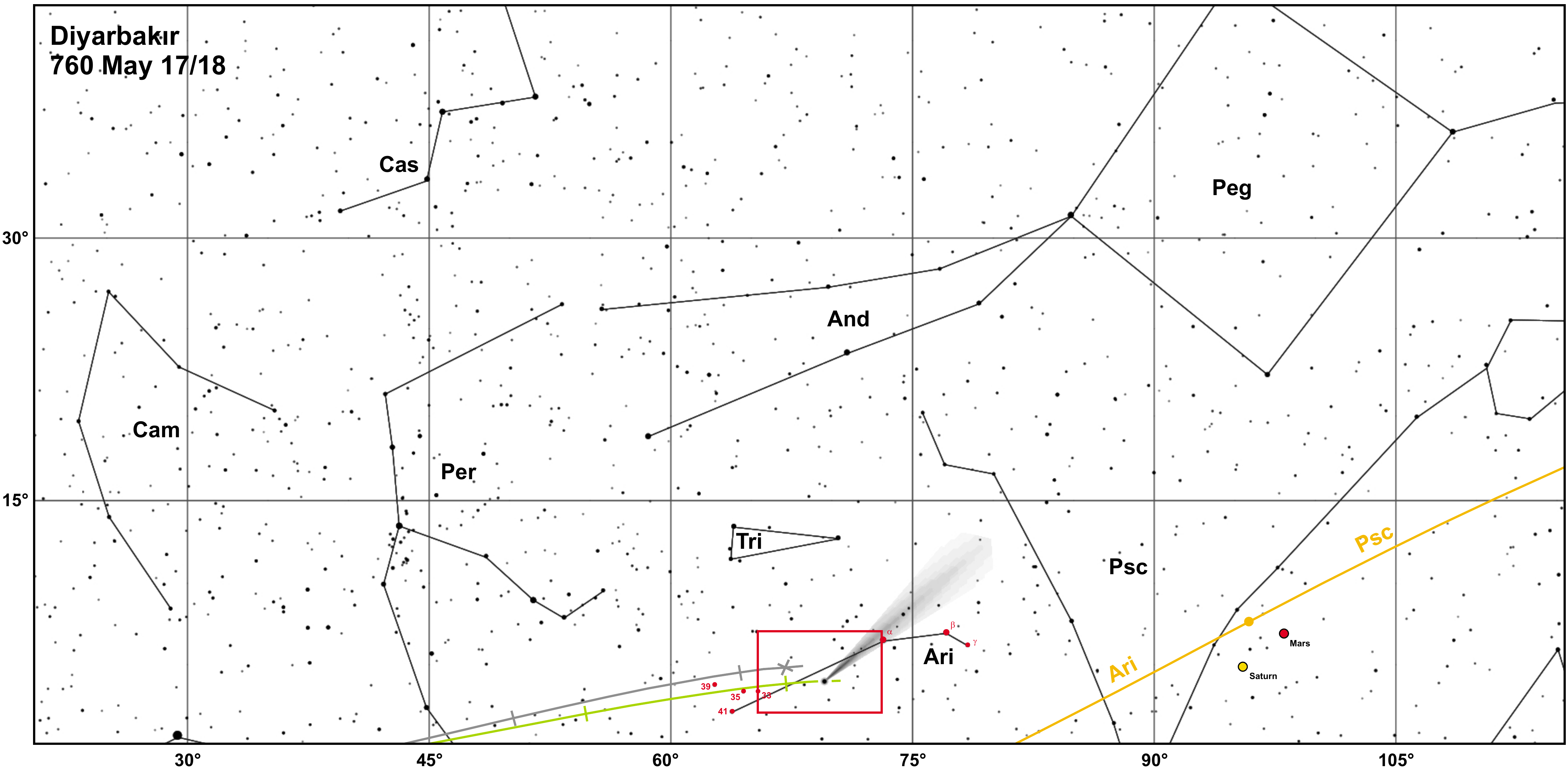}
\caption{
{\bf Horizon plot for Amida for AD 760 May 18 at 2:40h local time:}
{\rm The Chronicle of {\it Zuqn}\textit{\={\i}}{\it n} reported a
{\it white sign ... in ... Aries, to the north from these three stars in it, which are very shining ...
before early twilight},
the first dated position in Sect. 2 (Z1 in Table 1),
to the north of $\alpha, \beta, \gamma$ Ari on the horizontal system.
The comet with tail directed away from the Sun (in the NE below horizon) 
is indicated for May 18 at 0h UT on the new (green) orbit (as grey cross on the old orbit).
The expected positions on May 20.0 and 26.0 (UT) are indicated with green (and grey) tick marks.
Our positional error box is shown in red, the relevant stars in Aries as red dots with their names,
Mars as labeled red dot, Saturn yellow, and the ecliptic in orange.
}
}
\end{figure}

We can obtain the coordinate error box as follows: ``to the north from these three stars'' in
Aries means, in the Chronicler's horizontal system, an azimuth range from ${\alpha}$ Ari (the horizontally northernmost
star among ${\alpha}$, ${\beta}$, ${\gamma}$ Ari) to 33 Ari (the horizontally southernmost star among the set of stars
mentioned next: 33, 35, 39, 41 Ari, see below position 2); this gives an azimuth range of $65.3-73.1^{\circ}$. As altitude,
we use the full range from the lowest star of these two sets (41 Ari) to the highest (${\beta}$ Ari), $2-7^{\circ}$, see
Fig. 3. This converts then in certain ranges in right ascension and declination, used for our orbit fit.

The first detection of the comet was around the beginning of astronomical twilight or slightly
earlier (``before early twilight''), i.e. Sun $\sim 18^{\circ}$ below horizon; as uncertainty, 
we assume the time from astronomical to nautical twilight ($\pm 0.75$h centered around the beginning of astronomical twilight); 
at this time, the positional error box is rising above horizon.
We obtain as observing time May 18 at around 2:50h local time (0:10h UT), rounded to UT May 18.0 ($\pm 1$h).

\smallskip

(Z2) Later, on May 22, ``it was still in the same Aries at its edge/end/furthest
part''. Some translations give ``head'', the main meaning of the respective
Syriac word, but this is not the intended meaning here, because the comet was first near its three most shining stars,
which are clearly ${\alpha}$, ${\beta}$, and ${\gamma}$ Ari (see drawing and position Z1), usually considered the area
where the head of Aries is depicted.
However, since the Chronicler reported that the comet
``was going bit by bit to the North-East'' (i.e. toward azimuth $\sim 45^{\circ}$, Fig. 4) during the 15
nights since the first detection, and for May 22 that ``it was still in the same Aries'', the
observer must now mean the other end of Aries, namely 33, 35, 39, and 41
Ari\footnote{The relevant stars in Aries are all listed in Ptolemy's Almagest,
where only the ecliptic system for measurements and decriptions were used:
${\gamma}$ Ari as ``the more advanced of the two stars on the horn'' (Ptolemy: faint 3rd mag)
and ${\beta}$ Ari as ``the rearmost of them'' (3rd mag), the two first stars listed for
Aries, while ${\alpha}$ Ari is listed among the stars ``around Aries'', but still for
the same zodiacal sign, ${\alpha}$ Ari as ``the star over the head,
which Hipparch (calls) {\it the one on the muzzle}'' (bright 3rd mag); then also ``the four stars
over the rump'', namely 41 Ari as ``the rearmost, which is brighter (than the
others)'' (4th mag), 39 Ari as ``the northernmost of the other three, fainter
stars'' (5th mag), 35 Ari as ``the middle one of these three'' (5th mag), and 33
Ari as ``the southernmost of them'' (5th mag) (Toomer \& Ptolemy 1984, pp. 339-340).} (3.6-5.3 mag) -- the
{\it other} end compared to ${\alpha}$, ${\beta}$, ${\gamma}$
Ari.\footnote{Since the drawing (for a date around May 25) may imply that the comet head is roughly aligned with
${\alpha}$, ${\beta}$, and ${\gamma}$ Ari and the two planets Mars and Saturn, all at about the same
altitude, the star ${\delta}$ Ari cannot be meant, as it is at much lower altitude and even fainter than even
33 Ari.}

\begin{figure}
\includegraphics[angle=0,width=16cm]{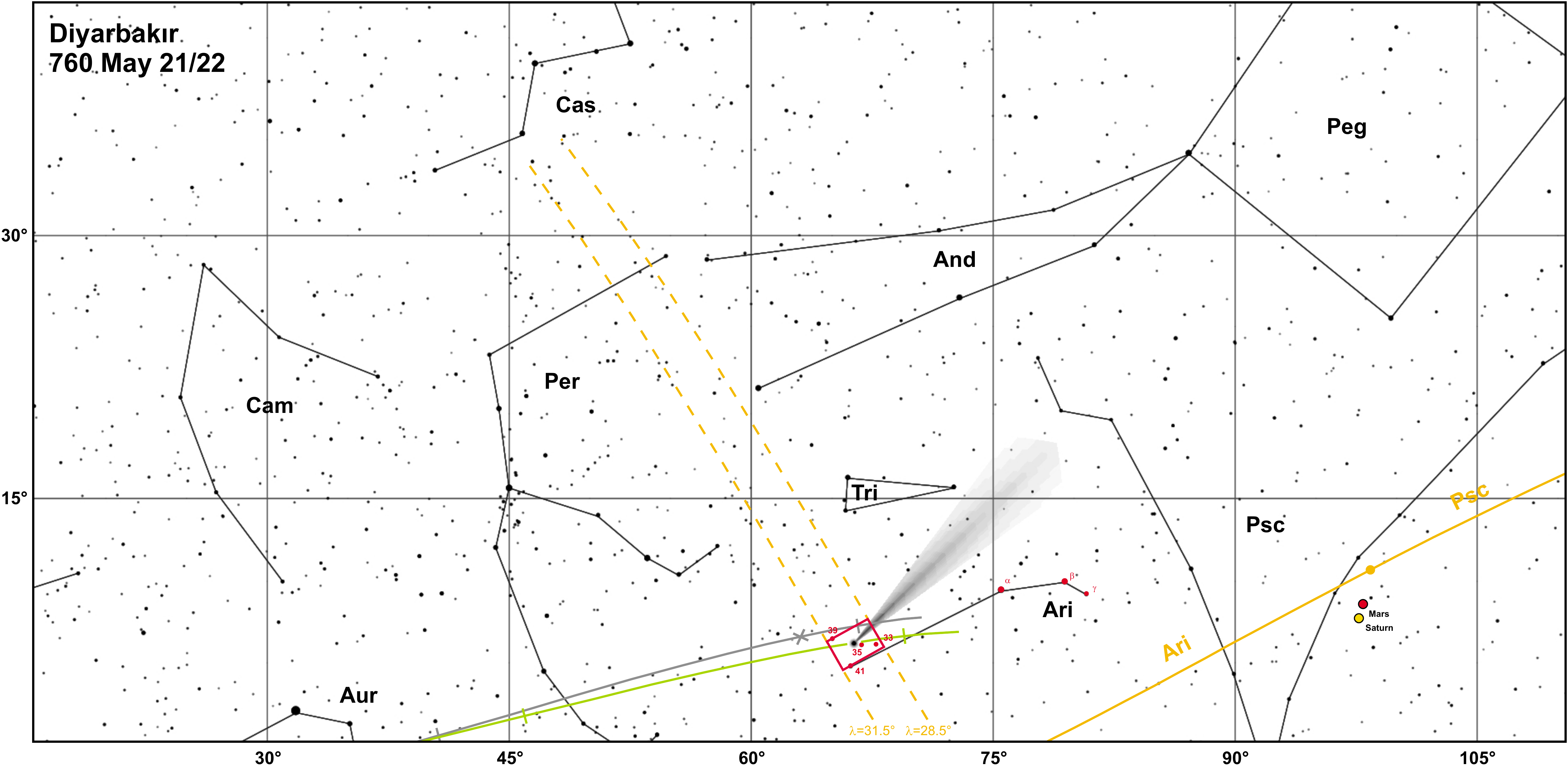}
\caption{
{\bf Horizon plot for Amida for AD 760 May 22 at 2:40h local time:}
{\rm The Chronicle of {\it Zuqn}\textit{\={\i}}{\it n} reported the comet to be 
{\it in/at the initial degree of} [the] {\it second} [sign] (i.e. Taurus) ...
{\it still in the same Aries at its edge/end/furthest part},
the second dated position in Sect. 2 (Z2 in Table 1).
This record constraints the comet position to be close to $30^{\circ}$ ecliptic longitude
(Taurus $0^{\circ}$) $\pm$ some uncertainty, estimated to be $\pm 1.5^{\circ}$
(from some other observations of the same author, see Sect. 2) --
indicated here by orange dashed lines with ecliptic longitude $\lambda$ given (ecliptic in orange).
The comet was at this longitude range and also still in Aries (at its end);
the end of Aries was built by 33, 35, 39, and 41 Ari, indicated as red error box,
while the head of Aries and its surrounding is made up by $\alpha$, $\beta$, and $\gamma$ Ari.
The comet is indicated for May 22.0 (UT) on the new (green) orbit (grey cross on the old orbit).
The expected positions on May 20.0 and 29.0 (UT) are indicated with green (and grey) tick marks.
}
}
\end{figure}

The meaning of the next wording may appear to be difficult: 
``in/at the initial degree [of] the second [sign] (i.e. Taurus) from these wandering stars,
Kronos and Ares, like somehow a bit to the
south''. The ``wandering stars, Kronos and Ares'' are Saturn and Mars, respectively,
which had a close conjunction in the night of AD 760 May 22/23 (a separation of only $40^{\prime}$ in ecliptic latitude when at the
same ecliptic longitude), when Mars overtook Saturn moving faster (from night to night) from the west to the east.
Given the Babylonian-Greek tradition, in which our Chronicler may be standing, moment and azimuth of rising are highly
significant -- and, indeed, ``somehow a bit to the south'' is already fulfilled just at their
rising: the azimuths of Mars and Saturn were just $1.5^{\circ}$ ``south'' of East on the horizontal system at their rising (so that we may
conclude that the Chronicler could obtain such a high positional precision); the two planets remained in the SE
quadrant until sunrise.

The wording ``still in the same Aries at its edge/end/furthest
part'' is descriptive meaning still in the constellation figure of Aries. 
The Syriac text then gives a colon indicating a more precise specification of the position:
``in/at the initial degree [of] the second [sign]'' points to the initial degree of the
ecliptic longitude range of the second zodiacal sign (or unit); given that the number zero was not yet available at
that time, ``in/at the initial degree'' of Taurus can be a longitude of around $0^{\circ}$ Taurus, i.e. an ecliptic longitude of
about $30^{\circ}$; the uncertainty range would be $\pm 1.5^{\circ}$, as obtained above for the positional precision of
the Chronicler; in Table 1, we then give an ecliptic longitude range of $\lambda=28.5-31.5^{\circ}$ (epoch of date).
(If the ``initial degree'' would instead be the longitude range Taurus $0-1^{\circ}$, this would be covered in our
uncertainty range.) The stars 33, 35, 39, and 41 Ari 
(``still in the same Aries at its edge/end/furthest part'') 
are located at about this ecliptic longitude for a correct
precession constant at around AD 760 (see Sect. 5 for a discussion of precession). 

The wording ``in/at the initial degree [of] the 2nd [sign] from these wandering stars, Kronos and Ares'' might mean that this
position is in the 2nd sign {\it after} the one with ``these wandering stars,
Kronos and Ares''; indeed, Saturn (ecliptic longitude 2$^{\circ}$) and Mars ($1^{\circ}24^{\prime}$)
were in the first zodiacal sign/unit, namely Aries, on May 22 (again showing the high positional precision of our Chronicler,
supporting the uncertainty assumed above);
in the sequence of the Zodiac, the comet was now in the 2nd sign ``from'' the planets, i.e. in Taurus (Fig. 4). 
It might be surprising that different systems (ecliptic, horizontal, and
descriptive) are used for specifying the position, but 
we find this also in the East Asian tradition, where positions are given descriptive and/or
with azimuth and/or with hour angles (equatorial), see Sect. 3.
The positions of 39 and 41 Ari (including 33 and 35 Ari) define the ecliptic latitude range ${\beta} = 10.4-12.4^{\circ}$
(Table 1, epoch of date).

This morning observation was on May 22 at around the start of astronomical twilight,
when the relevant objects were all visible at more than $5^{\circ}$ altitude; 
this is at 2:50h local time ($\pm 0.75$h as in position Z1, at the beginning of this time window,
our positional error box is rising), rounded to UT May 22.0 ($\pm 1$h).

\smallskip

The drawing, embedded into the text after the report on the morning sightings (before
reappearance in the evening) fits best for around May 25 given the relative position of Mars slightly east (left) of
Saturn (Figs. 1 and 2). Since the different separations between the objects shown in the drawing are not to scale, we
cannot derive exact coordinates of the comet from the drawing alone -- and we also cannot measure a realistic angular
tail length from the drawing. Hence, the drawing was not used for orbital reconstruction. 
In Figs. 2-8, a tail is shown by Cartes du Ciel directed away from the Sun like the plasma-tail,
while the drawing shows the (``white'') dust tail.

The mentioned ``tilting'' of the comet head toward the north and of
the tail toward the south, again meant in the horizontal system (``And [at] its one end/tip, the narrow
one, a very bright star was seen at its head/end/tip. And it was tilting to the north side, but the other wide and very
dark one was tilting to the south side'') is consistent with the comet path for those first 15
nights.\footnote{Hayakawa et al. (2017) interpret this description as two tails,
the plasma tail and the dust tail (in Hayakawa et al.'s section on their Drawing 6), but that is clearly wrong: at the
``end/tip'' that is given as ``the narrow one'', where ``a very bright star was seen'', is
obviously the comet head, and it ``was tilting to the north side'', correct for the head (not
for any tail): Aries was in the east (early morning around AD 760 May 25, Figs. 1 and 2), the comet head was pointing
toward the north side, while the tail was pointing (and drawn) toward the opposite direction (south side); the tail
is drawn roughly away from the Sun, which is about to rise at NE azimuth $61^{\circ}$ (observation {\it before early
twilight in the north-east}), as it should be both for a plasma ion tail and a dust tail blown by solar
wind (possibly with slightly different misalignments); given that the tail is described as
``white'', we deal here with the dust tail; Hayakawa et al. stated ``that 'one
end of it ... was turning to the north' sounds like the ion tail'', again wrong, because this observation
was in the morning in the East with the Sun about to rise from azimuth $61^{\circ}$ (NE), so that the ion tail must point away
from it to the south. Also, the drawing clearly shows one tail only, exactly as described in the text: at the head it
is narrow, and the opposite end (of the tail) is larger and wider (see Sect. 2.2). Furthermore, a comet with head and two tails has
three ends, while text and drawing show only two ends (head and one tail). The Hayakawa et al. interpretation is wrong
in respect of astronomy and philology as well as regarding the drawing. While Hayakawa et al. otherwise stress the
importance of drawings (and do not even consider those celestial observations in the Chronicle which come without
drawings), here they disregard and neglect the clear and fully realistic drawing and misinterpret the text. The
misinterpretation as {\it two-tailed comet} is even featured in the title of the Hayakawa et
al. paper as their main result.}
Indeed, as seen in Fig. 2, the comet head points toward
the northern half (``side''). That the tail is drawn and said to point toward the 
``south side'' would than mean that the tail is not strongly curved. 

\smallskip

(Z3) For the time until the morning of June 1 (``the sign itself remained for
fifteen nights, until dawn of the feast of Pentecost'', June 1), we are informed that the sign ``was going
bit by bit to the North-East''. This wording (``to the North-East'')
now implies a direction, while earlier the comet's appearance was described as seen ``in the north-east'' [quarter],
see position 1. Hence, the comet was last detected in the NE shortly after rising and shortly before sunrise, i.e. low
on the horizon. We conclude on an altitude of up to $15^{\circ}$ (Table 1, Fig. 5). Regarding the azimuth, NE is $45^{\circ}$ (east of
north), and since the Chronicler here uses an 8-point compass, the azimuth range is $45 \pm 22.5^{\circ}$. 
Since this is the last detection before it {\it ``became under the rays of the Sun''} (Michael of Syria, Sect. 4d),
the comet is in the early morning of June 1 already quite close to the Sun:
therefore, we restrict the observing time to the end of astronomical twilight at around 3:20h local time at the latest,
while the beginning of astronomical twilight at 2:40h is taken as the middle of the observing window,
so that we get June 1.0 ($\pm$ 0.7h) UT.

\begin{figure}
\includegraphics[angle=0,width=16cm]{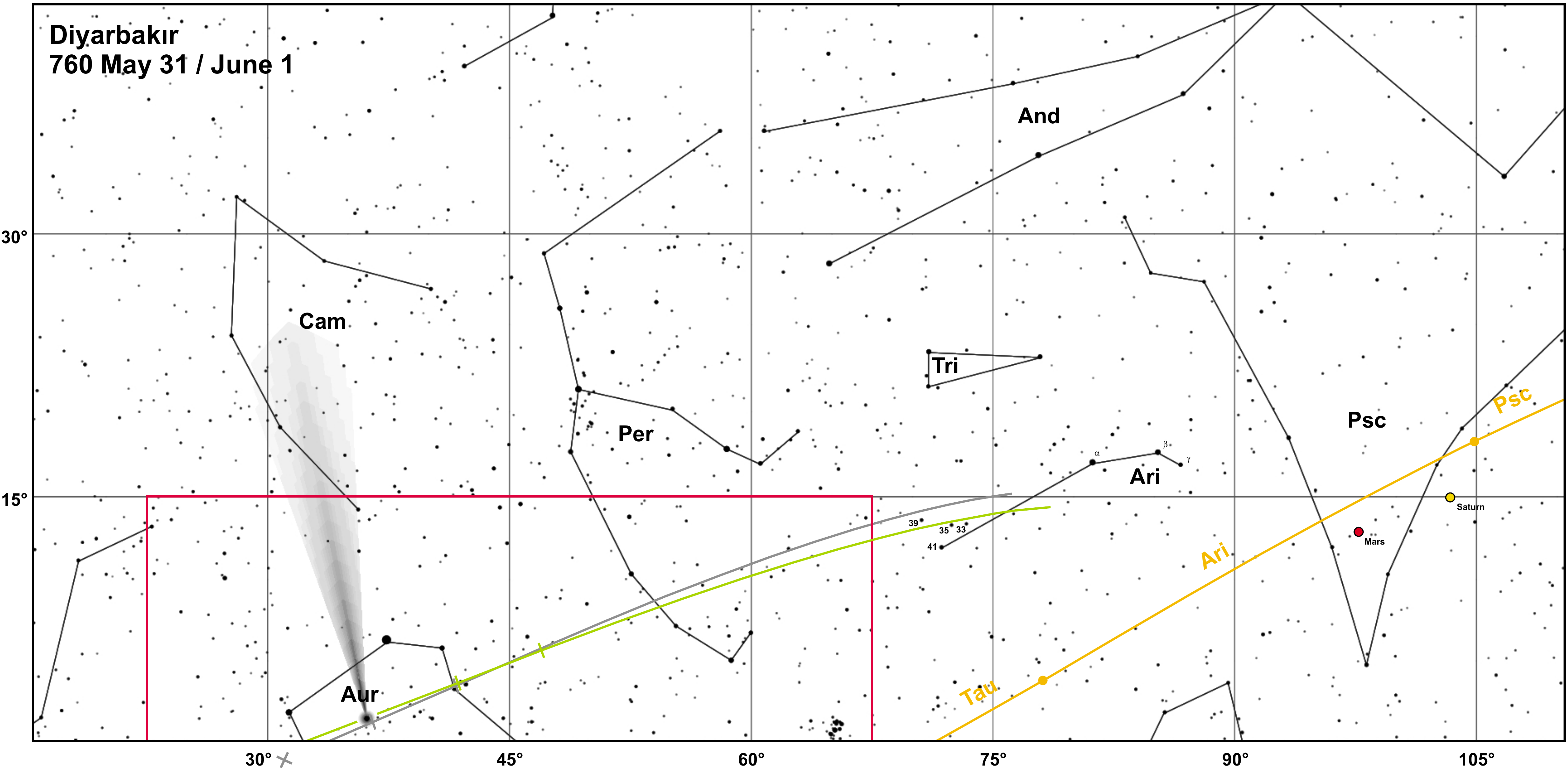}
\caption{
{\bf Horizon plot for Amida for AD 760 June 1 at 2:40h local time:}
{\rm The Chronicle of {\it Zuqn}\textit{\={\i}}{\it n} reported that the comet 
{\it was going bit by bit to the North-East} direction 
until dawn of Pentecost in the night May 31 to June 1,
the third dated position in Sect.~2 (Z3 in Table 1),
the red error box.
The NE direction is taken to be azimuth $45 \pm 22.5^{\circ}$.
The comet is indicated with tail directed away from the Sun for June 1 at 0h UT 
on the new (green) orbit (as grey cross on the old orbit slightly below the horizon).
On our new orbital solution, the comet was in conjunction with the 
Sun on June 1.8 (UT) with minimal elongation being $19.1^{\circ}$
(according to the standard JPL orbit, it was on May 31.9 with $18.5^{\circ}$ minimum elongation);
the last observation before comet-sun-conjunction 
as reported in the {\it Chronicle of Zuqn}\textit{\={\i}}{\it n} for {\it early morning}
of June 1 is consistent only with our new orbit regarding this inferior conjunction.
The expected positions on May 30.0 and 31.0 (UT) are indicated with green (and grey) tick marks.}
}
\end{figure}

\smallskip

(Z4) Then, after conjunction with the Sun, our Chronicler sighted it again ``at
the beginning of [the] third [day] after Pentecost {\dots} at evening time'' (Pentecost Sunday was on June
1): (a) our term {\it Tuesday} in Syriac language is called {\it third day} (as third day of the week), the oriental
{\it beginning} of it would be our Monday evening (June 2 evening) -- but then the Syriac wording would be slightly
different; (b) the most likely meaning of ``at the beginning of [the] third [day] after Pentecost'' is that the Chronicler
counted the days since after the last detection: 1st day after the last sighting is Monday, 2nd is Tuesday, 3rd is
Wednesday, which begins on our Tuesday after sunset (June 3 evening); (c) if the ``beginning'' is related to the start
of a new 24h-day after three full days, then he would mean our Wednesday (June 4) evening. 

With an additional text from Michael the Syrian (Sect. 4d), we can constrain the invisibility
of the comet to ``three days'' (i.e. 2.5-3.5 days), namely from June 1 in the early morning (last sighting) to June 3
or 4 in the evening.

Since the comet was last seen in the morning of June 1 in the NE and then again first in the
evening of June 3 or 4 in the NW, we can conclude that it was close to the Sun around June 2 (conjunction), i.e. not
seen due to the much brighter nearby Sun. Since we cannot independently derive the brightness of the comet 
around perihelion with sufficient precision due to possible cometary activity (Sect. 5), we refrain from 
estimating the area, where the comet was invisible due to the much brighter Sun.

\smallskip

(Z5) After conjunction with the Sun, the comet ``was seen again at evening time, from the
north-west [quarter] {\dots} for twenty-five evenings'', i.e. shortly after sunset and shortly before setting of the
comet, i.e. at low altitude of again up to $15^{\circ}$ (for first detectability) and an azimuth for the whole north-western
quarter ($270-360^{\circ}$), with an uncertainty for the full azimuth range of the whole NW quarter (Table 1, Fig. 6). 
The date of the observation was concluded above (Z4) to June 3 or 4, possibly already June 2. The Chronicler here says that this
observation was performed ``at evening time'', i.e. at between the start of nautical twilight and the end of
astronomical twilight, i.e. on June 3 ($\pm 1$d) between 19:50h and 21:10h local time (UT
17:10h-18:30h), i.e. on June 3.74 ($\pm 1$d) UT.

\begin{figure}
\includegraphics[angle=0,width=16cm]{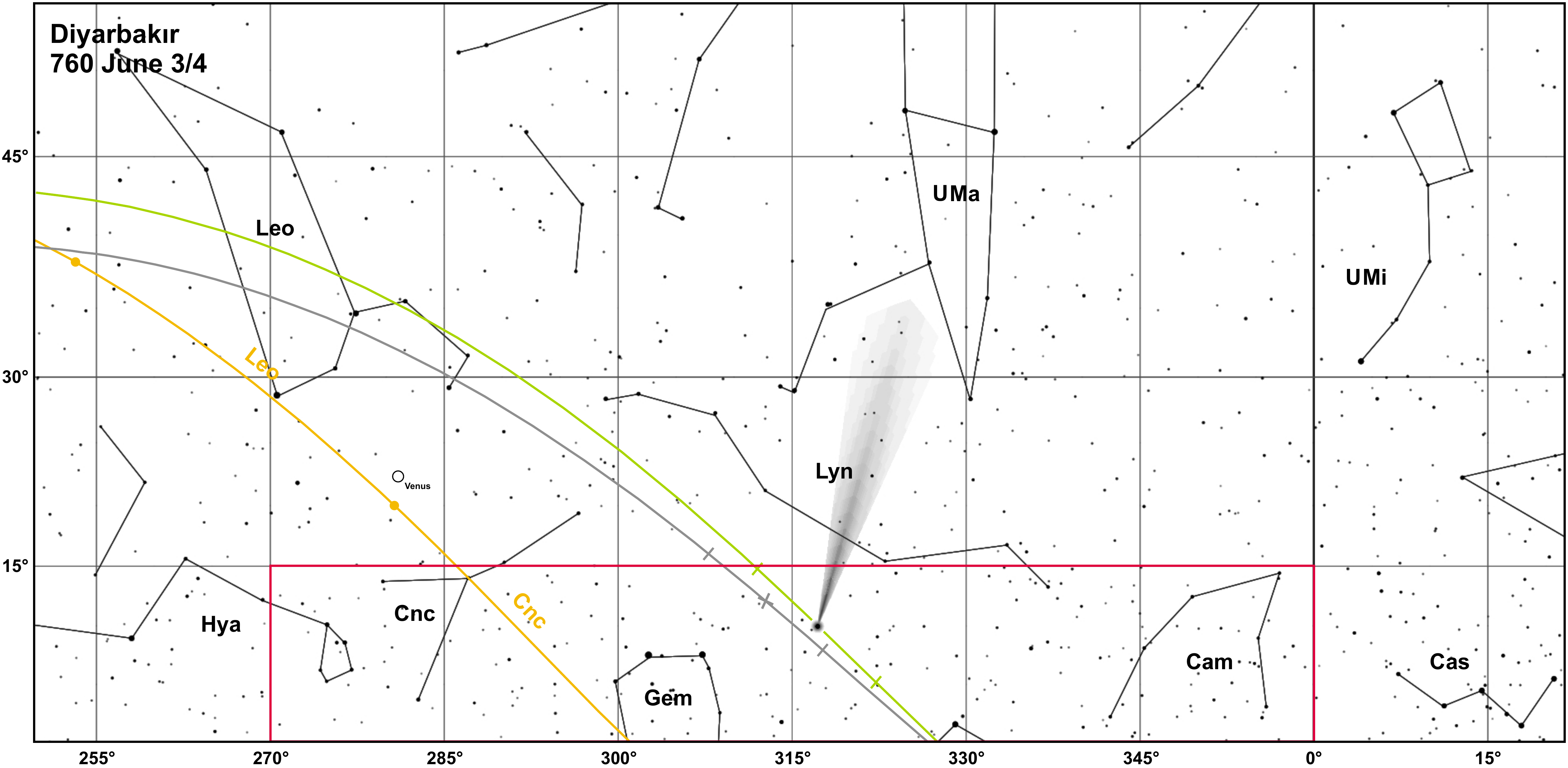}
\caption{
{\bf Horizon plot for Amida for AD 760 June 3 at 20:30h local time:}
{\rm The Chronicle of {\it Zuqn}\textit{\={\i}}{\it n} reported that the comet was seen
{\it from the north-west} [quarter] {\it at the beginning of} [the] {\it third} [day] {\it after Pentecost ... at evening time}
(Z4 and Z5 in Table 1, Sect. 2),
the red error box.
The comet is indicated with tail directed away from the Sun   
on the new orbit (cross on the old orbit).
Along our best fitting orbit of 1P/Halley the closest encounter of the comet with the Earth occurred 
on 760 June 3.6 (UT) with 0.37\,au (according to the JPL standard orbit, closest approach was on June 2.7 with 0.41\,au). 
The expected positions on June 2.74 and 4.74 (UT) are indicated with green (and grey) tick marks.}
}
\end{figure}

With an additional observation from China for June 9, we can constrain the azimuth even
further for that date, see below and Table 1: the {\it Chronicler of Zuqn}\textit{\={\i}}{\it n} 
has seen the comet in the NW quarter since June $3 \pm 1$ ``for 25 evenings'', 
so that it was in the NW on June 9. With an additional text from Michael the Syrian (Sect. 4d), 
we can constrain the first detection after conjunction to June 3 or 4 in the evening.

The comet ``was seen again at evening time, from the north-west [quarter] and it remained for
twenty-five evenings'' -- 
until June 27 or 28 (constrained with Michael the Syrian),
moving ``bit by bit to the south'' (horizontal system) and ``again disappeared''. 
When the author afterwards mentioned a reappearance in the SW quarter (``then it returned
[and] was seen in the south-west [quarter], and thus there it remained for many days''), 
he could indeed mean the fading comet (about 4th mag or fainter, Fig. 14),
which was difficult to be detected for a few nights due to the moon becoming too bright
(or, e.g., some obscuration along the horizon);
the given position, in the SW quarter, would be reliable for a detection around July 5
when the moon exited the evening sky.
However, the last position transmitted in the Chinese sources (near $\beta$ Vir around July 5, see C5 below)
would be `in the SW quarter' only at the beginning of astronomical twilight;
it may still be credible, because the Chronicler is otherwise quite precise, and the comet's brightness (Fig. 14) does allow such a detection.

We cannot exclude that the Chronicler meant a different
object\footnote{Chinese sources did report several more
{\it guest stars} after the perihelion of 1P/Halley in AD 837, see e.g. Xu et al. (2000).},
but it is not very likely -- the context (see Sect. 2.1) supports the former.
If the last object was again our comet, it must have been seen since at the earliest from June
29 on, and then for ``many days'', i.e. more than about three days, but less than a month. These considerations
do not yield a sufficiently precise dated position for the orbit determination -- we can compare it below with the
Chinese observations in July and our new orbital solution. Furthermore, the credible source Michael the Syrian
specified that the comet was seen for 40 days after conjunction with the Sun, which may be somewhat rounded, but would
be fully consistent with the Chinese and {\it Zuqn}\textit{\={\i}}{\it n} records, 
yielding until about or up to July 12 or 13 for the last sighting.

\section{The comet of AD 760 as reported in Classical Chinese sources}

\subsection{Chinese texts and text-critique}

Extant Chinese sources for this comet include Jiu Tang shu (JTS, Liu et al. 945;
in Ho 1962: CTS, using the older Wades-Gilles romanization), 
Tang hui yao (THY, Wang Pu et al. 961), and Xin Tang shu (XTS, Ouyang Xiu et al. 1061; in Ho: HTS) -- 
also, Wenxian tongkao (WHTK 286/23a, Ma Duanlin 1317) provides a copy of XTS 32.838 with one insignificant variant. 
We prefer here JTS as the basic text, because -- as will be shown -- it is the intrinsically most consistent and
detailed text (it is also the oldest one), 
but also consider the variants in XTS and THY; XTS omits certain details, e.g. the exact observing
time (``5th watch'') for the comet of AD 760.\footnote{Kiang (1972) uses only JTS and XTS
(after Ho no. 273); Stephenson \& Yau (1985) prefer the THY text.}

We present here our own new, technical, very literal translations, which aim to preserve the
detail and word order of the original Chinese, but have been slightly smoothed to present correct English sentences
(see appendix for the Chinese texts); significant variants in Ho (1962, no. 273 and 274), Xu et al. (2000), and
Pankenier et al. (2008) are mentioned in footnotes. First, we translate the oldest text from JTS (36.1324, and much
shorter in 10.258), counted as object no. 273 in Ho (1962), with some Chinese terms, explanations, and
significant variants from THY (43.767) and XTS (32.838, unless otherwise specified) in round brackets, 
our additions in square brackets (e.g. the day/night number in the 60-day-cycle), 
starting with the night AD 760 May 16/17, line breaks by us:
\begin{quotation}
``Tang Emperor Suzong (literal: Tang['s] Solem Ancestor) {\dots} 
{\it Qianyuan} [reign-period]\footnote{Xu et al. (2000) added here
``i.e. 1st year of the Shangyuan reign period'' -- in fact the
{\it Shangyuan} reign period started only at the beginning of the
4th intercalary month, after the {\it Qianyuan} reign period
had ended with the 4th month.} {\dots} 3[rd] year, 4[th] month,
{\it dingsi} (54) night 
(THY gives the lunar date: ``27[th] day'', XTS omitted ``night''), 
5[th] watch (``5[th] watch'' omitted in THY and XTS), 

\noindent [a] broom
({\it hui}) [star] (XTS: ``{\it hui xing''} for ``broom
star'') emerged (THY: ``seen at ({\it yu})'', XTS: ``there was ... at
({\it yu})'') east ({\it dong}) direction, colour being white, \\
length (JTS 10.258 adds: about) 4 {\it chi} (THY and XTS have color and length after the next phrase), \\ 
it was located/situated in ({\it zai}) {\it Lou}, [in] {\it Wei}\footnote{{\it Lou} (``Hillok'' or ``Lasso'') and {\it Wei}
(``Belly'' or ``Stomach'', see SK97 and Ho 1966) could be the asterisms
of that name (both in ``our'' Aries, i.e. the constellation as defined by the International Astronomical Union) or the
lunar mansions (right ascension ranges) named after these asterisms (LM 16 and LM 17, respectively) starting in the
west with the determinative star ${\beta}$ Ari for {\it Lou} and with 41 Ari for
{\it Wei}. See below for position C1.} 
for-a-while/space ({\it jian}), \\
it rapidly moved toward east
({\it dong}) north ({\it bei}) corner (THY omitted ``corner''; XTS has instead: ``east direction rapidly moved''), \\
passing through {\it Mao, Bi, Zui} (XTS: ``{\it Zuixi}''), {\it Shen}\footnote{Xu et al. (2000)
give ``Can'' here, which is a more common pronunciation of the Chinese character; however, in this context, the correct
pronunciation is ``Shen'', LM 21 and an asterism in Orion.}, {\it Jing}
(XTS: ``{\it Dongjing}''), {\it Gui} (XTS: ``{\it Yugui}''),
{\it Liu}\footnote{This list could point to either asterisms or LMs:
{\it Mao} (``Mane'', LM 18),
{\it Bi} (``Hunting net'', LM 19), 
{\it Zui} or {\it Zuixi} (``Beak'', LM 20),
{\it Shen} (``Triaster'' or ``Hunter'', LM 21),
{\it Jing} or {\it Dongjing} (``Eastern Well'', LM 22),
{\it Gui} or {\it Yugui} (``Spectral Carriage'', LM 23), and
{\it Liu} (``Willow'', LM 24); 
translations of asterisms here are the {\it Han} time interpretation, some have
changed later (SK97).} [and] {\it Xuanyuan}
(THY added ``{\it xiu}'' for ``lodge''\footnote{{\it Xuanyuan} (``Yellow Emperor'')
is usually only an asterism, which does not have the additional function as LM asterism; given that it seems to be listed here
as {\it xiu}, it may have some `lodge'-like function; {\it Xuanyuan} is meant as skeleton of 17 stars
in Leo and Lynx starting with $\alpha$ Leo close to the ecliptic.}), \\
reaching {\it Taiwei
Youzhifa}\footnote{{\it Taiwei} (``Great Tenuity Enclosure'' or ``Supreme Subtlety Palace'' or
``Privy Council'') is one of three asterisms, which are so-called ``enclosures''
({\it yuan}) with two ``walls'' each, {\it Taiwei} being a large area with 10
stars in Virgo and eastern parts of Leo (12 stars in {\it Tianguan
shu}, but then only 10 in the official {\it Shi Shi}, SK97); the determinative star of
{\it Taiwei} is {\it Youzhifa} ($\beta$ Vir) at
the southern end of {\it Taiwei's} western wall (SK97).} 7 {\it cun} position 
(THY: ``reaching {\it Taiwei} west ({\it xi}), {\it Youzhifa} west ({\it xi}) 7 {\it chi}''; 
XTS omitted ``{\it Taiwei}'' and has only ``reaching {\it Youzhifa} west ({\it xi})''), \\
in all more than 50 days, only then ({\it fang}) [it] disappeared
(THY very similar; XTS has ``in all more than 50 days, [it was] not seen'')''
(continued below).
\end{quotation}

We will discuss this transmission in detail below to obtain dated positions.

Next, we present additional relevant texts, not given in Xu et al. (2000) and Pankenier et al.
(2008). Ho (1962) cited under his no. 274 a record from JTS 36.1324 (Ho: CTS 36/8a), 
and gives two more texts, HTS 32/6b (=XTS 32.838) and, almost identical,
WHTK 286/23a (286/29b-30a in the Siku quanshu huiyao edition).
Here our own new literal translation of the JTS text (with variants from XTS and also from THY 43.767),
which follows immediately after the previous comet report:
\begin{quotation}
``Intercalary 4[th] (XTS omitted
``4[th]'') month, {\it xinyou} (58=May~20 with night 20/21), new-moon (THY: 
``{\it Shangyuan} reign-period, [initial] year, intercalary 4[th] month, 21[st] day'' (=June 9)),

\noindent [an] ominous star ({\it yao xing}) seen at ({\it yu}) south ({\it nan}) 
(THY: ``west ({\it xi})''; XTS: ``there was [a] broom star
({\it hui xing}) at ({\it yu}) west ({\it xi})'') direction, length several
{\it zhang}. \\
This time, since [the] beginning [of the] 4[th] month,
heavy fog [and] heavy rain,
reaching [the] end [of the] 4[th] intercalary month (i.e. the last 10 days),
only then ({\it fang}) [it, i.e. bad weather] stopped 
(instead of this whole sentence, THY and XTS have ``Reaching 5[th] month, [ominous star] disappeared'',
XTS adds: ``Only [when] reaching ...''). \\
This month, rebel bandit Shi Siming again captured [the] 
Eastern Capital (i.e. Luoyang). Grain prices leapt [up] in expense, 
{\it dou} (i.e. about 6 liters of rice) reaching eight hundred
{\it wen}. People ate each-other [and] corpses covered [the] ground.''
\end{quotation}

After reporting the disappearance of the comet, ``Only [when] reaching 5[th] month ...'', XTS (32.838) adds:
\begin{quotation}
``{\it Lou} corresponds to [the pre-imperial state of] {\it Lu}, {\it Wei} [and] {\it Mao} [and]
{\it Bi} correspond to {\it Zhao},
{\it Zuixi} [and] {\it Shen} correspond to {\it Tang, 
Dongjing} [and] {\it Yugui} correspond to
[the] capital city ({\it jingshi}) (meaning probably the historical capital of the Zhou dynasty)
allotment, [as for]
{\it Liu}, its half corresponds to [the] {\it Zhou} allotment. 
As-for-cases-in-which ({\it zhe}) two brooms seen in-succession, amassing disaster.
Moreover, {\it Lou}, {\it Wei} space ({\it jian}) 
[corresponds to] {\it Tiancang} (`Celestial Granary').''
\end{quotation}

The whole last paragraph is an astro-omenological interpretation of the comet report.
In Chinese astro-omenology, {\it Wei} (LM 17) governs granaries and warehouses,
as found in the {\it Jin shu} (Ho 1966, p. 100, Ho 2003, p. 147) --
and indeed, the term {\it Tiancang} means `Celestial Granary/ies'.
There is also an asterism {\it Tiancang}, which is however located mostly in LM {\it Kui} and 
only partly in LM {\it Lou}; there are further asterisms meaning {\it `Celestial Granaries'}
in LMs {\it Lou} and {\it Wei}, e.g. {\it Tianjun} (SK97), written {\it Tianqun} in Pankenier et al. (2008).
({\it Lou} governs cattle rearing and animal sacrifices, see Ho 1966, p. 100.)

In the past, it was considered that there were two comets in spring AD 760, e.g. Yeomans et al. (1986).
All sources for Ho no. 274 give ``several {\it zhang}'' as length, so that one could
consider that they mean the same object: the ``ominous star'' ({\it yao xing}) in the west 
in THY (June 9 evening) would fit with the comet path given in the previous text;
the object(s) in JTS, XTS, and WHTK for May 20/21 (morning) in the south or west are not consistent
with the path of comet no. 273, which was then still in the NE. If the previously cited JTS text refers
to the same object, a date correction would be needed -- it should be June 9 (as in THY) instead of May 20.
One explanation could be:
May 20 corresponds to the 58th day, {\it xin-you} in the 60-day-cycle, while June 9 is the 18th,
{\it xin-si}, so that only the 2nd part would have been
mistaken in JTS, XTS, and WHTK by a copying scribe ({\it you} for {\it si}); 
THY conserves the correct date as date in the lunar calendar (day 21 = June 9), converted from the 60-day-cycle as found in
its source. Note that the two dates (May 20 and June 9) pertain to the same Chinese lunar month (4th intercalary
month), just the day within the month is different. 
More reasonably, since ``new moon'', i.e. the first day of the lunar month, is given in JTS and XTS in
addition to {\it xin-you} (58), which is correct for May 20, a confusion between date and event might be just due
to a false concatenation in the compilation process; furthermore, it is plausible that the second comet
report, preserved correctly in the THY text, originates from another source and observing site, where, e.g.,
weather conditions did not allow a detection earlier than June 9.

To sum up, among the three texts for Ho object no. 274, the THY transmission appears to be the least
corrupt: sighting on June 9 (JTS and XTS: May 20/21), THY has west direction (XTS also west, but JTS has south).
That the information in THY is most reliable here, relies on the assumption that the ``two'' objects Ho no. 273
and 274 are one and the same comet; 
this is supported by the fact that the duration in the first comet report (about 50 days after May 17/18)
corresponds well with the disappearance in THY and XTS (``Reaching 5[th] month, [it] disappeared'').
This assumption is also supported by the following astro-omenological interpretation in XTS 32.838:
``As-for-cases-in-which ({\it zhe}) two brooms seen in-succession, amassing disaster''.
In the translation
``two separate broom stars appearing simultaneously'' (Stephenson \& Yau 1985), the word ``separate'' is added
(but not given in the Chinese text); the sense of the adverb in Classical Chinese ({\it reng}) suggests
repetition with close or immediate proximity in time (``appear one after the other'' or ``in quick succession'' or
``repeatedly'').\footnote{Stephenson \& Yau (1985)
and Yeomans et al. (1986) thought that, in addition to the comet seen since AD 760 May 16/17,
there would have been another comet seen in the south or west since May 20/21.} 
That it is only one comet is justified by further independent reports, 
where the conjunction with the Sun is explicitly reported, 
e.g. the {\it Chronicle of Zuqn}\textit{\={\i}}{\it n} (see above)
and several further East Mediterranean and West Asian reports (Sect. 4).

There is one more extant source, XTS 6.162-3, but the variant transmission gives only very short information:
\begin{quotation}
``4[th] month ... {\it dingsi} (54), there was [a] broom star,
emerged at ({\it yu}) {\it Lou, Wei},
{\it Jiwei} (56), Lai Zhen (died ca. AD 763) became
{\it Shannan} Eastern Circuit's Military Commissioner charged to
overcome [the rebellion of] Zhang Weijin.
Intercalary month (4[th] omitted) {\it xinyou} (58), there was [a] broom star,
emerged at ({\it yu}) west ({\it xi}) direction. ...
Jimao (16), [there was a] large amnesty, change [of] reign-period [title],
grant [of] civil [and] military office [and] rank. ... This month
[was a] large famine. Zhang Weijin surrendered.''
\end{quotation}
This late source shows how compilers work: XTS 6.162-3 concatenated input from
XTS 32.838, a source which is already shortened -- as one consequence, the comet's
position at the beginning is a bit corrupt.
This source, which belongs to the ``Basic Annals'' ({\it Benji}) section of the history
(a general chronicle of events during the reign of each emperor), rather than the technical
treatise, is only interested in the first appearance of the comet
(first sightings at the very beginning and after conjunction with the Sun) --
the main point is the connection to historical events on Earth.

\smallskip

The year 760 fell midway through the An Lushan rebellion (AD 755-763). The early years of the rebellion had witnessed
the abdication of an emperor who had reigned for more than forty years, the fall and subsequent recapture of the
main capital at Chang'an, and casualties reportedly numbering in the millions. In both JTS and XTS 6.162-3, close
chronological proximity associates the comet's appearance with politics, 
the rebellion and the famine that accompanied it;
XTS 32.838 reflects these in an astro-omenological interpretation. 

\smallskip

As quoted above, JTS reports the weather: ``This time, since [the] beginning [of the] 4[th] month
(new-moon on Apr 19/20), heavy fog [and] heavy rain, 
reaching [the] end [of the] 4[th] intercalary month 
(i.e. the last 10 days, new-moon on June 17/18), only then [it, i.e. bad weather] stopped.''
Monsoon typically arrives in May and may well end in June. 
In addition to shortenings and omissions in the compilation process, problems with weather and the rebellion may
also have influenced the observations and the data record (and might be partially responsible for the famine). Still,
since the beginning of the {\it Tang} dynasty (AD 618), there
are no better transmitted records for any comet before AD 760 (see Pankenier et al. 2008 for the texts).

\smallskip

The Korean ``Chronicle of the Three Kingdoms'' (Samguk sagi) briefly reported a ``{\it hui} comet'' sometime
during the lunar month AD 761 May 9 to June 7 (Ho 1962, no. 275); this work, compiled
AD 1142-1145 (Shultz 2004), is often off by a few years -- probably, our comet is meant. 
The Kingdom of Silla is traditionally dated BC 58 to AD 935. 
However, the Silla dynasty, which united the whole of peninsula, ran from AD 668 to 935.

\smallskip

From the Chinese observations also all five comet criteria mentioned above (Sect. 2) are
fulfilled. A ``broom {\dots} colour being white'' also points to a comet with dust tail.

\subsection{Dated positions from Chinese sources}

The observing times below are calculated for Chang'an
(today: Xi'an, China, longitude $108^{\circ} 57^{\prime}$ East, latitude $34^{\circ} 16^{\prime}$ North), the capital during the
{\it Tang}; however,
during the An Lushan rebellion AD 755-763, observations could have been obtained from the Eastern capital Luoyang
(longitude $112^{\circ} 27^{\prime}$, latitude $34^{\circ} 40^{\prime}$).
We can deduce the following dates and positions (Table 1):

(C1) When the Chinese give May 16 as date for an observation at the end of the night (JTS:
``{\it dingsi} [54] night, 5[th] watch''), 
they refer for the whole night (as usual) to the date at the start of
that observing night, so that the observation was in the morning of what we date May 17;
the 5th watch is the last fifth of the night, see below.
The discovery of the comet by the Chinese and the {\it Zuqn}\textit{\={\i}}{\it n} Chronicler at the
end of the nights of May 16/17 and 17/18, respectively, may have been facilitated by the nearby waning crescent moon at
that time: the comet and Moon (2 days before conjunction) were in Aries (Table 1, Fig. 7).

\begin{figure}
\includegraphics[angle=0,width=16cm]{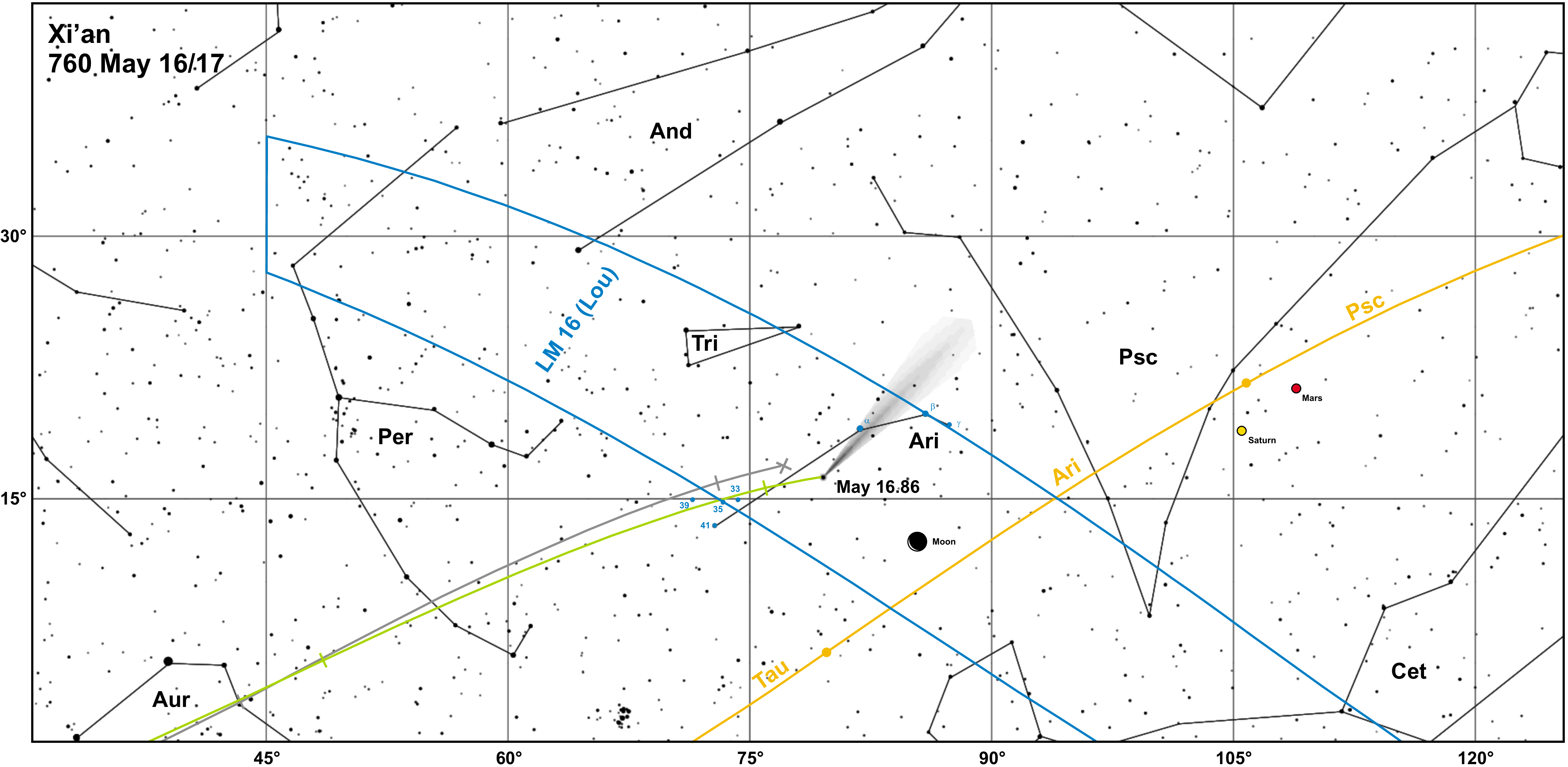}
\caption{
{\bf Horizon plot 
for Chang'an (now Xi'an, China), the {\it Tang} capital, for the night AD 760 May 16/17 at 4h local time:}
{\rm The waning crescent moon is seen in Aries. 
The Chinese reported a
{\it broom} [star] {\it emerged} {\it east direction ... it was located in Lou},
the first dated position in Sect. 3 (C1 in Table 1).
{\it Lou} is here Lunar Mansion 16 (LM 16), the right ascension range from $\beta$ Ari to 35 Ari
({\it Wei} is LM 17).   
The given {\it East direction} can be considered as azimuth $90 \pm 45^{\circ}$,
so that the given right ascension range (LM 16) is constrained in the NE by azimuth $45^{\circ}$,
but in the SE by the local horizon.
The positional error box is indicated by blue lines.
The star $\alpha$, $\beta$, and $\gamma$ Ari and 35, 39, and 41 Ari,
which make up the asterisms {\it Lou} and {\it Wei}, respectively,
are indicated as blue dots, but the asterisms are not meant here.
Our new and the previous (JPL/YK81) orbits are shown as green and grey lines, respectively,
from May 16/17 midnight onward to the east.
We draw a comet with plasma tail directed away from the Sun (in the NE below horizon) 
for its position on May 16.86 (UT) on the new (green) orbit -- and for the same date as cross on the old (grey) orbit.
In order to illustrate the motion of the comet along both paths the expected positions 
on May 20.0 and 30.0 (UT) are indicated with green and grey tick marks, respectively;
the ecliptic in orange.}
}
\end{figure}

The text then says ``[a] broom ({\it hui}) [star] emerged east
({\it dong}) direction''. The term ``east'' clearly marks a
direction, it does not mean ``morning'' -- the time of observation is given otherwise as ``5[th] watch''. 
The ``east direction'' given with the characters ``{\it dong fang}'' is
from a compass with at least four points.
(NB: the Chinese record from XTS 32.839 on Halley in AD 837 mentioned for the pointing of the tail, in turn, the directions
``west'' ({xi}), ``south'' ({\it nan}), ``north'' ({\it bei}), and then ``east'' ({\it dong}), always in 
combination with the character for ``pointing'' ({\it zhi})
(Pankenier et al. 2008) -- the pointing direction of the tail is not given in the record for AD 760.)
If the wording {\it dong fang} for ``east direction'' is from the
4-point-compass, it then yields an azimuth of $90^{\circ}$ (east of north) with an error bar of $\pm 45^{\circ}$;
while it could even be from an 8-point-compass (see below, C2), then with a
smaller error bar, we conservatively choose here the larger $\pm 45^{\circ}$ error bar from the 4-point-compass. 

\smallskip

The position is then given in JTS as
``{\it zai Lou}'' translated by us as ``it was located/situated in {\it Lou}''. 
The term {\it Lou} 
could in principle refer to LM 16 or the asterism
{\it Lou} with ${\alpha}$, ${\beta}$, and ${\gamma}$ Ari.
In almost all Chinese texts on comets (at least until AD 900), e.g. Halley in AD 837, as well as solar and lunar eclipses (as
quoted in Xu et al. 2000 and Pankenier et al. 2008)\footnote{also
noticed by J. Steele, priv. comm., as he told us during
the Leiden workshop on historical observations in Oct. 2019}, 
the character ``{\it zai}'' is related to a Lunar Mansion (or to one of
the three ``enclosures''), and it often comes with a
``{\it du}'' measurement or some other positional
specification.\footnote{However, compilers have also
concatenated the texts, so that ``{\it zai}'' can appear to
be connected to a non-LM asterism, e.g. comet Halley in AD 837: in the older JTS, there is a descriptive position
related to {\it Xuanyuan} (``emerged {\it Xuanyuan} of/from right'') without {\it zai} 
plus an equatorial hour angle (``{\it zai Zhang} (LM 26) 7 {\it du}''), which was then concatenated by XTS to
``{\it zai Xuanyuan} right'', i.e. {\it zai} with a non-LM asterism. NB: ``right'' could be a scribal error
for ''left'', ``{\it zai Zhang} (LM 26) 7 {\it du}'' is not consistent with ``{\it Xuanyuan} of/from right'' -- 
the Chinese characters for right and left are almost similar, such a mistake happens often.} 
In sum, the wording ``{\it zai Lou}'' translated as ``it was located/situated in {\it Lou}''
meant here Lunar Mansion {\it Lou} (LM 16).
 
The text continues with {\it ``Wei jian''}.
One of the principle meanings of the term {\it jian} is ``space''
by referring to a period of time.\footnote{Needham \& Wang (1959, p. 256, note g) pointed out that
the character transcribed {\it jian} originally showed the moon in a gate, later the Sun in a gate;
thus, it seems that {\it jian} is a fitting character for the space of a lunar mansion.
}
The information of our record could then be that the comet was first in LM {\it Lou} 
and then in the {\it Wei} space;
because shortly later, more LMs follow, the context and also the astro-omenological interpretation point to the fact 
that LM {\it Wei} is referred to here.\footnote{{\it Wei} (``Belly'') can in principle be either the
Lunar Mansion (LM) 17 or the asterism of the same name, both have
35 Ari as determinative star. There are three LMs, which are written in Chinese
characters all transcribed as ``{\it Wei}'';
and the same characters also refer to asterisms of the same names, in which the determinative star of the respective LM
lies. In this paper, we deal only with that asterism and LM {\it Wei},
whose determinative star is 35 Ari. The situation is similar for {\it Bi}:
there are two LMs whose Chinese characters are transcribed as
``{\it Bi}'', LM 14 and 19 (and also two asterisms of the same names); in this paper, we deal only with the
{\it Bi} that is LM 19 (``Hunting Net''), whose determinative star is ${\varepsilon}$ Tau (SK97).}
If we would interpret it as
``located/situated in LM {\it Lou}, asterism-space {\it Wei}'', i.e. around the asterism {\it Wei}, 
then the LM {\it Wei} would not have been mentioned as LM at all, even though
the comet did pass through this wide LM, and the following LMs are all given.
Although the rendering of {\it jian} as ``between'' (preposition) would be acceptable
from a linguistic perspective, it should be avoided, because it creates a false impression
that we are dealing with asterisms rather than lunar mansions.\footnote{The 
phrase ``{\it zai Lou Wei jian}'' was translated as ``between
{\it Lou} and {\it Wei}'' by Ho (1962), Stephenson \& Yau (1985), Xu et al. (2000), 
and Pankenier et al. (2008). However, this translation neglects
{\it zai} (as a technical term here, it is related to {\it du} measurements, see above), 
but focusses purely on {\it jian}, translated with the preposition ``between''.
The term `between' is ambiguous, e.g. `between A and B' may include or exclude A and B themselves, 
meaning something like either `among A and B' or `in between A and B'; see also note 33 
(in between two asterisms instead of between/amongst two LMs). Hence, we prefer the more literal ``space''.
}

One can also read {\it jian} as a verbal complement as ``for-a-while'' 
(see Wang Li 2000),\footnote{Wang Li (2000) gives a typical example; 
usually when used in this sense, {\it jian} is preceded by
the existential {\it you} ({\it you jian} -- lit. ``there was a while'').}
especially the best transmitted text, JTS, allows such a rendering in a grammatically plausible way:
``It was located/situated in ({\it zai}) [LM] {\it Lou}, [in LM] {\it Wei} for-a-while, 
[then] it rapidly moved toward east north corner, passing through {\it Mao}, {\it Bi}, ...''.
The general motion of this comet is consistent
with being first moving slowly in/through {\it Lou} (LM 16) and for a while through {\it Wei} (LM 17), 
but then more rapidly through Mao (LM 18) and Bi (LM 19), where the ``north east corner'' 
($45 \pm 22.5^{\circ}$) is situated (see below, Fig. 8).

In sum, the position statement ``{\it zai Lou Wei jian}'', both {\it Lou} and
{\it Wei} are most certainly Lunar Mansions (i.e. right ascension ranges).
For obtaining positions (and uncertainties) on sky, our interpretation of {\it Lou} as LM 
is also more conservative than as asterism, because we use large measurement uncertainties 
(the whole LM {\it Lou}, constrained to azimuth {\it east}).\footnote{Kiang (1972) used as position ``between
{\it Lou}-16 and {\it Wei}-17'', where the numbers point to LMs, but then
gave the positions of the two stars in the asterisms {\it Lou}
and {\it Wei}, namely right ascension $15^{\circ}$ and declination $17^{\circ}$
for {\it Lou} (which is almost exactly the 760.5 position of
${\alpha}$ Ari, the easternmost star of {\it Lou}, also according to Kiang 1972) and right ascension
$23^{\circ}$ and declination $22^{\circ}$ for {\it Wei} (the position of 35 Ari,
the westernmost star of {\it Wei}). The orbital fit by Kiang
(1972) indeed resulted in a position almost exactly in the middle between
${\alpha}$ Ari and 35 Ari.}

A position in LM {\it Lou} means that the right ascension is somewhere between the right ascension of the
determinative star of LM 16 {\it Lou}, namely ${\beta}$ Ari
with ${\alpha}$ = 0h 48m 4s, and the right ascension of the determinative star of LM 17
{\it Wei}, namely 35 Ari with ${\alpha}$ = 1h 33m 18s (epoch of date).

The position of the comet is then constrained by the LM {\it Lou} (right ascension range) and the azimuthal range
``{\it dong}'' for East ($90 \pm 45^{\circ}$), see Fig. 7. 
Such a position is fully consistent with the {\it Chronicle of Zuqn}\textit{\={\i}}{\it n}.
While the more professional Chinese
court astronomers detected the comet only one day before the {\it Zuqn}\textit{\={\i}}{\it n} chronicler, 
we have for the beginning (Z1) and for May 22 (Z2) two precise dated positions from the latter.
The compilations of the original Chinese sources transmitted mostly reduced information:
for astro-omenological purposes -- the addition of the XTS text (32.838) is a good example -- 
it was very important to record the motion through the lunar mansions.

Finally, we have to estimate the observing time: the ``5[th] watch''
specified by the Chinese corresponds to the last fifth of the night.\footnote{``A night (sunset to sunrise ...)
divided into five night-watches'' (Needham et al. 1986, p. 199); this was the general rule
for dividing the night into five watches with equal lengths per night, but varying during the year.
However, for cultural reasons and other applications, it seems that sometimes a certain amount of time (e.g. 2.5 {\it ke}
being 36 min, or different periods) were inserted between sunset and the beginning of the 1st watch (analogously in the morning);
the system changed over time and varied from location to location (Needham et al. 1986, pp. 199-205).
Since the habit at around AD 760 is uncertain, we used the general rule as default. If we would subtract the 2.5 {\it ke}
at dawn and dusk, the determination stars of {\it Lou} ($\beta$ Ari)
and {\it Wei} (41 Ari) would not yet be visible at the start of the (earlier) time window.}
On AD 760 May 17, the ``5th watch'' corresponds to 2h54m to 4h54m local time (Chang'an, now Xi'an, being 7:15h east of Greenwich)
or UT May 16.86 ($\pm$ 1h).\footnote{For morning
observations, Kiang (1972) changed the date given in
the Chinese source at midnight, but overlooked that practically, the reports do not give a new date for the second half
of the night (see introduction); for AD 760 May 16/17, he argued:
``The date correspondence to May 16, and in this case it seems more reasonable to reckon
the beginning of a day, not at midnight, but at daybreak, 6 a.m., say, hence the presumed time of observation is around
May 16.95 local time, or May 16.65 UT'', both incorrect, but the offset happens to be small; Kiang (1980) and YK81
already noticed that astronomical night reports use the date pertaining to sunset for the whole night; the observing
time for AD 760 May was not corrected by YK81, apparently because it happened to be almost correct in Kiang (1972)
and/or because it was considered as special case and ``unusually precise'' (Kiang 1972). YK81 did not revise the
historical Chinese observations for AD 760, but used those from Kiang (1972); by comparing their orbital solutions,
they found a difference of +1.83 day for their new perihelion time (AD 760 May 20.67) compared to Kiang
(1972).} 

Note that for these first few statements about the comet, the records from China and {\it Zuqn}\textit{\={\i}}{\it n}
are quite similar: (a) China ``night, 5[th] watch'' (May 16/17) / {\it Zuqn}\textit{\={\i}}{\it n}: ``before early twilight''
(May 17/18) (b) China ``east direction'' / {\it Zuqn}\textit{\={\i}}{\it n} ``north-east [quarter]'', (c) China: in
(LM) {\it Lou} / {\it Zuqn}\textit{\={\i}}{\it n}: in Aries, (d) China: ``colour being white'' /
{\it Zuqn}\textit{\={\i}}{\it n}: ``white sign'', (e) China: ``broom star'' ({\it hui
xing}) / {\it Zuqn}\textit{\={\i}}{\it n}: ``resembled in its shape a broom''.

\smallskip

(C2) After the comet discovery in the LM {\it Lou}, then LM {\it Wei}, the comet is next given to have
``rapidly moved toward east ({\it dong}) north ({\it bei}) corner'', without date; only the
oldest source, JTS, has ``corner'', while this detail was lost in the later THY and XTS; XTS
also omitted ``north''. The phrase ``toward east north corner'' is a clear direction on sky.
The azimuth is given here with the two terms ``east ({\it dong})'' and ``north
({\it bei})'' from the 4-point-compass (see above), but in
combination (and together with the term ``corner''), it is then a direction like from an 8-point-compass, i.e. an
azimuth of $45^{\circ}$ (east of north); the error bar would then be $\pm 22.5^{\circ}$. 
We cannot use this position for the orbit.

Next, our text provides a list of
LMs through which the comet has passed, which goes beyond azimuth NE, 
so that the motion toward the explicitly given azimuthal direction
``toward east north'' (JTS, THY) may have some significance here -- the 
compilation may point to the disappearance of the comet due to
conjunction with the Sun (LM 19, {\it Bi}). The {\it Chronicle of Zuqn}\textit{\={\i}}{\it n}
gives the same or a similar position (``it was going bit by bit to the North-East''), 
but with a date for the last sighting before conjunction with the Sun (June 1, see above).

\smallskip

(C3) Next, it is said that the comet is ``passing through'' the lunar mansions
{\it Mao} (LM 18) to {\it Liu} (LM 24), i.e. right ascension ranges (Fig. 8); the
comet crossed some of them before it reached the NE corner (before conjunction with the Sun). The extant record neither
gives any dates for these lunar mansions nor more precise positions. In the orbital fit, we require the comet to go
through these LMs. The Chinese wording ``passing through''
({\it li}) indicates a transit without necessarily contacting
an asterism; hence, if the LM asterisms (with the same name as the LMs) were meant, 
the comet would not need to travel a very curvy path from asterism to asterism,
inconsistent with a real comet orbit. 
Indeed, our final orbit does not cross the asterisms
{\it Mao} to {\it Liu}, but it passes through the lunar mansions LM 18-24. 
This is the first time in the {\it Tang} dynasty that so many LMs are
given for a comet path, see Pankenier et al. (2008) for the texts.

\smallskip

(C4) After crossing LM 24, the comet was in ``{\it Xuanyuan''}; 
THY explicitly wrote ``{\it Xuanyuan xiu}''; 
the term ``{\it xiu}'' is normally translated to ``lunar mansion'' or ``lodge''. 
However, there is no lunar mansion with that name, {\it Xuanyuan} is just an asterism consisting of stars in Leo
and some in Lynx, but ``lodge'' could emphasize a position in {\it Xuanyuan};
the same THY source gave credible information
for a comet on June 9 in the west (possibly in {\it Xuanyuan},
see below). Alternatively, the characters for
``{\it Xuanyuan}'' and ``{\it xiu}'' (``lodge'' or plural ``lodges'') could have been transposed, so that
``{\it xiu}'' would be connected to the LMs listed just before (however, to use ``{\it xiu}'' in this way is rare,
e.g. in AD 178, a comet is reported to ``pass through more than 10 {\it xiu}'' by meaning 10 LMs, 
see Pankenier et al. 2008, Ho 1962 no. 106).

We can constrain the position further: since the object is mentioned to have
{\it passed} LM 24, but not LM 25, we can exclude those stars
of {\it Xuanyuan}, which lie in LM 24; since the list of LMs
ends with LM 24, but not with LM 25, we can also exclude LM 26 (otherwise LM 25 would have been mentioned to have been
crossed). A location in {\it Xuanyuan} and in LM
25 yields a right ascension range of 8h 26m 25s ($\alpha$ Hya, determinant star LM 25) to 8h 51m 54s
(upsilon Hya, determinant star LM 26), while the declination range is given by the stars of
{\it Xuanyuan}, which are also inside LM 25 ($\varepsilon$, $\mu$, and 15 Leo), 
i.e. $\delta = 31.9 \pm 3.1^{\circ}$ (always epoch of date); we did exclude omicron Leo
here, which is much further south, but this star was known well as
``{\it Xuanyuan} right horn'', mentioned otherwise often and being somewhat significant.

We now have a position (in {\it Xuanyuan}), but without a date. 
The sources of the second comet report give here further information:
THY reported that an {\it ominous star} ``seen at west ({\it xi}) direction'' on
June 9 (the most credible source for Ho no. 274, see Sect. 3.1); 
this is a clear date, but the position (``west direction'') is not
well constrained. From the {\it Chronicle of Zuqn}\textit{\={\i}}{\it n} and Michael the Syrian (Sect. 4d), 
we know that the comet was seen in the evenings in the NW
quarter since June 3 or 4, so that we may assume that the Chinese date (June 9) can be
combined with the position in {\it Xuanyuan}, which was at
that time indeed in the west. 
The sources for what was considered the second comet suggested that the
observation on June 9 was a first detection of a new comet:
it is plausible that this transmission originated from a different source and/or place,
where due to rainy and foggy weather (see JTS), the observers did not see the comet before conjunction:
maybe, the exact position provided ({\it Xuanyuan}) was later lost or got separated from the date --
{\it Xuanyuan} is still present in all sources of the first report.
This would then be a newly derived dated position, which was not considered
before, in particular not for solving the orbit. The observation in
{\it Xuanyuan} then took place at the beginning of the night June 9/10 
at Chang'an, now Xi'an, mainly during astronomical twilight and a bit later,
around June 9.58 UT ($\pm 1$h). Still, because the combination of location and date is
not fully certain, we will use in the orbit fit only the position, not the date. (The orbital fit below, Sect. 5, then
provides a date for the position in {\it Xuanyuan}, namely indeed June 9.)

The text in THY gives the west azimuth:
``intercalary 4[th] month, 21[st] day (June 9), [an] ominous star seen at west
({\it xi}) direction''; XTS also gives the {\it west} azimuth. Therefore, we can constrain the position for
June 9 to azimuth $270 \pm 45^{\circ}$.\footnote{Stephenson \& Yau
(1985) did not consider these data for the orbital reconstruction.}
Together with the {\it Chronicle of Zuqn}\textit{\={\i}}{\it n}, 
which gives NW quarter for this time span, we can reduce this azimuth range to $292.5 \pm 22.5^{\circ}$. 
(The above determined position in {\it Xuanyuan}, probably on June 9, would also be within this
azimuthal range.)

\smallskip

(C5) The final position is given as ``reaching {\it Taiwei Youzhifa} 7 {\it cun} position'' (JTS); 
THY: ``reaching {\it Taiwei} west ({\it xi}), {\it Youzhifa} west ({\it xi}) 7 {\it chi}''; XTS omitted
``{\it Taiwei}'' and the separation, it has only ``reaching {\it Youzhifa} west ({\it xi})''. 
The Chinese term for ``reaching'' ({\it zhi}) is regularly used for specifying the last position
(see, e.g., comets up to AD 800 in Pankenier et al. 2008). The JTS record gives a precise position, just 7
{\it cun} within $\beta$ Vir, the star {\it Youzhifa} in the western wall of the enclosure
{\it Taiwei} (``Privy Council'', the ``Spring Palace of {\it Huangdi}'', SK97); 7
{\it cun} correspond to only about 0.7 to $1^{\circ}$, indeed, the
Chinese could give precisely an angle as small as about 0.7 to $1^{\circ}$ (Kiang 1972, Kiang 1980, Stephenson \& Green 2002);
another such case is found in AD 821 March 7 (JTS): ``about 7
{\it cun} from the first star'' (Pankenier et al. 2008). 
Conservatively, we use a position of $1^{\circ}$ around $\beta$ Vir\footnote{Stephenson \& Yau
(1985) use $7^{\circ}$ west of $\beta$ Vir, based on THY,
partly because the YK81 orbit would not allow a much closer approach, but the YK81 orbit is based on the positions in
Kiang (1972), who used $1^{\circ}$ west of $\beta$ Vir on
``around July 9'' (the 54th day of those ``more than 50 days'') in their orbit fit leading to ``rather more than $1^{\circ}$''
(namely $3.5^{\circ}$) separation from $\beta$ Vir on July 9.
The orbital fit by Kiang (1972) for July 9 ended up at right ascension $158.2^{\circ}$ and declination $7.3^{\circ}$, 
but this would be closer to upsilon Leo, the brightest star in the asterism
{\it Mingtang}, so that this would have been mentioned in the
historical report; upsilon Leo does not fit to 7 {\it chi} nor
7 {\it cun} off or west of $\beta$ Vir. In Stephenson \& Yau (1985), the closest
approach to $\beta$ Vir would be $3.7^{\circ}$ on July 12, and
a position of $7^{\circ}$ west of $\beta$ Vir would be reached
on July 3, considered as ``last visibility'', but this is less than ``more than 50 days'' since May 17; a separation of
$3.7^{\circ}$ would not favour a figure of $7^{\circ}$ off $\beta$ Vir
from THY (compared to about $1^{\circ}$ from JTS).} -- the oldest text (JTS) did not specify that it
was west of $\beta$ Vir. We will test the $1^{\circ}$ error circle in our independent orbit fit for AD 760 below.

Several practical and text-critical arguments speak in favour of such a small and precise
value instead of 7 {\it chi} for about $7^{\circ}$ in the later THY.
If the comet would have been $7^{\circ}$ west of $\beta$ Vir, it would have been closer to $\sigma$ Leo (in Chinese
``{\it Xi shangjiang}'' for ``west upper general'', SK97),
both on the horizontal and the equatorial system, so that this star would have been mentioned in comparison. In the THY
text, not only the separation ``7 {\it chi}'' is corrupt,
but it also brings ``west'' twice: ``reaching {\it Taiwei} west, {\it Youzhifa} west 7 {\it chi}''. 
While $\beta$ Vir indeed is in the western
part of the {\it Taiwei} enclosure, it would not be necessary
to specify this, because the Chinese name of $\beta$ Vir,
``{\it Youzhifa}'', literally means ``right law
administrator'' (SK97) or ``right gate keeper'' or ``enforcer of the right'' gate of
{\it Duanmen}, ``right'' ({\it you}) on sky-view is ``west''; the two stars $\beta$
Vir and $\eta$ Vir at the southern ends of the western and eastern
{\it Taiwei} walls, respectively, were very significant as the
two gate keepers or enforcers of the gate {\it Duanmen} (SK97). 
The hint ``{\it Youzhifa} west'' in THY and XTS may mean 7 {\it cun} west of
{\it Youzhifa}, and it is indeed included in our $1^{\circ}$ error
circle around $\beta$ Vir. 

Furthermore, the {\it Taiwei} enclosure had particular importance in Chinese astro-omenology:
while the north polar region was considered the palace, so that moving 
object (e.g. comets) or (other) guest stars could be used as portents for the court, 
the planets and the Moon of course do not move through the north polar region; hence, an additional
area on sky was considered also relevant for emperor and court, which was close enough
to the ecliptic, namely the {\it Taiwei} enclosure (Ho 2003, p. 142). The movement of the comet toward
$\beta$ Vir, the right gate keeper of this enclosure, was therefore certainly closely watched --
and we can trust precise measurements.

Down to which magnitude could the Chinese astronomers detect the comet so close to $\beta$ Vir?
Good naked-eye observers can resolve Mizar and Alcor (2.3 and 4.0 mag, respectively) at $0.2^{\circ}$
separation (both known to pre-telescopic Arabic astronomers).
In Ptolemy's Almagest star catalog (Toomer \& Ptolemy 1984), there are also a few close faint pairs,
e.g. ``the nebulous and double star at the eye [of Sgr]'', 
which are $\nu ^{1}$ (V=4.9 mag) and $\nu ^{2}$ Sgr (V=5.0 mag) at a separation of only $0.2^{\circ}$ 
at the epoch of the Almagest star catalog.
Furthermore, Bedouine observers considered $\nu$ Cap (4.75 mag)
as the {\it sheep} of $\alpha$ Cap ($\alpha^{1}$ Cap has 3.55 mag), see Kunitzsch (1961, p. 101),
their separation at epoch AD 760 was 1/3 degree only.
While it may be uncertain whether {\it Tang} dynasty astronomers did resolve these pairs,
they did present other close pairs separately on the {\it Dunhuang} maps:
{\it Zui} with $\phi ^{1}$ Ori (4.4 mag),
$\phi ^{2}$ Ori (4.1 mag), and $\lambda$ Ori
(3.7 mag) with mutual separations from 0.4 to $0.7^{\circ}$, and also $\lambda$ Sco (1.6 mag) and $\upsilon$ Sco (2.7 mag)
in {\it Wei} at $0.6^{\circ}$ separation (epoch AD 700). Hence, the
Chinese court astronomers should have been able to resolve a comet down to about 5.5 to 6 mag
separated by about $1^{\circ}$ from $\beta$ Vir (3.6 mag), even if without tail.

The duration of visibility is given in JTS as ``in all more than 50 days,
only then [it] disappeared'' (XTS: ``[it was] not seen'' instead of ``disappeared`''). In
other observing records of comets, we find the wording of ``more than'' in combination with
days in 10-day steps (Pankenier et al. 2008), e.g. 50 or 60 days; hence, ``more than 50
days'' means somewhere from 51 to 59 days. The 51st day since May 17 is July 6, so that the comet disappeared some time
from July 6 to 14. This is consistent with the specification in XTS 32.838, ``only [when] reaching 5[th] month, 
[it] disappeared'' (similar in THY), which means that it went out of sight sometime during the 5th
lunar month ending on July 16 (new-moon on June 17 and July 17). 
(This is also consistent with the {\it Chronicle of Zuqn}\textit{\={\i}}{\it n}, see Z5.)

While the number of days in connection with ``in all'' often are precise to the very day
(e.g. Pankenier et al. 2008), the text seems to indicate that around the 50th day since May 17, the brightness of the
object was still sufficient to measure a position as precise as ``7 {\it cun}''. 
The exact date may not matter much, because
normally comets do not move much at the end of their visibility; we use July $5 \pm 2$ days ($1 \sigma$
error) as date for the last measurement for the orbit fit. The
observation in the early evening of July 5/6 ($\pm 2$d) at Chang'an, now
Xi'an, was then at around July 5.6 ($\pm$ 2d) UT.

We note that the first Chinese sighting on May 17 fits well with the Chronicle of {\it Zuqn}\textit{\={\i}}{\it n}
(May 18), also the motion toward the NE corner ({\it Zuqn}\textit{\={\i}}{\it n}: until the night May 31/June 1), and the specifications
toward the end of visibility: for June 29 or a bit later, the comet was newly detected in the SW quarter ({\it Zuqn}\textit{\={\i}}{\it n}) --
$\beta$ Vir and, hence, the nearby comet were in the SW [quarter] at the beginning of astronomical twilight.

\section{The comet of AD 760 as reported in further East Mediterranean and West Asian chronicles}

In addition to the Chronicle of {\it Zuqn}\textit{\={\i}}{\it n}, the comet of AD 760 was also mentioned as
eastern and western comet in four other East Mediterranean and West Asian chronicles:

(a) Theophanes (died AD 817 in Byzantium) wrote his world chronicle in his last years: \\
``In the same year [AD 760/1] a brilliant apparition appeared in the east for ten days and again in the west
for twenty-one'' (Turtledove 1982).

A duration of 10 days in the east (before conjunction with the Sun) and 21 days in the west
(after conjunction) is slightly shorter but consistent with the reports from {\it Zuqn}\textit{\={\i}}{\it n} (and China) for the comet of AD 760. 
Theophanes' chronology is sometimes uncertain by 1-2 yr (Mango \& Scott 1997)
-- here one year, since the Byzantine year runs from AD 760 Sep 1 to 761 Aug 31.

\smallskip

(b) Nu$^{c}$aym ibn \d{H}amm\=ad (died AD 843 in Baghdad, Iraq): \\
``We saw the comet rising in {\it Mu\d{h}arram} in the year [Anno Hijra, AH 145
= AD 762/3] with the dawn from the east, and we would see it during the dawn for the rest of
{\it Mu\d{h}arram}; then it disappeared. Then we would see
it after the sunset in the twilight, and afterwards between the north and the west for two
month or three. Then it disappeared for two or three
years.''\footnote{This
report by Nu$^{c}$aym ibn \d{H}amm\=ad then continues with: ``Then we saw a mysterious star with blazing fire the length
of two degrees, according to what the eye saw, near Capricorn, orbiting around it like the orbit of a planet during the
months of Jumada [July] and [some of the] days of Rajab [Oct] and then it disappeared'' (as translated by Cook 1999
with his additions in brackets, dated by him to AH 145 = AD 762/3). This and the following reports definitely do not
belong to the comet in AD 760.}

Given other dating errors in this quite apocalyptic Hadith collection, it may be dated to AH
143, i.e. AD 760/1 (Cook 1999). With new-moon on AD 760 Apr 20 and May 19, the month of
{\it Mu\d{h}arram} would run from AD 760 about Apr 21 to May
20 ($\pm 1$ or 2 days depending on the first detection of the 
crescent moon), but the comet of AD 760 did not
disappear at around May 20. However, the source used by Nu$^{c}$aym ibn \d{H}amm\=ad could have given the date on a
western calendar system, e.g. as May, which would have been converted loosely to
{\it Mu\d{h}arram}, probably based on a Christian source
using e.g. the West Syrian Seleucid calendar as, e.g., the Chronicle of {\it Zuqn}\textit{\={\i}}{\it n}.
A scribal error is then required
only for the year number (AH) ``145'', which should be 143.
Then, the text would be fully consistent with the Chronicle of {\it Zuqn}\textit{\={\i}}{\it n}:  
seen first since some time in the month of May of AD 760 in the morning dawn (``with the dawn'', {\it Zuqn}\textit{\={\i}}{\it n}:
``{\it \v{s}afr\=o}'') in
the east and also like that for
the rest of that month ``during the dawn'' ({\it Zuqn}\textit{\={\i}}{\it n}:
``{\it n\=ogah}'') -- instead of ``rest
of {\it Mu\d{h}arram}'', we should read ``rest of
{\it iyy\=or}/May''; the Chronicle of {\it Zuqn}\textit{\={\i}}{\it n}
reported the last visibility before
conjunction with the Sun for the early
morning of the night May 31/June 1 (``Pentecost'').
Then, according to Nu$^{c}$aym the comet was seen after conjunction
``after the sunset in the twilight {\dots} between the north and the west for two month or three'', i.e. again similar
as in {\it Zuqn}\textit{\={\i}}{\it n} (for 25 evenings in the NW and later again for ``many days''), after conjunction the comet was
definitely seen in two different months (June and July). When Nu$^{c}$aym ibn \d{H}amm\=ad mentioned a reappearance ``two
or three years'' later, he could either
mean some other comet or transient object, or he could have interpreted the
text in the Chronicle of {\it Zuqn}\textit{\={\i}}{\it n}, which is found in the report for SE 1075 (AD 763/4),
which is, however, again about the comet of AD 760:
``The effect of 'the broom' seen a short while before, was clearly seen in reality, as it
swept the world like a broom that cleans the house'' (Harrak 1999), see Sect. 2.1 for full citation
(given that the Chronicle of {\it Zuqn}\textit{\={\i}}{\it n} does
not mentioned any other comet or celestial sign in between the comet report in AD 760 and this short
statement later, it is likely that the latter short note points to the comet of AD 760). Therefore, given all the
similarities (except the offset by 2 years), it is likely that the (direct or indirect) source used by
Nu$^{c}$aym ibn \d{H}amm\=ad is the Chronicle of {\it Zuqn}\textit{\={\i}}{\it n} --
this would be the first hint that our chronicle was active before been buried in a Sinai monastery in the 9th century.

\smallskip

(c) Agapius of Manbij (died AD 941/2, Melkite bishop of Manbij (Syria), author of a world
chronicle running until the 770ies): \\
``In this year [AD 760] the star with a tail appeared, and it was in
Aries before the Sun, and the Sun was in Taurus. It proceeded until it was under the rays of the Sun, then went behind
it and stayed 40 days'' (Cook 1999, Vassiliev 1911), from which Cook remarks:
``This observation is probably from Theophilus of Edessa himself''. (Note that Cook (1999)
incorrectly gave ``and the Sun was in Leo'', while the text clearly gives Taurus, see Vassiliev 1911.)
We will discuss
this text with the next one, because they both depend on the same source(s).

\smallskip

(d) Michael the Syrian (AD 1126-1199, a patriarch of the Syriac Orthodox Church since AD 1166,
author of a world chronicle): in the French translation of Michael's Chronicle by Chabot (1899-1910),
this comet report was related to the
{\it Zuqn}\textit{\={\i}}{\it n} report on a comet in SE 1080 (AD 768/9), see Harrak (1999, p. 226) --
presumably dated SE 1076 by Michael. The comet of SE
1080 in {\it Zuqn}\textit{\={\i}}{\it n} was actually seen in AD 770 according to other well-dated oriental sources,
while the comet record by Michael, presumably dated to SE 1076, obviously means the comet of AD 760 May (SE
1071):\footnote{Our translation is based on Ibrahim (2009, pp. 477-478);
for a French translation, see Chabot (1899-1910).}

``And in this year in the month of {\it iyy\=or} (May), a comet star
[{\it kawkb\=o qumi\d{t}us} -- the latter obviously from the Greek {\it kometes}] was seen before the Sun in Lamb
(Aries), when the Sun was in Taurus. It looked like a pillar/column
[{\it `\=om\=ud\=o}] and was extending its rod\footnote{What is translated as ``rod'' points to the tail. Chabot
translated it as ``chevelure'', i.e. ``lock of hair'' (Chabot, p. 524 n.3). The 16th century Edessan manuscript
emended the term {\it \v{s}abuq\=o} (``rod'') to {\it \v{s}\=obq\=o} (``emission''),
and Chabot interpreted this word as (curled) hair.} to the south. It moved
close to the Sun for twenty days, 
and became below the rays of the Sun for three days. Thereafter, it was behind the
Sun for forty days. Due to its appearance, fear gripped everyone.''

The expression ``and in this year'' refers to SE 1076 (AD 764/5), if related to the preceding account
which deals with an earthquake in Khorasan. However, chapter 25 of Michael the Syrian, in which the comet report appears, covers
the period from SE 1066 to 1076 -- this cast doubt about the expression ``and in this year''. When Michael the Syrian
quotes large texts, he names his sources, but when he gathers information to include in a Chapter, he picks and copies,
but not necessarily in chronological order.

This text by Michael the Syrian on a comet has
many similarities to the previously studied reports, the relationship to Agapius of Manbij is obvious: Michael ``a comet star'',
Agapius ``the star with a tail''; Michael ``before the Sun in Lamb'', Agapius ``in Aries before the Sun''; Michael
``moved close to the Sun for twenty days'', Agapius ``It proceeded''; Michael ``became below the rays of the Sun'',
Agapius ``it was under the rays of the Sun''; Michael ``Thereafter, it was behind the Sun for forty days'', Agapius
``then went behind it and stayed 40 days''. Agapius says for the first sighting ``the Sun was in Taurus'' (before comet
conjunction with the Sun), and Michael says the same: ``when the Sun was in Taurus''. There is otherwise no
information in Agapius that is not found in Michael. However, Michael has a few extra details: ``month of
{\it iyy\=or} (May)'', ``looked like a pillar/column and was
extending its rod to the south'' (c.f. Chronicle of {\it Zuqn}\textit{\={\i}}{\it n}: ``one was tilting to the south side''),
``close to the Sun for twenty days'', ``below the rays of the Sun for three days'', and the final astrological
interpretation (``fear''). All information, in particular also the ``3 days'' of invisibility due to conjunction with
the Sun, are fully consistent with the Chronicle of {\it Zuqn}\textit{\={\i}}{\it n}, and actually confirms both the Chronicle and our close
reading, e.g. regarding the first sighting in {\it iyy\=or}
(May) and the number of days of invisibility due to conjunction with the Sun.

The specification ``It moved close to the Sun for twenty days'' for the time before
conjunction with the Sun (June 1) could point to a slightly earlier discovery than by all other observers (China since May 17;
{\it Zuqn}\textit{\={\i}}{\it n} May 18). This consideration is supported by the
specification ``Sun in Taurus'', which is an alternative dating, namely the month when the Sun was thought to be in the
zodiacal sign of Taurus, which was since antiquity set to Apr 17 to May 17 (afterwards in
Gemini);\footnote{This rule is mentioned by, e.g., Pliny in his {\it Natural
History}, namely that the Sun is located in a zodiacal sign from the 15 Calends to 16 Calends of the
next month, also otherwise still in use in the 8th/9th century, e.g. by the Carolingians based on Bede citing
Pliny (Wallis 1999, p. 86). According to the 10th century Calendar of Cordoba, Al-Batt\=ani (AD ca. 858-929)
would have said that the Sun enters Aries on Mar 16 (McCluskey 1998, p. 167). According to Byzantine practice
(John of Damascus, AD 675/6-749), the Sun would be in Taurus Apr 23 to May 23 (Tihon 1993).} if
the same rule was applied with this date range, the
observation started before May 18. The period of invisibility ``for three days'' during conjunction with the Sun (last
seen by {\it Zuqn}\textit{\={\i}}{\it n} the night May 31/June 1 in the morning) points to a reappearance on
June 3 or 4 in the evening (see Sect. 3.6, position Z5).
A visibility of 40 days after conjunction with the Sun is consistent with {\it Zuqn}\textit{\={\i}}{\it n}
(and China); the duration of 20 and 40 days could, however, be somewhat rounded.

According to Cook (1999), the text by Agapius (and then also parts from
Michael) are probably based on an observation by Theophilus of Edessa, a Maronite Christian
astrologer/astronomer, AD 695 (Edessa) to 795 (Baghdad). The transmission to both Agapius and Michael could also
originate from the otherwise lost Chronicle of Dionysius of Tell-Ma\d{h}r\=e (died AD
845), a major source for Michael the Syrian, whom he extensively quoted for the 8th and the first part of the 9th
century.

The comet of AD 760 has been reported and transmitted extensively by Christian scholars,
and Nu$^{c}$aym ibn \d{H}amm\=ad, the Muslim author of a Hadith collection,
may had one of those Christian records as his source, namely the Chronicle of {\it Zuqn}\textit{\={\i}}{\it n}.

\begin{landscape}

\begin{figure}
\begin{center}
\includegraphics[angle=0,width=22cm]{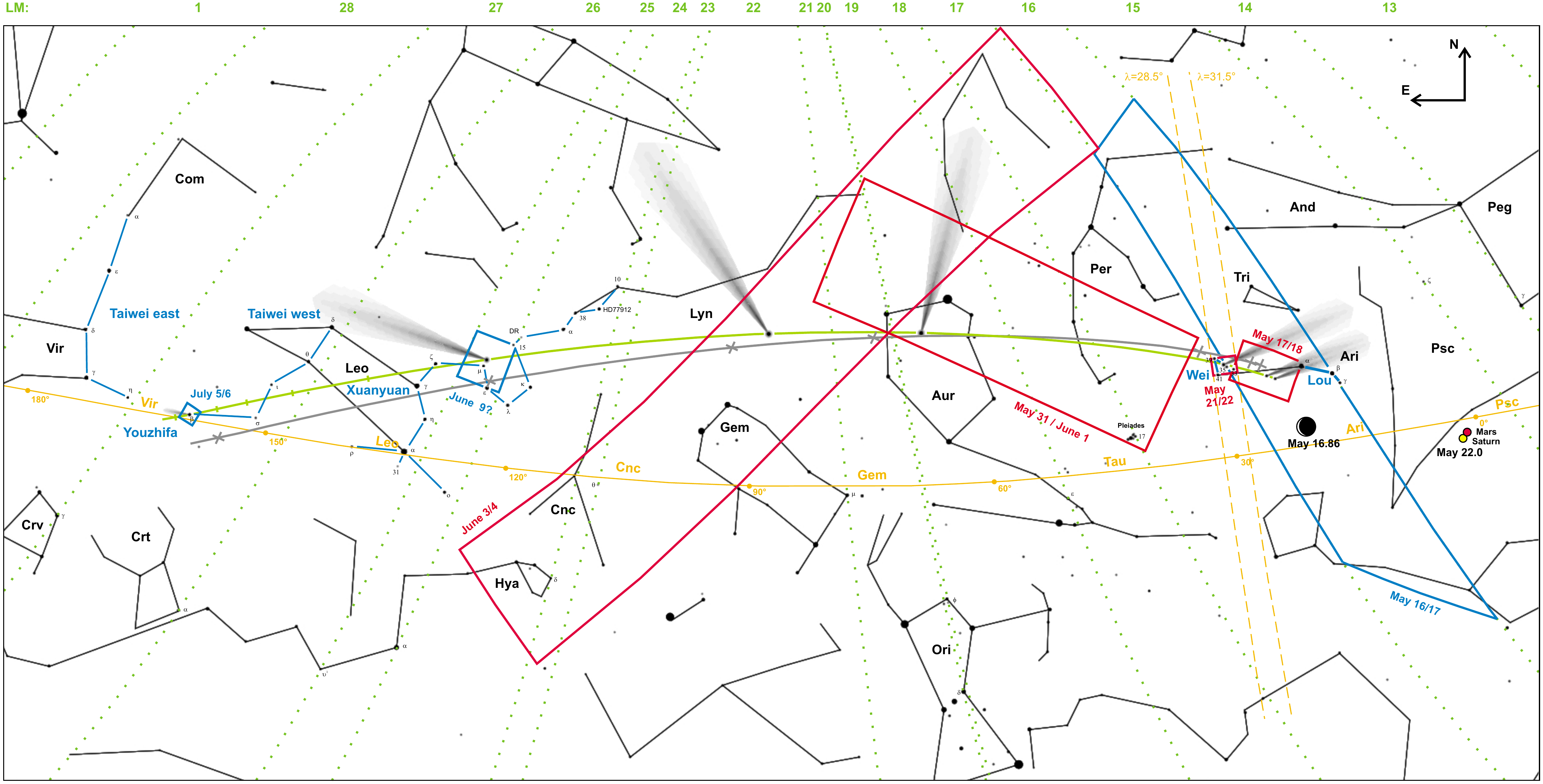}
\end{center}
\caption{
{\bf This equatorial plot shows the whole comet path for the AD 760 perihelion}
{\rm (north up, east left):
borders of the Chinese lunar mansions (right ascension ranges) are indicated as
dotted lines (numbers on top) always going through their determinative star.
The Chinese asterisms Lou, Wei, Xuanyuan, and Taiwei east and west (with Youzhifa in Taiwei west)
are shown as blue skeletons. The ecliptic is shown as orange line with dots every
$30^{\circ}$ to indicate the borders of zodiacal signs.
The planets Mars and Saturn are shown in close conjunction for May 22 at 0h UT, 
the waning crescent moon for May 16.86 (UT). 
The positions derived here for the comet (Table 1) are shown as coloured boxes.
Our new orbit is shown in green, the previous (YK81/JPL) orbit in grey, both
from May 17 at 0h UT (in Aries, right) until July 15 at 0h UT (in Virgo, left), 
comet positions are shown as {\it comet symbols} on our new orbit on the dates of observation,
and as crosses on the old orbit; 
the software {\it Cartes du Ciel} v3.10 shows the plasma tail directed away from the Sun.
We can see that on June 9.58 (UT), the comet is indeed in Xuanyuan.
Tick marks are shown on the new orbit 
on June 14.0, 19.0, 24.0, 29.0, and July 10.0 (UT) showing that the comet slowed down
and got too faint (Sect. 5).}
}
\end{figure}

\end{landscape}

\begin{landscape}

\begin{table}
\caption{{\bf Dated positions of the comet in AD 760 from historical observations} {\rm (coordinates for epoch of date)}}
\begin{tabular}{llllll}
\hline
Loc. & Date 760  & Text regarding position and date (1) & \multicolumn{2}{c}{Coordinates / Obs. time (UT)} \\ \hline
Chin. & May 16/17 & C1: east direction ... located/situated in {\it Lou} (LM 16) & $\alpha$=00:48-01:33 & azi=$90 \pm 45^{\circ}$ \\
(2)   & Fig. 7  & {\it Tang} Emperor Suzong ... {\it Qianyuan} [reign-period] ... & ($\alpha$ in hh:mm) & alt $\ge 0^{\circ}$ \\ 
      &         & ... 3[rd] year, 4[th] month, {\it dingsi} (54) night, 5[th] watch & \multicolumn{2}{c}{(3)~~~~~May 16.86 ($\pm$ 1h)} \\ \hline
Zuq. & May 17/18 & Z1: Aries, to the north from these three stars in it ... very shining & azi=65.3-73.1$^{\circ}$ & alt=2-7$^{\circ}$ \\
(4)  & Fig. 3 & [SE] 1071 iyy\=or\,...\,before early twilight\,...\,15 nights, until dawn\,...\,Pentecost & \multicolumn{2}{c}{(5)~~~~~May 18.0 ($\pm$ 1h)} \\ \hline
Zuq. & May 21/22 & Z2: in/at the initial degree [of] the second [sign] (i.e. Taurus) ... & $\lambda$=28.5-31.5$^{\circ}$ & $\beta$=10.4-12.4$^{\circ}$ \\
     & Fig. 4    & ... still in the same Aries at its edge/end/furthest part & & \\
     &  & [SE]\,1071 iyy\=or (May) ... on [day] 22 in the same month & \multicolumn{2}{c}{(6)~~~~~May 22.0 ($\pm$ 1h)} \\ \hline
Chin. & after May 17 & C2: moved toward east north corner & azi=22.5-67.5$^{\circ}$ & & \\ \hline
Zuq. & May 31/Jun 1 & Z3: it was going bit by bit to the North-East [direction] & azi=$45 \pm 22.5^{\circ}$ & alt=0-15$^{\circ}$ \\
     & Fig. 5       & sign itself remained for 15 nights, until dawn of the feast of Pentecost & \multicolumn{2}{c}{(7)~~~~~June 1.0 ($\pm$ 0.7h)} \\ \hline
Zuq. & \multicolumn{2}{l}{Z4: June 1/2, probably also 2/3 (possibly also 3/4): not seen, i.e. close to the Sun} & \multicolumn{2}{c}{close to the Sun} \\ 
\multicolumn{6}{l}{Michael of Syria: below the rays of the Sun for three days, i.e. not seen at least June 1/2 and 2/3} \\ \hline
Zuq. & June 2/3,  & Z5: seen ... from the north-west [quarter] & azi=270-360$^{\circ}$ & alt=0-15$^{\circ}$ \\
     & 3/4, or 4/5 & at the beginning of [the] 3rd [day] after Pentecost ... at evening time & \multicolumn{2}{c}{(8)~~~~~June 3.74 ($\pm$ 1d)} \\ 
     & Fig. 6    & \multicolumn{3}{l}{(June 2/3 excluded from Michael of Syria)} &  \\ \hline
Chin. & (9) Fig. 8 & C4: ... {\it Xuanyuan} (but not in LM 24 nor 26) & \multicolumn{3}{l}{$\alpha$=8h39$\pm 13$m (9) $\delta$=$31.9 \pm 3.1^{\circ}$} \\ \hline
Zuq.  &             & \multicolumn{3}{l}{June 3 or 4 to 28 or 29: seen ... from the north-west [quarter] ... for 25 evenings (see Z5)} & \\
      &             & \multicolumn{3}{l}{June 2/3 excluded from Michael of Syria (see Z4 and Z5)} &  \\
Chin. & June 9/10 & C4: THY: ominous star seen at west direction & azi=270-315$^{\circ}$ & alt=0-90$^{\circ}$  \\
 & \multicolumn{2}{l}{THY: intercalary 4[th] month, 21[st] day} & \multicolumn{2}{c}{(10)~~~~~June 9.58 ($\pm$ 1h)} \\ \hline
Chin. & July 5/6  & C5: reaching {\it Taiwei Youzhifa} 7 {\it cun} position ($=0.7^{\circ}$ off $\beta$ Vir) & $\alpha$=10h46$\pm 4$m & $\delta$=$8^{\circ}38^{\prime} \pm 1^{\circ}$ \\
      & Fig. 8    & [since May 17] in all more than 50 days, only then [it] disappeared & \multicolumn{2}{c}{(11)~~~~~July 5.6 ($\pm$ 2d)} \\ \hline
\end{tabular}

{\rm \raggedright Remarks: \\
\raggedright (1) Section, where we explain this dated position, is indicated
by, e.g., Z1 or C1 meaning first {\it Zuqn}\textit{\={\i}}{\it n} or first Chinese dated position, respectively. \\
\raggedright (2) {\it Tang} dynasty capital (now Xi'an, China, $l=108^{\circ} 57^{\prime}$, $b=34^{\circ} 16^{\prime}$), JTS text unless otherwise stated. \\
(3) Right ascension ($\alpha$) range for LM 16; azimuth (azi) for {\it east direction}; altitude (alt) above horizon.\\
(4) Monastery of {\it Zuqn}\textit{\={\i}}{\it n} near Amida (now Diyarbak\i r, 
Turkey, $l=40^{\circ} 13^{\prime}$, $b=37^{\circ} 55^{\prime}$). \\
(5) Azimuth from $\alpha$ to 33 Ari and altitude from 41 to $\beta$ Ari. \\
(6) Ecliptic longitude {\it initial degree [of] the 2nd [sign]} (Taurus) would be $\lambda \simeq 30^{\circ}$,
with an error of $\pm 1.5^{\circ}$; ecliptic latitude range for 41 to 39 Ari. \\
(7) Altitude above {\it Zuqn}\textit{\={\i}}{\it n}'s horizon at azimuth NE during astronomical morning twilight or earlier. \\
(8) Altitude above {\it Zuqn}\textit{\={\i}}{\it n}'s horizon at azimuth NW during astronomical or nautical evening twilight on June 3 or 4. \\
(9) Most likely June 9 (see C4), but date not used for orbital fit; right ascension ($\alpha$) range for LM 25, declination ($\delta$) range for the stars
of {\it Xuanyuan}, which are also in LM 25, i.e. 15, $\mu$, and $\epsilon$ Leo. \\
(10) Western azimuth for June 9 from China and NW azimuth from {\it Zuqn}\textit{\={\i}}{\it n} ``for 25 evenings''
since ``third [day] after Pentecost'' (including June 9) together. \\
\raggedright (11) Conservatively, $1^{\circ}$ around {\it Youzhifa} ($\beta$ Vir).}
\raggedright
\end{table}

\end{landscape}

\section{Reconstruction and discussion of cometary orbit for AD 760}

To summarize, the author of the Chronicle of {\it Zuqn}\textit{\={\i}}{\it n} gives five, maybe even six dates with
positions (mornings of May 18 and 22 relatively precise).

Modern studies of the Chinese records mention only two dates with positions, namely for the
first and last observation (e.g. Kiang 1972). Our close reading and historical-critical interpretation of the Chinese
records yield corrected dates and positions as well as constraints on further positions.

In total, we can derive from the Chinese and Syriac three dated positions with relatively small error bars 
(May 17/18, 21/22, and July 5/6), four
with less well constrained positions, plus one well-constrained position without a certain date (in
{\it Xuanyuan}, possibly on June 9). 
This may allow to fit an orbit for this perihelion passage only from historical sources (see Tables 1-3).

We used the software {\it find$\_$orb} (version Nov 6, 2017, projectpluto.com/find$\_$orb.htm)
for fitting an orbit based on the dated positions in Table 1.
Due to relatively large astrometric ($\pm 1^{\circ}$ and more) and timing ($\pm 0.7$h and more)
uncertainties of these observations, a bundle of Keplerian orbital solution exists, which all fit the given astrometric
data of the comet, but yield very different orbital properties. In order to explore the characteristics of these
well-fitting orbits, in numerous runs the astrometric position as well as the observing time of the comet were randomly
chosen for all observing dates within their given astrometric and timing uncertainties. In each run the best fitting
Keplerian orbital solution was then determined by least squares fitting, which yielded the orbital elements of this
orbit together with their uncertainties, taken from the derived covariance matrix. Thereby only runs were considered
for further orbit characterization, which yielded orbital solutions with
$\chi^{2}_{\rm red}$ {\textless} 2, and 1 million such runs
were carried out in total. 

The vast majority of all runs ($99\%$) results in non-periodic orbital solutions (eccentricity $e \geq 1$).
The Keplerian elements obtained for the non-periodic solutions (e=1) are
displayed in Fig. 9 and compared in Table 2 with those for periodic solutions and the
current JPL orbits for AD 760 and 1986.

If the comet had an eccentricity $e \geq 1$, it would have had only one perihelion passage in the past
(in AD 760). It is considered that the vast majority of naked-eye comets are long-periodic comets.
The time span with particulary good data is relatively short,
and one can see some potential in pre-telescopic comet records (see Introduction),
so that it may be feasible to find more intermediate-period comets among them.
In the context of our test case (comet of AD 760 with orbital solution from historical observations),
we have to clarify under which conditions an identification with a known and, hence, periodic comet is possible.

According to the classical approach of Halley (1705, 1749), the orbital elements for
perihelion passages have to be similar (except $\omega$ and $\Omega$, which can change substantially
from perihelion to perihelion, also of course $T$, see e.g. Yeomans \& Kiang 1981 for 1P/Halley) -- or, even more,
the periods between perihelia should be similar. Halley first tried parabolic solutions and then,
since three had similar elements, he tried eccentric orbits and predicted the next return (Halley 1705, 1749).

In our test case of comet AD 760, the parameters for parabolic and periodic solutions are fully consistent
with the elements of 1P/Halley according to the JPL orbit (Table 2).

Nota Bene: For a comet with some 77 yr period, it is principally problematic to solve for the orbit with
data from only a few months. Also Edmund Halley must have had this problem.
In a similar approach, we have also solved for the orbit of the comet of AD 837,
which turned out to be fully consistent with those of AD 760 and, again, comet 1P/Halley
(D.L. Neuh\"auser et al., in prep.; first results in D.L. Neuh\"auser et al. 2018c and Mugrauer et al. 2018).
The time between AD 760 and 837 is also fully consistent with the known period range of 1P/Halley.
With the orbital elements q and e for closed orbits (Table 2), we estimate the period from the
AD 760 data to be $76.5 \pm 6.7$ yr.
(Furthermore, the light curve estimated with the absolute brightness and activity parameter
as previously suggested for 1P/Halley, e.g. from recent telescopic
observations, yield intrinsically consistent results for the AD 760 observations, see Fig. 14 below.)

Even when such conditions are fulfilled (and were also fulfilled in E. Halley's calculations),
it cannot be totally excluded that a very unlikely coincidence happened, namely
that the comet in AD 760 and also that one in AD 837 were both (different) non-periodic comets
with very similar parameters -- and similar to 1P/Halley (which then had to remain unobserved in AD 760 and 837).
Even a confirmed prediction of a next return cannot prove the opposite.
The comet cannot be observed all the time, but only near perihelion.

Since the comet of AD 760 shows clear indications of being consistent with a periodic comet,
we will now consider only those solutions with eccentricity $e < 1$.

\begin{figure}[ht]
\begin{center}
\includegraphics[angle=0,width=16cm]{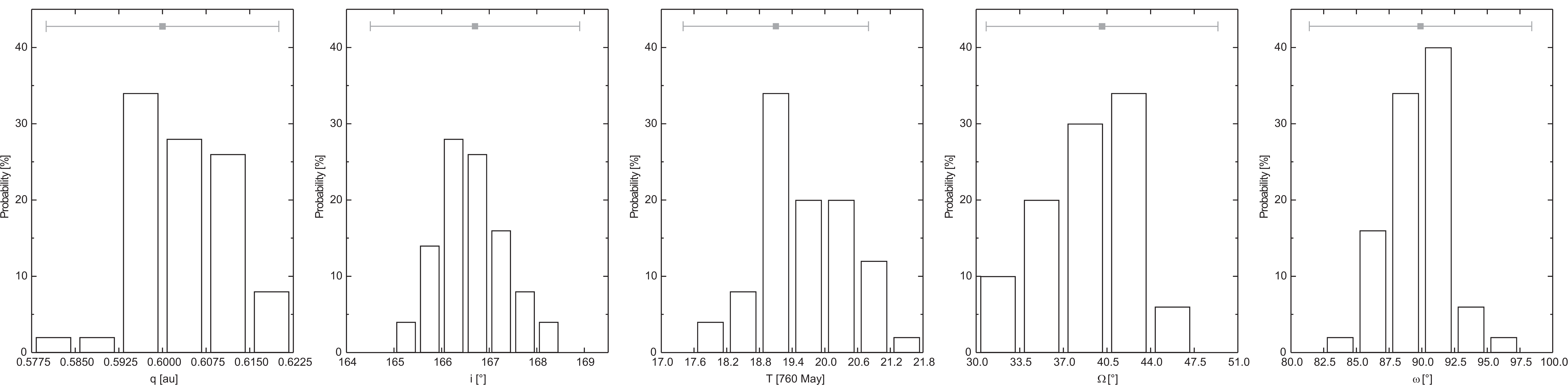}
\end{center}
\caption{{\bf Keplerian elements for non-periodic solutions} (eccentricity e=1):
They can be compared to (and are fully consistent with) the parameter ranges for periodic solutions
with semi-major axes 17-19 au (Table 2, Fig. 13)
shown as grey data points with error bars in the upper parts of the graphs.}
\end{figure}

In 12349 runs periodic Keplerian orbits were obtained, all
with perihelion distances q larger than the solar radius, i.e. all are possible periodic orbital solutions around the
Sun. Most of these closed orbits ($\sim 90\%$) are eccentric (0.33 {\textless} e {\textless} 0.99) and exhibit small
semi-major axes in the range between 0.44 and 5 au (up to 11 yr period). The remaining orbital solutions are wider
orbits (semi-major axis a {\textgreater} 5 au) which are all highly eccentric (e {\textgreater} 0.87). 

\begin{figure}[ht]
\begin{center}
\includegraphics[angle=0,width=12cm]{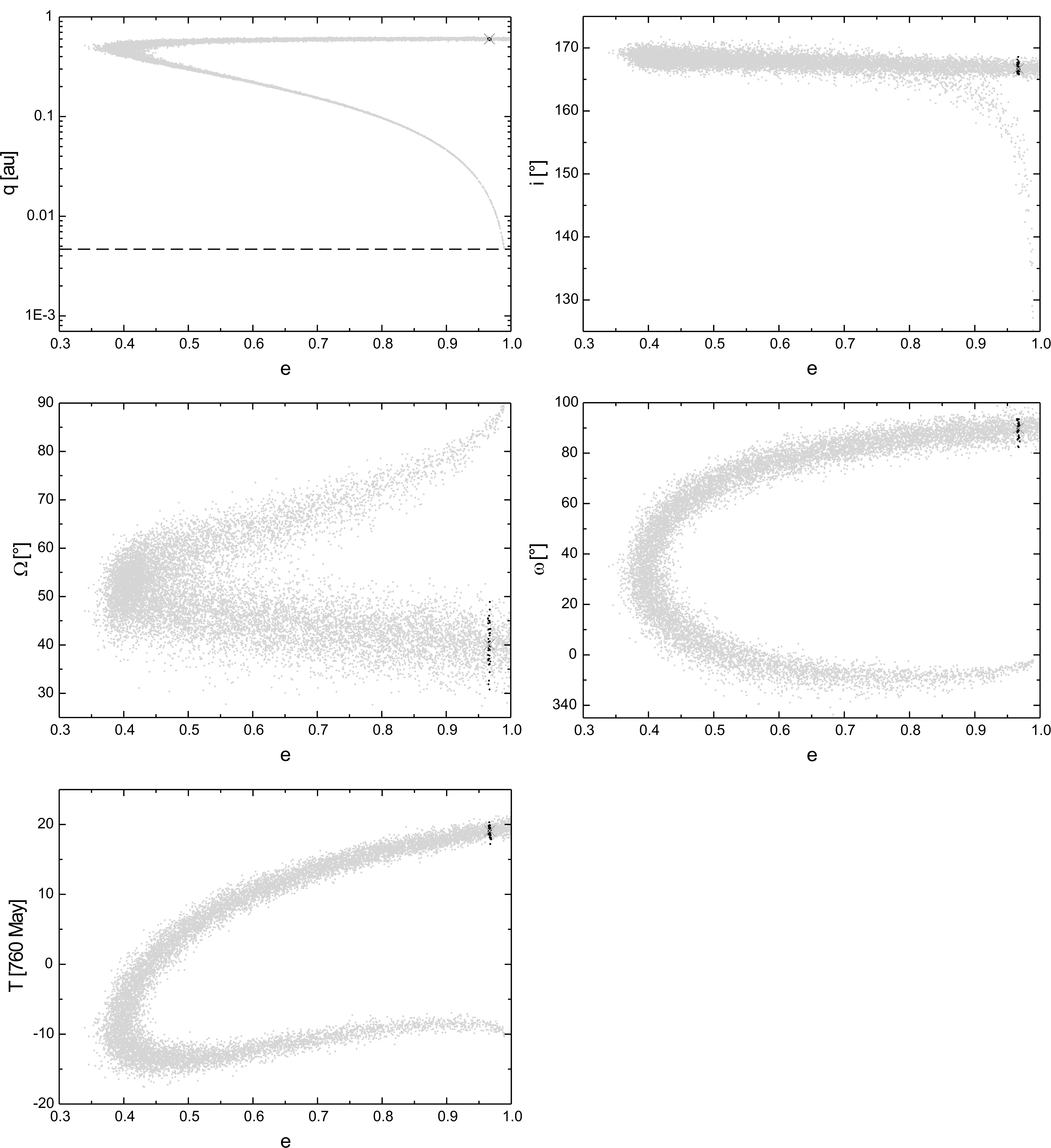}
\end{center}
\caption{{\bf Results from fitting the orbit:} 
{\rm Correlations between the Keplerian elements,
shown as light grey points, of all closed solutions found: 
perihelion distance $q$,
inclination $i$,
longitude of ascending node $\Omega$,
argument of perihelion $\omega$, and
perihelion time $T$ (from top left to bottom).
In the upper left, for $q$, we indicate the solar radius as dashed line.
Since most solutions for perihelion distance $q$ and inclination $i$ cluster
within small ranges, we can use these two to identify the comet of AD 760, see next two figures.
Having identified the comet as 1P/Halley (Fig. 12), we constrain the solution to
those with semi-major axes from 17-19 au as 1P/Halley -- 
these are shown here as dark points, and the best fit among them as cross.}
For eccentricity $e=1$, one can see the parameter range for non-periodic solutions.
}
\end{figure}

\begin{figure}[h]
\begin{center}
\includegraphics[angle=0,width=13cm]{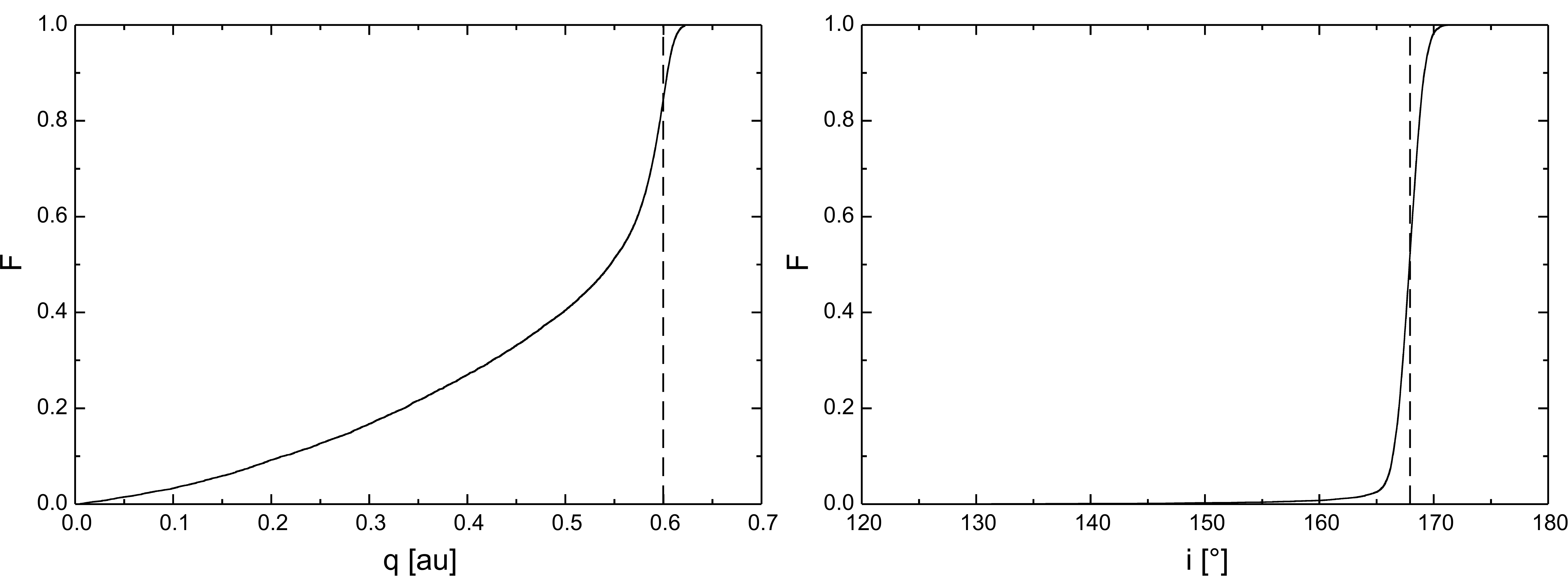}
\end{center}
\caption{
{\bf The cumulative distribution functions for q (left) and i (right panel)}
{\rm (all solutions as in the previous figure): The highest slopes (i.e. the peaks in the probability density distribution)
are indicated by vertical dashed lines and are located at $q \simeq 0.60$ au and $i \simeq 168^{\circ}$ (i.e. retrograde).
We use these values for identifying the comet (Fig. 12).}
}
\end{figure}

In Fig. 10, we show the orbital elements for all closed solutions found. Most solutions for
perihelion distance $q$ and inclination $i$ cluster within small ranges. The
cumulative distribution functions of the Keplerian elements q and i from all solutions are shown in Fig. 11, their
highest slopes (i.e. the peaks in the probability density distribution) are at $q \simeq 0.6$ au and
$i \simeq 168^{\circ}$ (i.e. retrograde).

We can now use these values for perihelion distance and inclination 
to clarify whether this combination of parameters is fulfilled by a known comet.
As we show in Fig. 12, there is no comet known other than 1P/Halley with similar values for $q$ and $i$ as our comet
of AD 760, so that we can thereby identify the comet of AD 760 as 1P/Halley,
see Fig. 12 caption for details.

Having identified the comet as 1P/Halley, we can constrain the solution to
those with periods or semi-major axes as 1P/Halley --
here conservatively to semi-major axes from 17-19 au.
The best-fit solution among them and many similar ones exhibit a set of orbital elements,
which is, within the uncertainties, the same as found for 1P/Halley, see Table 2. 

\begin{table}
\caption{{\bf Keplerian elements of our best fitting orbital solution}
{\rm for the AD 760 perihelion passage of comet 1P/Halley
(heliocentric ecliptic J2000.0) --
first our best solution, then the JPL orbits for AD 760 and 1986 for comparison,
than finally our parameters for non-periodic solutions. Our solutions are based on six
astrometric data points observed within a span of time of 50 days.
The best closed orbit has $\chi^2_{red}=0.09$, clearly indicating that the given
astrometric and timing uncertainties are overestimated.
}}
\begin{tabular}{lll|l|l}  \hline
Orbital parameters                & our new orbit               & JPL orbit (1)    & JPL orbit (2)   & non-periodic    \\ \hline
Epoch                             & 760 June 2.0                & 760 June 2.0     & 1986 perihelion & our solutions   \\ 
                                  & JD=1998800.5                & JD=1998800.5     & (JD=2449400.5)  &  for AD 760      \\ \hline
perihelion time            $T_{p}$  & 760 May $19.1 \pm 1.7$    & 760 May 20.671   & 1986 Feb 5.895317(5) & 760 May 18.1-21.6 \\
eccentricity               $e$      & $0.9667 \pm 0.0016$       & 0.96785          & 0.967142908(5)  & (e=1) \\
perihelion distance        $q$      & $0.60 \pm 0.02$\,au       & 0.58184\,au      & 0.58597812(9)\,au & 0.58-0.62 au  \\
inclination                $i$      & $166.7 \pm 2.2^\circ$     & $163.443^\circ$  & 162.262691(7)   & 165.2-168.2$^{\circ}$ \\
argument perihelion      $\omega$ & $89.9 \pm 8.5^\circ$        & $99.997^\circ$   & 111.33249(1)$^{\circ}$ & 84.7-95.8$^{\circ}$\\
longitude asc. node      $\Omega$ & $40.1 \pm 9.3^\circ$        & $44.687^\circ$   & 58.420081(9)$^{\circ}$ & 31.3-46.7$^{\circ}$ \\ \hline
period                    P        & $76.5 \pm 6.7$ yr (3)     & $77.0$ yr        & 75.3 yr & - \\ \hline
\end{tabular}

\vspace{.25cm}

{\rm Remarks: (1) based on YK81, but precessed to J2000.0 (Marsden \& Williams 2008),
as given on ssd.jpl.nasa.gov without error bars;
Yeomans \& Kiang (1981) gave error bars on perihelion times for 9 other perihelia
from AD 141 to 1301 as $\pm 0.05$ to $\pm 1.7$ days;
the perihelion time for AD 760 in Yeomans \& Kiang (1981) was fixed by historical observations,
but just those from Kiang (1972) without any revision;
Kiang (1972) fixed the perihelion time for AD 760 with two historical observations
(revised by us) and obtained AD 760 May 22.5. 
(2) https://ssd.jpl.nasa.gov (error of last digit in brackets).
(3) Calculated from q and e.}
\end{table}

\begin{figure}[ht]
\begin{center}
\includegraphics[angle=0,width=12cm]{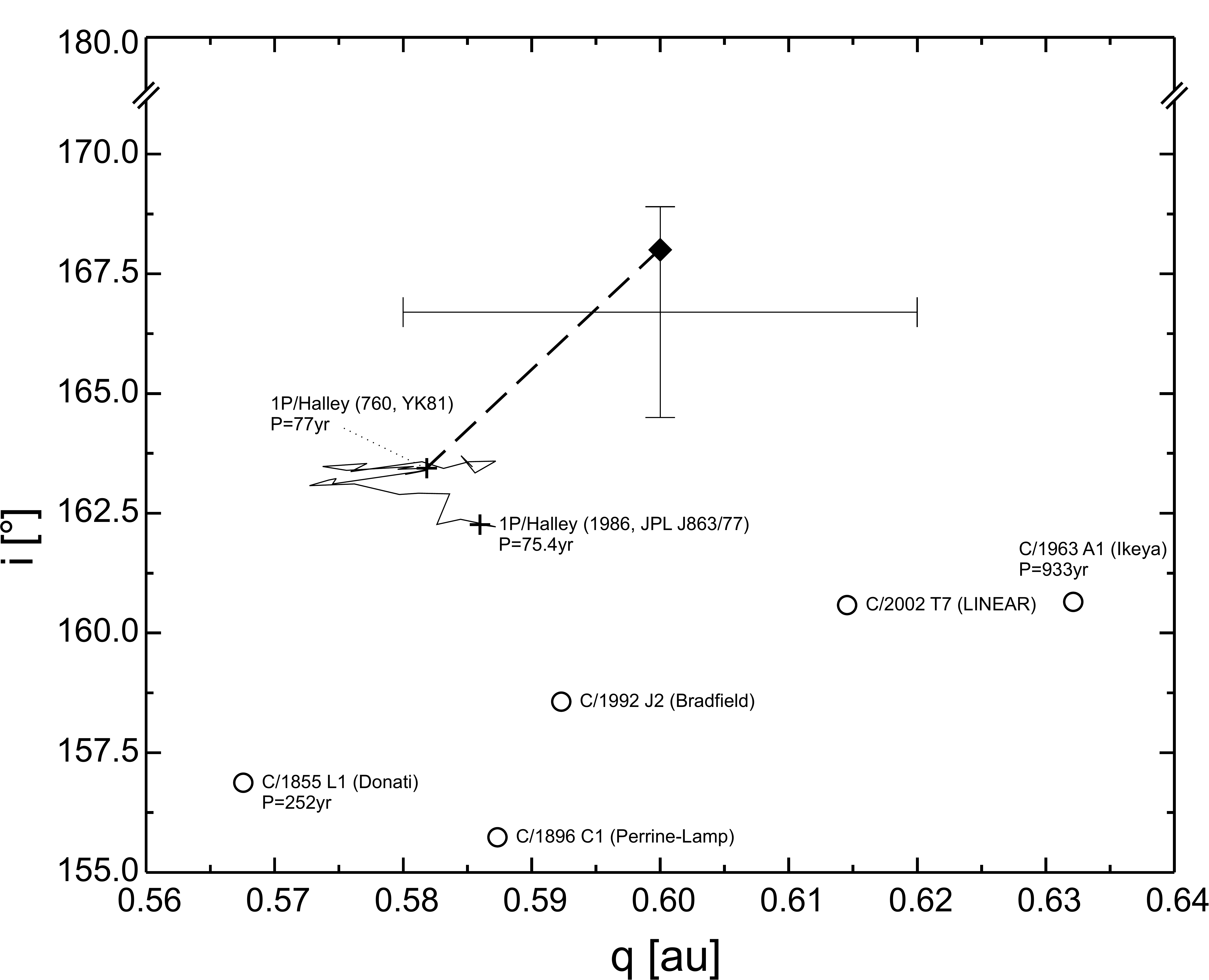}
\end{center}
\caption{
{\bf Perihelion distance q versus inclination i.}
{\rm the comet of AD 760 (as constrained to $q \simeq 0.6$ au and $i \simeq 168^{\circ}$, Fig. 11,
as diamond in the upper part) 
is compared to other presently known comets with values in the range of $i=155-180^{\circ}$ and $q=0.56-0.64$ au:
C/1855 L1 (Donati) has a very uncertain period of about 252 yr, it was observed only for 14 days,
C/1963 A1 (Ikeya) has a well-known orbit, but was not visible in AD 760;
the three other comets shown as circles in the lower center are unperiodic.
For C/1992 J2 (Bradfield) and C/1896 C1 (Perrine-Lamp) there is no evidence
that they have periods less than many thousands of years, the case is similar for C/2002 T7
(eccentricity $e=1.00048565(39)$, JPL 144, i.e. non-periodic).
There are no comets with $i=170-180^{\circ}$ in the plotted range for q.
For comet 1P/Halley, we show the data pair for its perihelion in AD 1986 (plus sign)
and all pairs from the orbital solutions in Yeomans \& Kiang (1981) connected by a line
(their data for perihelion AD 760 as plus), obtained from the JPL small-body database, 
precessed to 2000.0; the orbital elements from telescopic observations scatter more 
than those from extrapolation to pre-telescopic time.
In the upper part, we show $q$ and $i$ 
of the best fitting solution for 17-19 au semi-major axes for the comet of AD 760:
they are best consistent with comet 1P/Halley (YK81 did not specify error bars).
Therefore, the identification of the comet in AD 760 as comet 1P/Halley 
is justified -- for the first time for this perihelion only from historical data.}
}
\end{figure}

The residuals of this orbital solution are summarized in Table 3. The best fitting orbit is
shown in Figs. 3-8. We compare our new orbit with the one by YK81, which was, however, published without error bars.
All the orbital elements are consistent within our $1 \sigma$ uncertainties with YK81,
except the inclination (consistent within $1.5 \sigma$) -- but YK81 should have had similar if not larger error bars
as our's (less historical constraints). 

\begin{table}
\caption{{\bf Residuals (O-C) of the best fitting orbital solution} {\rm of comet 1P/Halley for its perihelion passage in AD 760
with their significances, listed in brackets, which take into account astrometric as well as timing uncertainties
(we give the modern names of the towns, where the comet was observed, see Table 1).}}
\begin{tabular}{llrr}  \hline
Location  & Dates 760       & $\Delta$ RA~~~~  ($\sigma$)~~ & $\Delta$ Dec~~~~  ($\sigma$)~~~\\ \hline
Xi'an        & May 16.86      & $-0.95\,^\circ~~(0.18)$       & $+1.75\,^\circ~~(0.05)$\\
Diyarbak\i r & May 18.00      & $-0.23\,^\circ~~(0.09)$       & $+0.62\,^\circ~~(0.20)$\\
Diyarbak\i r & May 22.00      & $+0.02\,^\circ~~(0.01)$       & $-0.13\,^\circ~~(0.14)$\\
Diyarbak\i r & Jun 1.00 & $-10.80\,^\circ~~(0.66)$      & $-1.19\,^\circ~~(0.08)$\\
Diyarbak\i r & Jun 3.74 & $-0.29\,^\circ~~(0.01)$       & $-3.28\,^\circ~~(0.12)$\\
Xi'an        & July 5.60 & $+0.01\,^\circ~~(0.01)$          & $+0.05\,^\circ~~(0.04)$\\ \hline
\end{tabular}
\end{table}

\begin{figure}[ht]
\begin{center}
\includegraphics[angle=0,width=12cm]{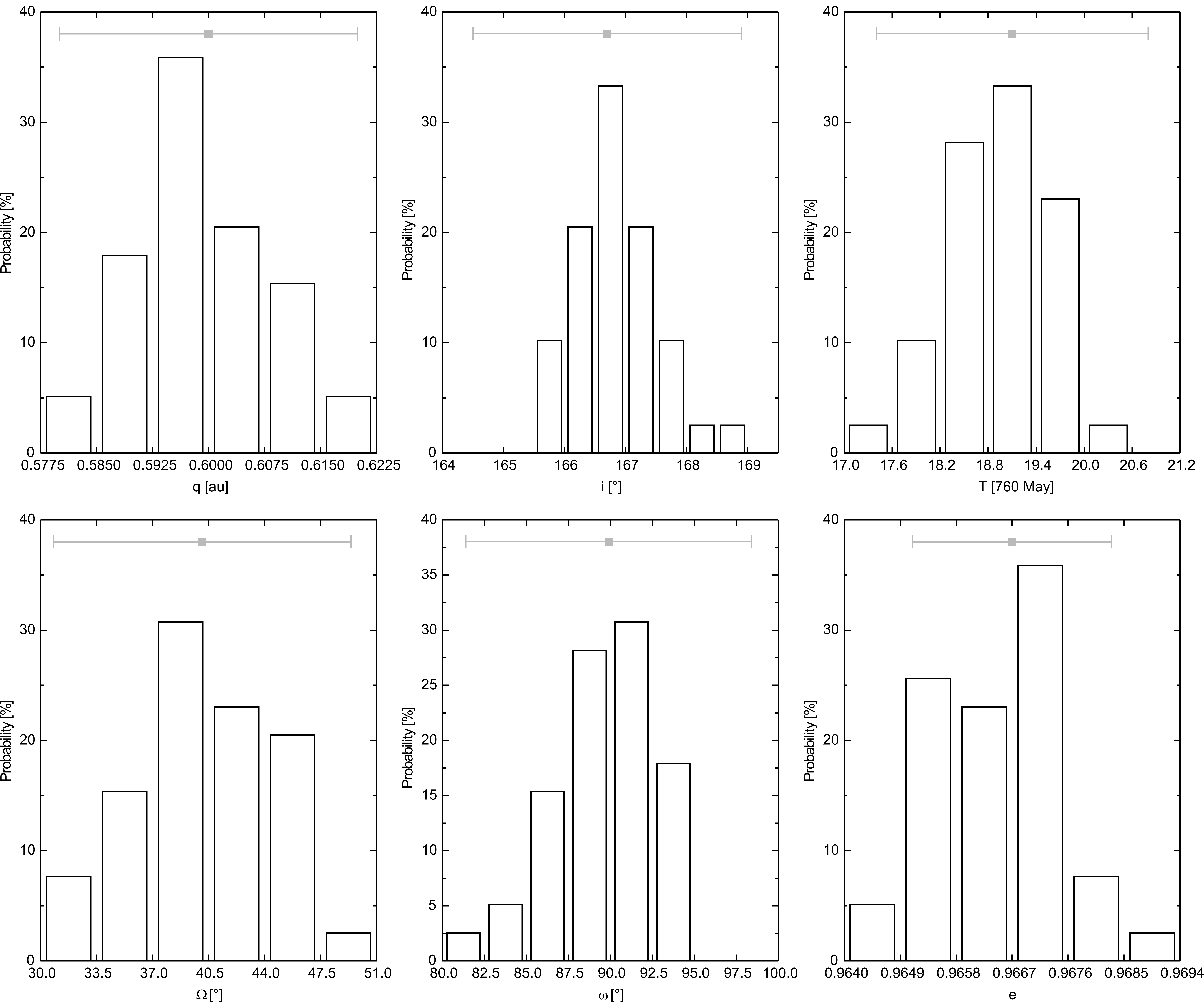}
\end{center}
\caption{{\bf Probability distribution of the Keplerian elements} 
{\rm as found in the semi-major axes range of 17-19 au.
Small grey squares with error bar indicate the best fitting solution (Table 2).
The distributions of the elements correspond well with the elements of the best fitting solution --
within their uncertainties.}
}
\end{figure}

For the three dates with best positions (AD 760 May 17/18 and 21/22, and July 5/6), the historically
reported positions agree to within less than $1^{\circ}$ with our orbital fit. Also for the other positions with large
uncertainties, the new orbit passes through them. The position in {\it Xuanyuan} is indeed on June 9, as considered. The new
orbit is well consistent with a position within $1^{\circ}$ (within $0.2 \sigma$) around $\beta$ Vir on July 5/6 (so that the
text can indeed be 7 {\it cun} instead of 7 {\it chi}).

According to our new orbital elements, 1P/Halley has an orbital inclination of around
$166.7^{\circ}$ in AD 760; it was observed while being around inferior conjunction with the Sun. 
The {\it Chronicle of Zuqn}\textit{\={\i}}{\it n} reports the last observation before (June 1.0, UT) 
and -- together with Michael the Syrian -- the first after (June 3.74 or 4.74, UT) comet-sun conjunction, 
and both are consistent only with our new orbit, while the conjunction 
is off by one day in the old (YK81) orbit --
the precise dates of the invisibility of the comet at (inferior) conjunction turns out to be the most
critical test of the orbital solutions (Fig. 14 lower panel).
In the AD 760 perihelion, the comet had its perihelion passage on May $19.1 \pm 1.7$ (UT), its
minimum solar elongation on AD 760 June 1.8 (UT) with
$19.1^{\circ}$ (unobserved the nights June 1/2 and 2/3), 
and then, on June 3.6 (UT), it had its closest encounter with Earth, namely 0.37 au.

Our new perihelion time (AD 760 May $19.1 \pm 1.7$) is consistent with backward extrapolated orbits,
e.g. perihelion on May 20.7 by YK81, May 20.7 by Landgraf (1986), and May 20.9 or 20.5 by Sitarski (1988),
his table 5, for constant and parabolically changing non-gravitational forces, respectively.
Offsets between the computed perihelion times (or conjunction with the Sun) and historically 
derived perihelion passages (or observed non-detection due to conjunction with the Sun) could be considered to be 
due to non-gravitational forces.
However, since our perihelion time has an uncertainty of $\pm 1.7$ day (probably similar on the YK81 orbit),
the difference is not yet significant;
a precise determination of the conjunction with the Sun (e.g. explicitly reported non-detection close
to the Sun as here in the Chronicle of {\it Zuqn}\textit{\={\i}}{\it n} and by Michael the Syrian for two nights)
can be considered as constraint on non-gravitational forces.

\smallskip

We have shown that it is possible to fit the orbit for this perihelion passage just with
historical data, i.e. without extrapolating backward from modern telescopic observations. Given that all our dated
positions as derived from historical texts are located quite centrally within the positional error boxes, our error
bars are overestimated, i.e. conservative. The so-called standard orbit derived by YK81 from an extrapolation of
telescopic data and then just fixed by problematic perihelion dates (first and last observation), 
is not inconsistent with our new, purely historically determined orbit, but there are also differences. 
We list here important results regarding the orbit:
\begin{itemize}
\item May 16/17 (C1): while the old orbit for this date is inside our reconstructed positional error
box, the previously derived position between the asterisms of {\it Lou} and {\it Wei} may be correct only by coincidence, 
as we interpreted {\it Lou} and {\it Wei} as lunar mansions.
(This is one of the two dated positions used by YK81 for fixing their orbit).
\item May 17/18 (Z1): a new relatively precise dated position -- the old orbit is consistent with it.
\item May 21/22 (Z2): a new very precise dated position, the position of the old orbit is outside of our error box.
\item May 31/June 1 (Z3): a new dated position -- the old orbit is outside of this error box and
located slightly below the horizon at the observational time derived by us
(the YK81 orbit has the sun-comet conjunction in this night).
\item June 3/4 or 4/5 (Z5): a new dated position, both consistent with the old (and new) orbit (Fig. 6).
\item June 9/10 (C4): a new relatively precise position; the Chinese record did not explicitly gave the
date for this position -- probably June 9 was meant, which is consistent with both the old and new orbit.
\item July $5/6 \pm 2$ (C5): 
our orbit fits the precise constraint from the most reliable Chinese text to be $0.7-1^{\circ}$ (``7 {\it cun}'') 
around $\beta$ Vir, while the old orbit is too far off $\beta$ Vir,  
and also not consistent with variant readings (Fig. 8). (This is one of the two dated position used by
YK81 to fix their orbit extrapolated backward from telescopic observations.)
\item Very slow motion since the beginning of July. 
\item Until June 27/28 or 28/29 (see Z5), the comet was located in the NW quarter since the start of
astronomical twilight or earlier as recorded in the Chronicle of {\it Zuqn}\textit{\={\i}}{\it n}.
\item Around July 5, the comet was in the SW quarter at the start of astronomical twilight, which is
probably reported in the Chronicle of {\it Zuqn}\textit{\={\i}}{\it n}.
\end{itemize}

\begin{figure}
\begin{center}
\includegraphics[angle=0,width=8cm]{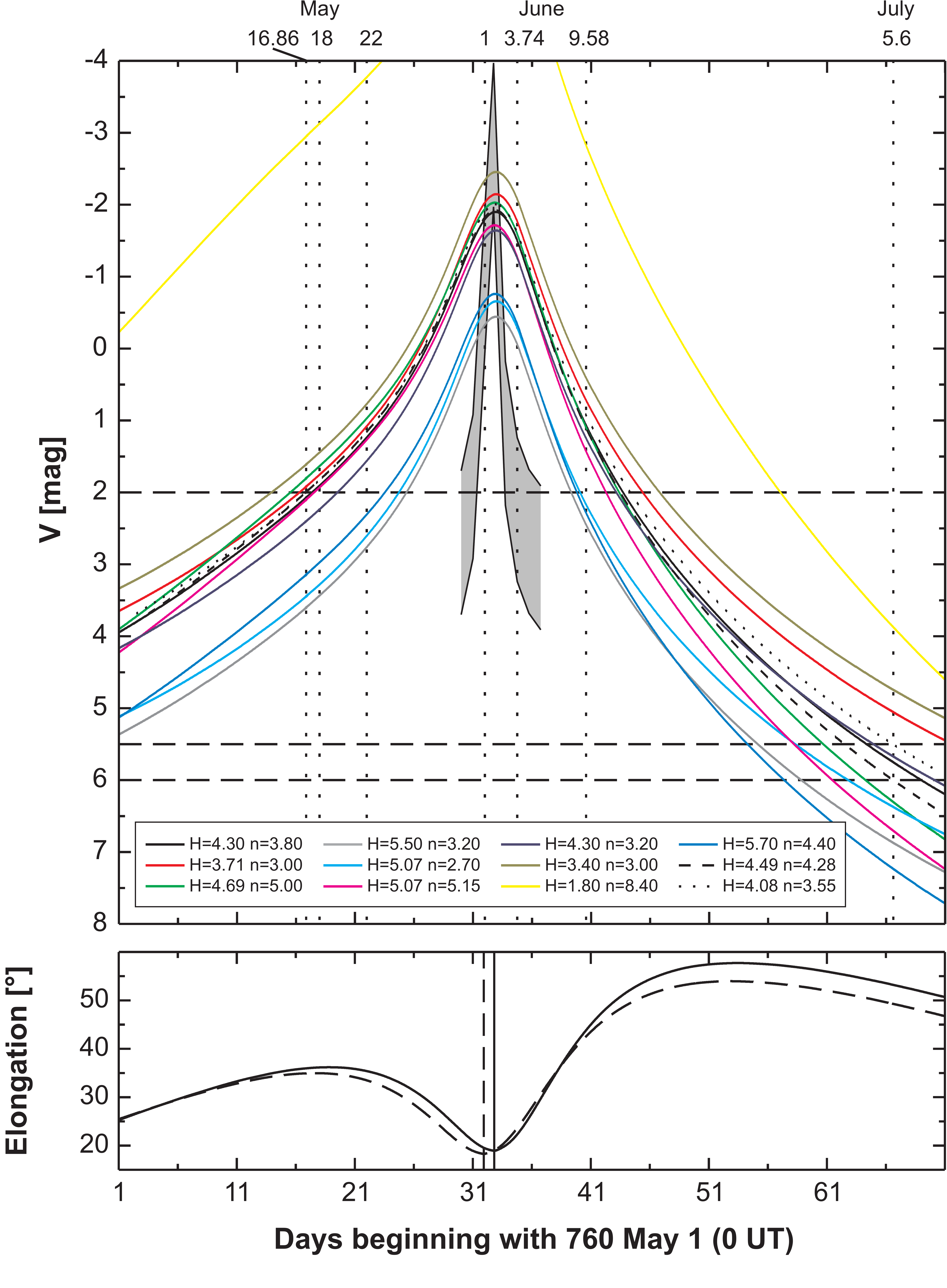}
\includegraphics[angle=0,width=8cm]{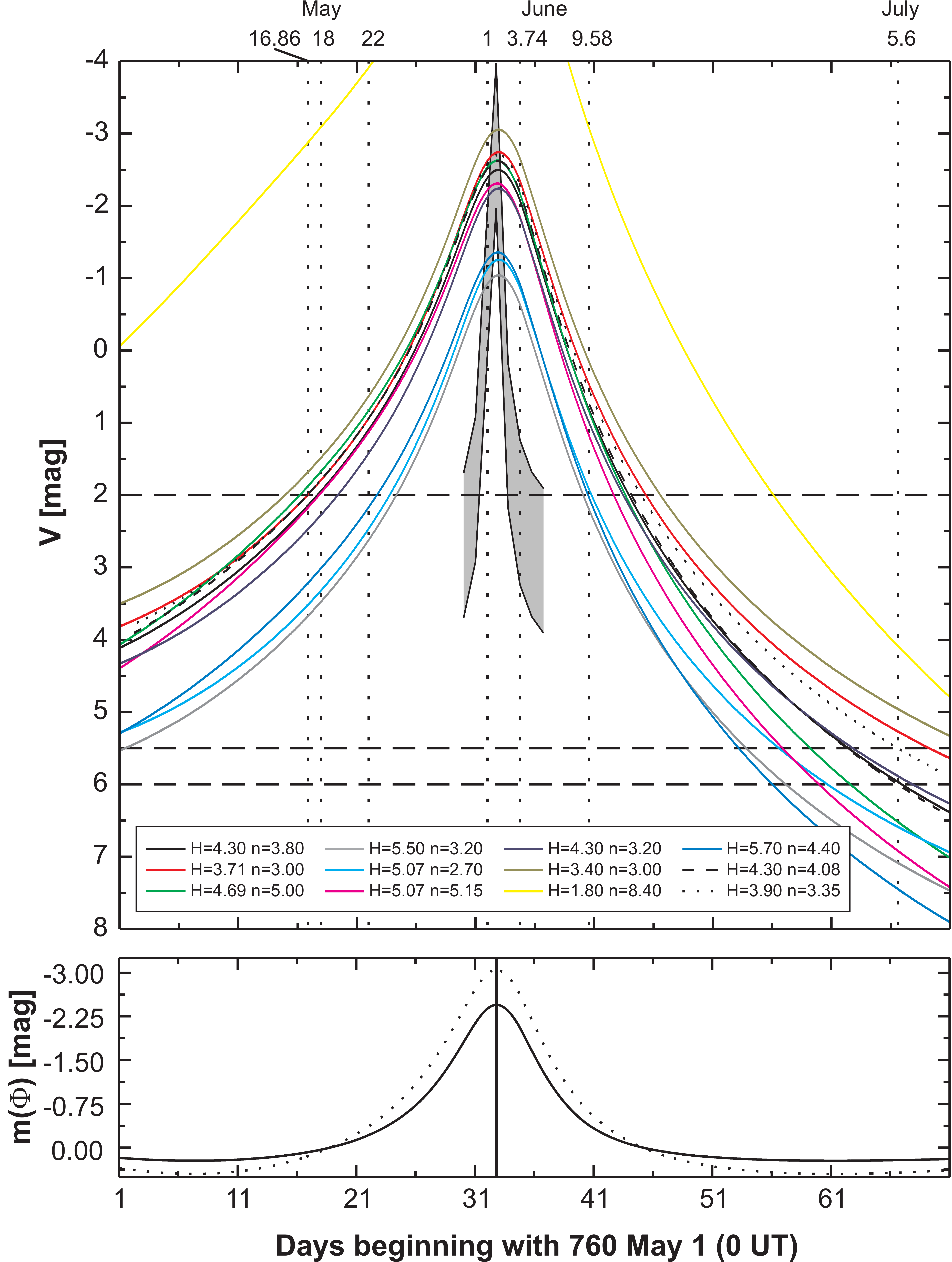}
\end{center}
\caption{{\bf Apparent brightness evolution for a usual comet, upper left panel}:
{\rm apparent brightness of comet 1P/Halley for our best fitting orbit
during the perihelion passage AD 760 using Equ. 1
for usual comets with $\delta_{90}=1$ (Marcus 2007a,b) in the phase function.
We plot apparent V-band magnitude m (in mag) versus date until mid July
(the difference between our new orbit and the old YK81/JPL orbit in terms of
estimated brightness is only $\le 0.1$ mag).
We plot the estimated magnitudes for all ten published parameter sets
of absolute magnitude $H$ and activity parameter $n$ listed in Sect. 5.
The comet was brightest around conjunction with the Sun early June.
The vertical dotted lines indicate the observing dates (Table 1).
The grey shaded area ($\pm 1 \sigma$ error range) around conjunction displays the estimated background sky brightness at the
location of the comet according to our new orbit -- consistent with non-detection in the
night June 1/2 (and probably also 2/3) as reported in the {\it Chronicle of Zuqn}\textit{\={\i}}{\it n}
and by Michael the Syrian.
Two (horizontal) dashed lines are for 2 and 6 mag, which are the likely apparent brightness for
first discovery ($\sim 2$ mag) and last detection ($\sim 6$ mag);
assuming these two values and that both H and n would be constant during from
first to last detection in AD 760, we fit $H$ and $n$ for this perihelion passage and
obtain $H=4.49$ mag and n=4.28 (bold dashed line) --
but if we assume 5.5 mag for the last detection (very close to $\beta$ Vir),
we would obtain $H=4.08$ mag and n=3.55 (dotted line).
The fit from Broughton (1979) assuming first detection at 3-4 mag is shown as
full pink line (one of the three bottom lines, H=5.7 mag, n=4.4).}
{\bf Apparent brightness evolution for a dusty comet, upper right panel}: 
{\rm light curve as in upper left, but for dusty comets with
$\delta_{90}=10$ in the phase function (Marcus 2007a,b).
Our best fit with 2 mag at discovery and 6 mag at last is now $H=4.30$ mag and n=4.08 (bold dashed line),
and $H=3.90$ mag and n=3.35 (dotted line) for 5.5 mag for the last detection.
The comet is now almost 1 mag brighter, even better consistent with detection until May 31/June 1
and non-detection for about the next two nights.}
{\bf Elongation of comet 1P/Halley, lower left panel}:
{\rm The full line is for our new orbit, the dashed line for the previous orbit (YK81) --
the previous orbit, where conjunction happened on AD 760 May 31.9 UT (dashed vertical line)
is inconsistent with the explicite last detection on June 1.0 UT,
as reported in the {\it Chronicle of Zuqn}\textit{\={\i}}{\it n},
but consistent with our new orbit, where conjunction is on June 1.8 UT (full vertical line).
The first detection of this comet
actually happened close to the maximal elongation ($\sim 36^{\circ}$) before perihelion passage.}
{\bf Brightening of comet 1P/Halley due to light scattering by dust, lower right panel}:
{\rm we plot the brightening $m(\Phi)$ (see Equ. 1) due to light scattering by dust 
for a usual comet ($\delta_{90}=1$, full line) and a dusty comet ($\delta_{90}=10$, dotted line).}
}
\end{figure}

Then, both for the old orbit (YK81, JPL ephemeris) and our new orbit, we can estimate the
apparent brightness $m$ of the comet following
\begin{equation}
m = H + 5~mag \cdot \log d + 2.5~mag \cdot n \cdot \log r + m(\Phi) 
\end{equation}
with absolute brightness $H$ in mag (defined as brightness $m$ as seen from the Sun at a distance of 1 au), 
distance $d$ in au between Earth and comet (well-known due to the orbit), 
activity parameter $n$ depending on cometary and solar activity (n=2 for pure reflection), 
distance $r$ in au between comet and Sun,
and $m(\Phi) = -2.5 mag \cdot \log \Phi$ as brightening due to light scattering on cometary dust with the phase function $\Phi$, 
which is a compound Henyey-Greenstein function with $g_{f}=0.9$, $g_{b}=-0.6$, $k=0.95$, 
and $\delta_{90}=1$ for {\it usual} comets, but $\delta_{90}=10$ for dusty comets (Marcus 2007a,b),
which becomes relatively large in our case due to the large phase angle of the comet (up to $\sim 149^{\circ}$, lower conjunction) 
resulting in significant forward scattering by cometary dust. When we reconstruct the light curve below, 
we will consider both cases, an usual and a dusty comet (Fig. 14).
For the absolute brightness $H$, values from 1.8 to 5.7 mag
have been estimated for 1P/Halley, and for the activity parameter $n$, values from 2.70 to 8.40 have been obtained:
\begin{itemize}
\item H=4.30 mag and n=3.80 (lightcurve. narod.ru/curves/0001p.html),
\item H=3.71 mag and n=3.00 (Kronk 1999),
\item H=4.69 mag and n=5.00 (Kronk 1999),
\item H=5.50 mag and n=3.20 (ssd.jpl.nasa.gov),
\item H=5.07 mag and n=2.70 (Yazdi 2014),
\item H=5.07 mag and n=5.15 (Yazdi 2014),
\item H=1.80 mag and n=8.40 for r=5.2 to 1.5 au for pre-perihelion
(Green \& Morris 1986, 1987),
\item H=4.30 mag and n=3.20 for r=1.5 to 0.6 au also for pre-perihelion
(Green \& Morris 1986, 1987),
\item H=3.40 mag and n=3.00 for r=0.6 to 2.9 au for post-perihelion
(Green \& Morris 1986, 1987),
\item H=5.7 mag and n=4.4 (Broughton 1979, fit to 7 best historical returns).
\end{itemize}
Note that the parameters obtained for the 1986 return do not necessarily apply to the 760 perihelion
and that the activity parameter $n$ can be different before and after perihelion as observed 1986
(Green \& Morris 1986, 1987).
Furthermore, 1P/Halley is known to have experienced major outbursts, e.g.
five years after the 1986 perihelion (West et al. 1991, Sekanina et al. 1992), and also in 1835
ten weeks after perihelion (Sekanina 2008).

For these ranges of parameters and the known orbit, we can estimate the
apparent brightness of 1P/Halley, see Fig. 14. It was brightest during conjunction with the Sun,
partly just due to forward scattering of light by cometary dust.

Except the highest and the three lowest light curve reconstructions in Fig. 14, all would indicate a brightness of
around 2 mag for May 16/17 and 17/18, the time when it was detected first 
in China and by the Chronicler of {\it Zuqn}\textit{\={\i}}{\it n}
(according to Michael the Syrian, the comet may have been discovered as early as May 13). There is no historical
supernova known, where the estimated brightness was fainter than about 1.5 to 2 mag when discovered by the naked eye
(Strom 1994, Clark \& Stephenson 1977). While comets are more readily detectable than novae or supernovae, because they
often display tails and often emerge first near the ecliptic, Broughton (1979) assumed that the first detection of a
comet would be possible at 3-4 mag, which may be very ambitious; Broughton's (1979) solution appears to be too faint
around both the first and last detection (Fig. 14). 

Once a new celestial object is discovered (or detected anew after conjunction with the Sun),
one can follow it until about 6 mag. Indeed,
most models consistently show that the comet was observed until it was as faint as
5-6 mag. If we assume that the comet was discovered at 2 mag and detected last when at either 5.5 or
6 mag, respectively, we can then obtain the best fit for the two unknown parameters in Equ.
1, and we obtain for this 
perihelion passage H=3.90-4.49 mag and n=3.35-4.28 for 2 mag at discovery and 5.5-6.0 mag at the last dated position. 
Please note that in principle the parameters H and n can change even during a time as short as
the period of visibility in AD 760 May-July (e.g. during the close approach to the Sun), so that they would have
different values before and after perihelion passage.

\smallskip

For the time around conjunction with the Sun (late May and early June), we also computed the sky brightness for
the location of the comet according to our new orbit (considering Moon and Sun) at May 30.0, 31.0, June 1.0, and June 2.0,
all UT (astronomical morning twilight for Diyarbak{\i}r)
and for June 1.74, 2.74, 3.74, 4.74, and 5.74, all UT (i.e. around 20:45h local time Diyarbak{\i}r) -- using the {\it Skycalc} code
(with the exact Julian dates as input, for $10^{\circ}$C temperature, $20\%$ humidity, Snellen ratio 1,
average experience, 50 years observer age, and a height of 675 meters as for Diyarbak{\i}r now;
the extinction coefficient is then 0.27 mag).
The uncertainties of the limiting visual magnitudes are then around $\pm 1$ mag.
As we see in Fig. 14, indeed, the limiting visual magnitude (or sky brightness) is fainter than most of the
comet brightness reconstructions for before June 1 and after June 3, so that the comet was detectable,
while in the nights June 1/2 and probably also 2/3, the comet was fainter than the limiting visual magnitude 
(hence, remains undetectable).
There is a good consistency between the observing reports on non-detection, our orbit, and the brightness reconstruction. 
The comet is detected on May 31/June 1 (at June 1.0) -- even though of the relatively bright sky -- only
when considering forward scattering by cometary dust: we can see in Fig. 14 left (usual comet) that the expected
comet brightness on May 31/June 1 (and also 2 or 3 nights later) is just about the sky brightness, 
while in Fig. 14 right (dusty comet), the comet is expected to be brighter than the sky -- this may possibly 
be seen as evidence that 1P/Halley was quite dusty around the AD 760 perihelion.
(This might have been partly facilitated by cometary and/or solar activity, the latter was high anyway around AD 760,
see Neuh\"auser \& Neuh\"auser 2015a,b.)

\smallskip

Since both the {\it Chronicle of Zuqn}\textit{\={\i}}{\it n} and the Chinese records reported the ``sign'' and
``tail'' to be ``white'', we deal with an observed dust tail. 
The Chinese reported two dust tail lengths: ``4 {\it chi}'' ($\sim 4^{\circ}$) for AD 760 May 17 and ``several
{\it zhang}'' (tens of degrees) most certainly for AD 760 June 9 (in the west,
in {\it Xuanyuan}). 
The drawing in the {\it Chronicle of Zuqn}\textit{\={\i}}{\it n} 
(Fig. 1) shows for around May 25 as impressive tail,
and the tilting is explicitely reported for the time before conjunction with the Sun.
The software Cartes du Ciel v3.10, as used in our figures, calculates and 
plots the {\it plasma} tail directed away from the Sun.

\section{Excursus: The precession constant as used implicitly}

We can consider whether and which observational technique and precessional shift was applied
by the author of the {\it Chronicle of Zuqn}\textit{\={\i}}{\it n} when he 
specified so precisely the ecliptic longitude of comet 1P/Halley on
May 22: ``while it was still in the same Aries at its edge/end/furthest part:
in/at the initial degree [of] the second [sign] (i.e. Taurus) from these wandering stars,
({\it kawkb\=e}) Kronos (Saturn) and Ares (Mars)'', i.e. at the same time in the
initial degree of the sign Taurus ($\lambda = 30^{\circ}$) and still in Aries (at its end). Indeed, for the correct
precessional shift at epoch 760.5, the stars at the end of Aries (33, 35, 39, and 41 Ari) had an ecliptic longitude of
about $30^{\circ}$;
also the fact that he mentioned Mars and Saturn in Aries on May 22 (Z2), correct to within $2^{\circ}$,
is consistent with a correct precessional shift.

Given that he knows well the three brightest stars of Aries
\footnote{$\alpha, \beta, \gamma$ Ari were also
listed as the three brightest stars in Aries in Ptolemy's Almagest, but $\alpha$ Ari was described to be
located outside the Aries constellation figure.
Yet, possibly following the Almagest, for the chronciler of
{\it Zuqn}\textit{\={\i}}{\it n}, these three (brightest) stars were particulary prominent.
See also footnotes 7 and 15.} 
and the other end of Aries and
that he specifies a location in degree within an ecliptic sign, he may have had some basic knowledge of Ptolemy's
Almagest (from the 2nd century): it describes in detail the setup of an armillary sphere, how to use it to measure
positions of stars, and gives positions with ecliptic coordinates for some 1000 stars. A copy of this work could have
been available in the monastery library in its original language Greek; there is no evidence that the author of the
{\it Chronicle of Zuqn}\textit{\={\i}}{\it n} could read Greek, and he did not even call the comet of AD 760 with the Greek word
``kometes'', as he did for some other comets reported by him before his own lifetime from
other sources (where he probably found in his sources that the Greek would call them
``kometes''). The author could speak Arabic (Harrak 1999), but the first Arabic translation
of the Almagest was probably undertaken later.\footnote{The Almagest was translated to Arabic first 
under Ya\d{h}y\=a b. Kh\=alid (AD 786-803 vizier of Caliph H\=ar\=un al-Rashid, died 805), 
probably around AD 791 (Sezgin 1978, p. 18),
then by or under Ab\=u \d{H}ass\=an and Salm (director of the Bayt al-\d{h}ikma, some kind of science and translation academy, 
under Caliph al-Ma'm\=un, reigned AD 813-833), both lost, and then by al-\d{H}ajj\=aj ibn Ma\d{t}ar (AD 786-833), 
all mentioned by Ibn al-Nadim; al-\d{H}ajj\=aj's version is extant as an almost complete 11th century copy, 
from which al-Kind\textit{\={\i}} cited the Almagest; and there are also two MSS with copies of an older version, 
namely by Is\d{h}\=aq b. \d{H}unayn (AD ca. 830-910) extant in Th\=abit b. Qurra's version, late 11th century
(Kunitzsch 1974, pp. 17-41).}

There was also an earlier Syriac translation of the Almagest (Kunitzsch 1974, p. 59).
E.g., the Syriac scholar Severus Sebokht of Nisibis (AD 575-667), a Christian bishop of Kennesrin, south of
Aleppo, Syria (see, e.g., Sezgin 1978, p. 111), has written books about the
constellations and the armillary sphere in Syriac (Nau 1910); in the former, he mentioned the Almagest as well as other
books by Ptolemy (Nau 1929-30), so that it may well be possible that he even translated the Almagest to Syriac
(or used an older version).

Even some 100 years before Severus Sebokht, there was already the work ``On the
Use and Construction of the Astrolabe'' by Johannes Philoponos (AD 490-575), who lived in Alexandria as a
monophysite like the monks of {\it Zuqn}\textit{\={\i}}{\it n}. Jacob of Edessa (AD 633-708), a student of Severus Sebokht and a known
source of the {\it Chronicle of Zuqn}\textit{\={\i}}{\it n}, has written in Syriac about Ptolemy's Almagest and works by Philoponos (Wilks
2008, p. 233). These works could have been available at {\it Zuqn}\textit{\={\i}}{\it n}. 
The monastery library was well known to scholars in the area with valuable books by, e.g., Eusebius
and Socrates, sources of parts I and II of the {\it Chronicle of Zuqn}\textit{\={\i}}{\it n} (Palmer \& Brock 1993, p. 70).

Since the Chronicler otherwise never
specifies any celestial position in degrees (nor any device or other observational tools), it appears unlikely that he
used an armillary sphere. Of course, we can hardly exclude that he got this observational report 
(and drawing?) with the precise
longitude measurement from some other source, e.g. a visiting astronomer, but the wording, vocabulary, and grammar in
the Halley story is all very typical for the author of the {\it Chronicle of Zuqn}\textit{\={\i}}{\it n}.

One can of course also specify the initial degree of Taurus without an armillary sphere: even
if one would not know the location of the start ($0^{\circ}$) of Taurus on sky, one can measure it easily (e.g. with a Jacob's staff)
when knowing the start of Aries, i.e. the location of the vernal equinox (Aries $0^{\circ}$) and that the zodiacal signs were
$30^{\circ}$ long. The fact that the ecliptic longitude of the stars in the end of Aries (33, 35, 39, 41 Ari) agrees with the
initial degree of Taurus for the correct precessional shift in AD 760 is still surprising; we can conclude that
knowledge of the current location of Aries $0^{\circ}$ and Taurus $0^{\circ}$ was 
available at {\it Zuqn}\textit{\={\i}}{\it n} at around AD 760. George, the
so-called Bishop of the Arabs (about AD 686-724), did write about precession in Syriac, namely that ``all fixed stars
fall back by $1^{\circ}$ in 100 years'' from Ptolemy's Almagest, where the limit from Hipparch was given (see Ryssel 1893,
p. 53-54 and Sezgin 1978, p. 112-114); this value could have been known to our Chronicler, but he did not adopt it.

In principle, one could also just take the ecliptic longitudes of the stars 33, 35, 39, and 41
Ari from Ptolemy's Almagest and then apply some precessional shift to the epoch of AD 760.5. To shift 33, 35, 39, and
41 Ari from Ptolemy's longitudes ($\lambda = 19^{\circ}10^{\prime}, 19^{\circ}40^{\prime}, 21^{\circ}20^{\prime}$, 
and $21^{\circ}40^{\prime}$, respectively) to $\lambda = 30^{\circ}$
from the epoch given by Ptolemy himself (``beginning of the reign of Antonius'', i.e. AD
137/8, see Toomer \& Ptolemy 1984, p. 340) to AD 760.5, i.e. $9.5 \pm 1.2^{\circ}$ in 623 yr with 33 Ari (9.1
$\pm 1.1^{\circ}$ in 623 yr without 33 Ari), one would need to apply a precessional constant of 
$1^{\circ}$ in 65.6 yr (or
$1^{\circ}$ in 68.5 yr without 33 Ari), close to today's best value ($1^{\circ}$ in 71.6 yr).

{\it Al-\d{S}\=uf}\textit{\={\i}} (AD 903-986) wrote in his ``Book on the Fixed Stars'' (in around
AD 964) about precession that ``the authors of al-Mumta\d{h}en tables and those who came after Ptolemy
confirmed it to be $1^{\circ}$ every 66 yr'' (translation to English in Hafez 2010, p. 86).
The star catalog {\it al-Mumta\d{h}en Zij} was composed
under Caliph al-Ma'm\=un (caliphate AD 813-833) by Yehy\=a b. Abi Mans\=ur (died AD 830 in Aleppo, today Syria). With a
value of $1^{\circ}$ in 66 yr, one would shift the longitudes of 33, 35, 39, and 41 Ari given by Ptolemy (adopted for AD
137/8) to 28.8 to $31.1^{\circ}$ by AD 760.5, consistent with the initial degree of Taurus, i.e.
about Taurus $0^{\circ}$. Hence, those rather precise values given in the 
{\it Chronicle of Zuqn}\textit{\={\i}}{\it n} confirm information from
{\it al-\d{S}\=uf}\textit{\={\i}},
namely that the value of $1^{\circ}$ in 66 yr was known since long ago (maybe from India, e.g. Sezgin 1978). 

\section{Summary and future perspective}

Our main results and conclusions are as follows:
\begin{itemize}
\item We have solved the orbit of the comet in AD 760 for the first time just with
dated positions from historical observations.
The application of text-critical methods for the understanding of historical sources
leads to substantial improvements -- as our test case shows.
\item We independently identify the comet of AD 760 as comet 1P/Halley;
the backward extrapolation by YK81 is not inconsistent with our new orbital solution from
historical observations alone -- e.g., the comet-sun-conjunction on May 31.9 (UT) on the YK81/JPL
orbit is not consistent with the detection on June 1.0 (UT), 
as reported in the {\it Chronicle of Zuqn}\textit{\={\i}}{\it n},
but it is consistent with our new orbit.
\item We re-visited the Chinese records in detail, revised the dated positions and found new ones;
the two dated positions used by YK81 to fix their orbit are problematic, the Chinese sources give a less precise
position at the beginning, and a more precise one at the end -- only our new orbit can fit all dated positions.
\item We obtained a precisely observed perihelion time (AD 760 May $19.1 \pm 1.7$) -- this could be used
for fixing further backward extrapolations of the comet orbit.
\item We argue that only one comet in AD 760 was transmitted in the Chinese sources -- not two, as considered in previous papers.
\item We studied the comet's brightness evolution and determined its absolute brightness and activity --
it was first detected at around 2nd mag, brightest at comet-sun-conjuntion, and lost at around 6th mag;
the observations are best consistent when including brightening due to light scattering on cometary dust.
\item We have used for the first time the detailed observations in the {\it Chronicle of Zuqn}\textit{\={\i}}{\it n}
for the orbit determination of this comet; we found that the position given 
in the {\it Chronicle of Zuqn}\textit{\={\i}}{\it n} for May 22 is not only
precise (within 1-2$^{\circ}$), but also implicitly uses a surprisingly accurate precession constant --
hence, this knowledge was (somehow) available. 
\item The positions and dates from China and {\it Zuqn}\textit{\={\i}}{\it n} are mutually consistent with each other and
with Keplerian motion, further sources from the Mediterranean and West Asian area were re-visited 
(e.g. Michael the Syrian) and support our two main sources -- therefore, the transmissions have high credibility.
A text in a Hadith collection by Nu$^{c}$aym ibn \d{H}amm\=ad may be based on 
the {\it Zuqn}\textit{\={\i}}{\it n Chronicle} -- hence, this Chronicle was not only buried.
\end{itemize}

There are further perspectives for future studies:
\begin{itemize}
\item With our improved methods, orbital elements of further periodic and non-periodic comets can 
be determined just from historical records (both from known and new sources);
and for comets, where not enough historical positions are available, it may be possible
to identify them by linking them to well-known comets. Then, also links to meteor showers can be revisited.
\item For comet 1P/Halley, it may well be possible to refine orbital elements from other historical
observations, e.g. AD 837 (in prep.), maybe even to quantify non-gravitational forces.
\item The multidisciplinary approach and our methods of very literal technical translation, source-
and text-critique, as well as close reading can yield improvements in the correct understanding of historical records
about celestial observation in general -- in particular for positions of other transient phenomena like supernovae and novae.
\end{itemize}


\smallskip

\noindent {\bf Acknowledgments.} We would like to thank
Peter Stein (U Jena) for his additional help with two details regarding the Syriac text. We acknowledge the Vatican
Library for providing digital scans of the {\it Chronicle of Zuqn}\textit{\={\i}}{\it n}. We are grateful to Jean
Bonnet-Bidaud (CEA Paris) for further information on the {\it Dunhuang} maps and to the late Paul Kunitzsch (LMU M\"unchen) 
for advice on the Syriac and Arabic versions of the Almagest.
We also acknowledge advice from Alden Mosshammer (UC San Diego) on the Easter dating problem in AD 760.
We used orbital data for comet 1P/Halley from JPL (ssd.jpl.nasa.gov), the software {\it find$\_$orb} for fitting a new orbit,
and John Thorstensen's Skycalc code (www.k3pgp.org/star.htm).
We are grateful for very competent reviewing of two anonymous referees.

\appendix

\section{Appendix}

In Fig. 15, we present the Syriac text from the {\it Chronicle of Zuqn}\textit{\={\i}}{\it n} in transliteration.
For the Syriac hand writing, see Fig. 1.
A translation to English is found in Sect. 2.

\begin{figure}
\includegraphics[angle=0,width=14cm]{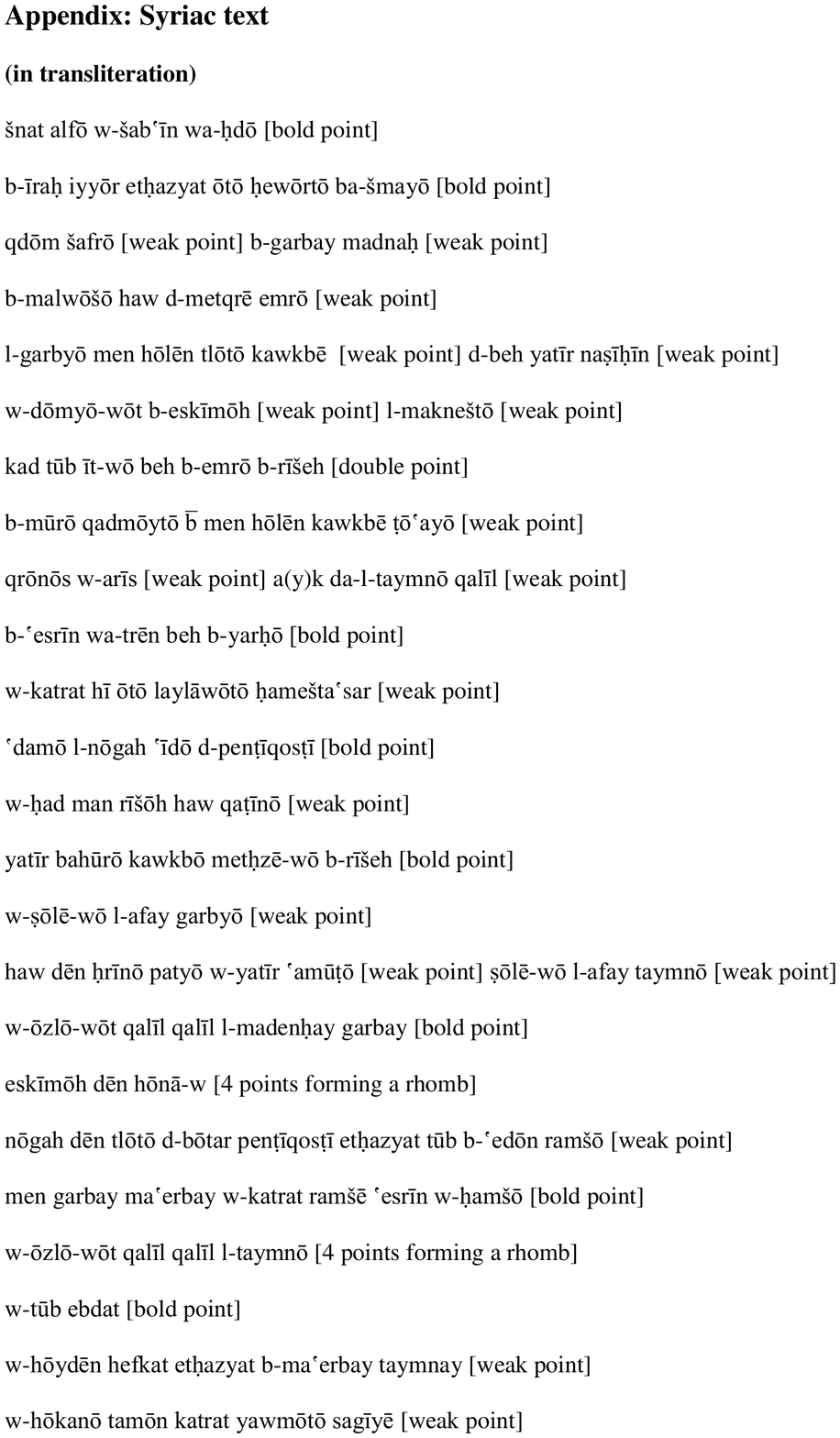}
\caption{{\bf Transliteration of the Syriac text on the comet of AD 760} 
{\rm (see also Fig. 1 and Sect. 2).}}
\end{figure}

In Fig. 16, we present the Chinese texts.
Translation to English with partial transcription are found in Sect. 3.

\begin{figure}
\includegraphics[angle=0,width=16cm]{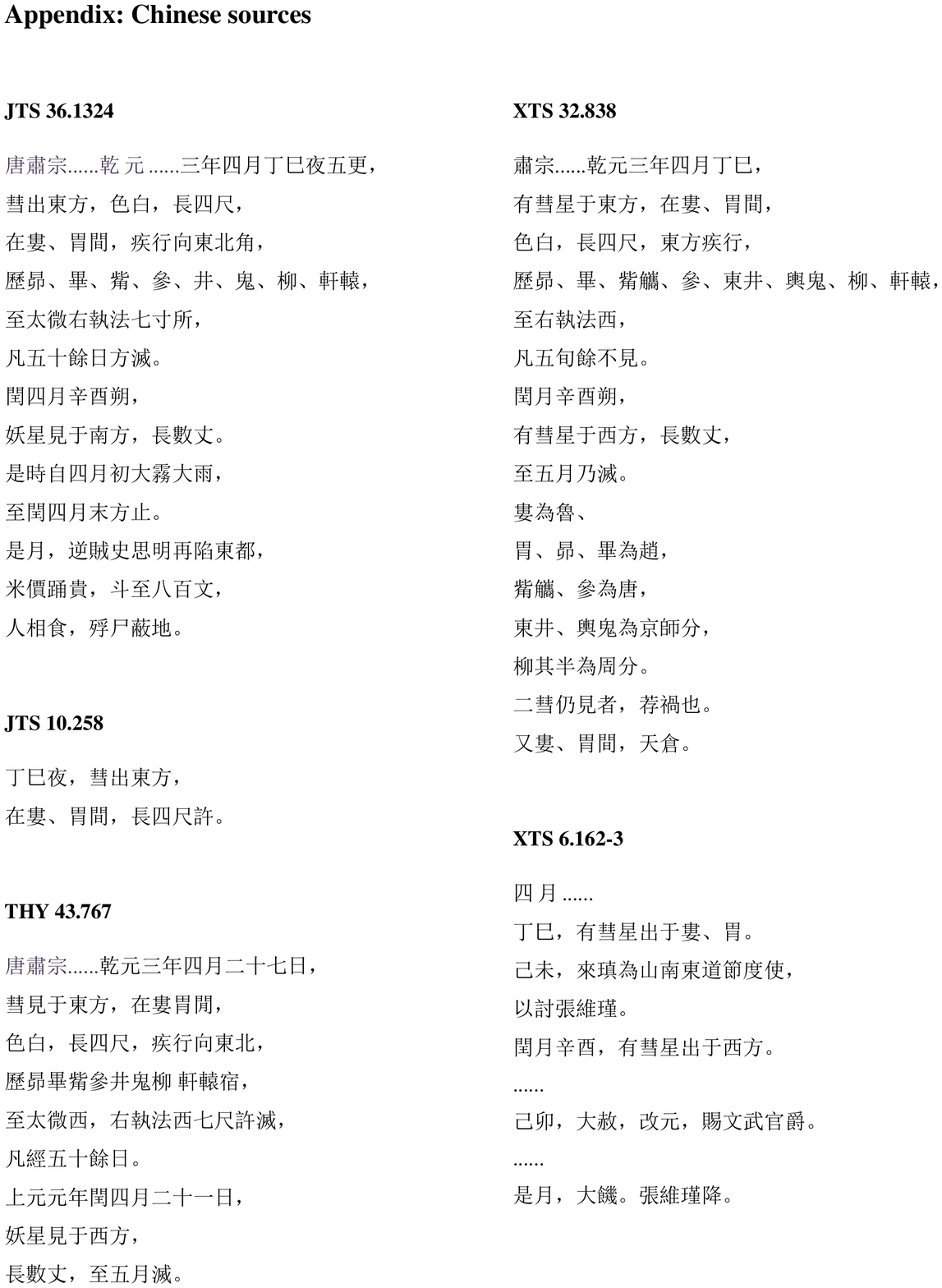}
\caption{{\bf The Chinese texts from JTS, THY, and XTS on the comet of AD 760} {\rm (see Sect. 3).}}
\end{figure}

\end{document}